\documentclass[12pt,titlepage]{article}
\usepackage{axodraw,epsfig}

\def\_{\rule{.3em}{.15ex}}

\newcommand{\be}{\begin{equation}}
\newcommand{\ee}{\end{equation}}
\newcommand{\bea}{\begin{eqnarray}}
\newcommand{\eea}{\end{eqnarray}}
\newcommand{\f}{\frac}
\def\slash#1{\setbox0=\hbox{$#1$}#1\hskip-\wd0\dimen0=5pt\advance
       \dimen0 by-\ht0\advance\dimen0 by\dp0\lower0.5\dimen0\hbox
         to\wd0{\hss\sl/\/\hss}}

\def\ra{\rightarrow}
\def\epsk{\mbox{$\epsilon_K$~}}
\def\kk{\mbox{$\bar{K}^0K^0$~}}
\def\bb{\mbox{$\bar{B}B$~}}
\def\bbd{\mbox{$\bar{B}_dB_d$~}}
\def\bbs{\mbox{$\bar{B}_sB_s$~}}
\def\bsg{\mbox{$b\ra s\gamma$~}}
\def\TeV{\mathrm{TeV}}
\def\GeV{\mathrm{GeV}}
\begin{document}

\begin{titlepage}

 \begin{flushright}
  {\bf IFT/99/10\\
       hep-ph/9906206\\
       June 1999}
 \end{flushright}

 \begin{center}
  \vspace{2cm}

\setlength {\baselineskip}{0.3in}
{\large \bf Constraints on Phases of Supersymmetric Flavour Conserving
Couplings}
\vspace{10mm} \\
\setlength {\baselineskip}{0.2in}

{S. Pokorski$^{a}$, J. Rosiek$^a$ and C. A. Savoy$^b$}\\

\vspace{5mm}
{\it $^a$Institute of Theoretical Physics, Warsaw University,\\ Ho{\.z}a 69,
00-681 Warsaw, Poland\\}
\vspace{5mm}
{\it $^b$Service de Physique Theorique, CEA Saclay,\\
91191 Gif-sur-Yvette CEDEX, France}

\vspace{2cm} 
{\bf Abstract} \\ 
\end{center} 
{\small In the unconstrained MSSM, we reanalyze the constraints on the
phases of supersymmetric flavour conserving couplings that follow from
the electron and neutron electric dipole moments (EDM). We find that
the constraints become weak if at least one exchanged superpartner
mass is $>{\cal O}(1~\TeV)$ or if we accept large cancellations
among different contributions. However, such cancellations have no
evident underlying symmetry principle. For light superpartners, models
with small phases look like the easiest solution to the experimental
EDM constraints. This conclusion becomes stronger the larger is the
value of $\tan\beta$. We discuss also the dependence of $\epsilon_K$,
$\Delta m_B$ and \bsg decay on those phases.}  \vfill

\setlength {\baselineskip}{14pt}
\noindent \underline{\hspace{5cm}}\\ {\footnotesize This work was
  supported in part by the Polish Committee for Scientific Research
  under the grant numbers 2~P03B~052~16 and 2~P03B~030~14 (J.R.) and
  by French-Polish Programme POLONIUM.}

\end{titlepage} 

\setlength{\baselineskip}{18pt}

\section{Introduction}

In the Minimal Supersymmetric Standard Model (MSSM) there are new
potential sources of the CP non-conservation effects. One can
distinguish two categories of such sources.  One is independent of the
physics of flavour non-conservation in the neutral current sector and
the other is closely related to it. To the first category belong the
phases of the parameters $\mu$, gaugino masses $M_i$, trilinear scalar
couplings $A_i$ and $m_{12}^2$, which can in principle be
arbitrary. They can be present even if the sfermion sector is flavour
conserving. Not all of them are physically independent.

The other potential phases may appear in flavour off-diagonal sfermion
mass matrix elements $\Delta m^2_{ij}$ and in flavour off-diagonal LR
mixing parameters $A_{ij}$. These potential new sources of CP
violation are, therefore, closely linked to the physics of flavour
and, for instance, vanish in the limit of flavour diagonal (in the
basis where quarks are diagonal) sfermion mass matrices. It is,
therefore, quite likely that the two categories of the potential CP
violation in the MSSM are controlled by different physical
mechanisms. They should be clearly distinguished and discussed
independently.

Experimental constraints on the ``flavour-conserving'' phases come
mainly from the electric dipole moments of electron~\cite{EDM_E_EXP}
and neutron~\cite{EDM_N_EXP}\footnote{Status of the new experimental
limit on the neutron EDM is still under discussion~\cite{EDM_N_DISC}.}:
\bea
E_e^{exp}<4.3\cdot 10^{-27} e\cdot cm\nonumber\\
E_n^{exp}<6.3\cdot 10^{-26} e\cdot cm\nonumber
\label{eq:edm_en_exp}
\eea

The common belief was that the constraints from the electron and
neutron electric dipole moments are strong~\cite{STRONGCP1,STRONGCP2}
and the new phases must be very small. More recent calculations
performed in the framework of the minimal supergravity
model~\cite{OLIVE,NATH,BARTL} indicated the possibility of
cancellations between contributions proportional to the phase of $\mu$
and those proportional to the phase of $A$ and, therefore, of weaker
limits on the phases in some non-negligible range of parameter space.
The possibility of even more cancellations have been reported in
ref.~\cite{KANE} in non-minimal models. For instance, for the electron
dipole moment, the coefficient of the $\mu$ phase has been found to
vanish for some values of parameters. Since the constraints on the
supersymmetric sources of $CP$ violating phases are of considerable
theoretical and phenomenological interest, in this paper we reanalyze
the electric dipole moments with the emphasis on complete
understanding of the mechanism of the cancellations.

The new flavour-conserving phases in the MSSM, beyond the
$\theta_{QCD}$ present in the SM, may appear in the bilinear term in
the superpotential and in the soft breaking terms: gaugino masses and
bi- and trilinear scalar couplings - see
eqs.~(\ref{eq:superpot-cp},\ref{eq:lsoft-cp}). We define them as:
\bea
e^{i\phi_{\mu}} = \frac{\mu}{|\mu|}
\hskip 15mm
e^{i\phi_i} = \frac{M_i}{|M_i|}
\hskip 15mm
e^{i\phi_{A_I}} = \frac{A_I}{|A_I|}
\hskip 15mm
e^{i\phi_H} = \frac{m_{12}^2}{|m_{12}^2|}
\label{eq:zpmphi}
\eea
Not all of those phases are physical. In the absence of
terms~(\ref{eq:superpot-cp},\ref{eq:lsoft-cp}) the MSSM Lagrangian has
two global $U(1)$ symmetries, an $R$ symmetry and the Peccei-Quinn
symmetry~\cite{DUGAN}. Terms~(\ref{eq:superpot-cp},\ref{eq:lsoft-cp})
may be treated as spurions breaking those symmetries, with appropriate
charge assignments.
Physics observables depend only on the phases of parameter
combinations neutral under both $U(1)$'s transformation. Such
combinations are:
\bea
M_i \mu (m^2_{12})^{\star} \hskip 2cm A_I \mu
(m^2_{12})^{\star} \hskip 2cm A_I^{\star} M_i
\label{eq:phasinv}  
\eea
Not all of them are independent. The two $U(1)$ symmetries may be used
to get rid of two phases. We follow the common choice and keep
$m_{12}^2$ real in order to have real tree level Higgs field VEV's and
$\tan\beta$\footnote{Loop corrections to the effective potential
  induce phases in VEV's even if they were absent at the tree level.
  Rotating them away reintroduces a phase into the $m_{12}^2$
  parameter.}. The second re-phasing may be used e.g. to make one of
the gaugino mass terms real - we choose it to be the gluino mass term.

A particularly simple picture is obtained assuming universal gaugino
masses and universal trilinear couplings $A_I = A$ at the GUT scale.
In this case $U(1)$'s invariant parameter
combinations~(\ref{eq:phasinv}) contain at that scale only two
independent phases. Defining $\phi_g\equiv\phi_1=\phi_2=\phi_3$,
$\phi_A\equiv\phi_{A_i}$ we can write them down as
\bea
\delta_1 = \phi_A -\phi_g, \hskip 3cm 
\delta_2 = \phi_g + \phi_{\mu} - \mathrm{arg} (m_{12}^2)
\eea
The re-phasing freedom may be used in this case to make all $M_i$
simultaneously real. The RGE for $M_i$ at one loop does not introduce
phases once they are set to zero at GUT scale. The second $U(1)$
rotation can be used again to remove phase from $m_{12}^2$ already at
$M_Z$ scale. Then only $\mu$ and $A_I$ parameters remain complex at
electroweak scale. Phases of various $A_I$ parameters are not
independent and can be calculated from the RGE equations. 

In most of the calculations in the next sections we keep in general
$\mu, M_1, M_2$ and $A_I$ complex. As one can see, all the physical
results depend explicitly only on the phases of parameter
combinations~(\ref{eq:phasinv}), as follows from the general
considerations above.

In Section 2 we discuss in detail the electron electric dipole
moment. First, we present the results of an exact calculation, which
is convenient for numerical codes. For a better qualitative
understanding, we also perform the calculation in the mass insertion
approximation. The results of the two methods can be compared by
appropriately expanding the exact results for some special
configurations of the selectron and gaugino masses.  After those
technical preliminaries we discuss in Section 2 the magnitude of
various contributions to the electric dipole moment and investigate
the pattern of possible cancellations. The first observation we want
to emphasize is that, even without any cancellations, there are
interesting regions in the parameter space where the phases are weakly
constrained. Secondly, we do not find any symmetry principle that
would guarantee cancellations in the regions where the phases are
constrained. Such cancellations are, nevertheless, possible by proper
tuning of the $\mu$ and $A$ phases to the values of the soft masses.

In Section 3 we analyze the neutron electric dipole moment with
similar conclusions. In Section 4 we discuss the role of the $\mu$
phase in the $\epsilon_K$ measurement and in the \bsg decay. The
necessary conventions, Feynman rules and integrals are collected in
the Appendices.

\section{Electric  dipole moment of the electron}
\label{sec:electron}

\subsection{Mass eigenstate vs. mass insertion calculation}

The electric dipole moments (EDM) of leptons and quarks, defined as
the coefficient $E$ of the operator
\be
{\cal L}_E = - \frac{i}{2} E \bar{\psi} \sigma_{\mu\nu} \gamma_5 \psi
F^{\mu\nu},
\label{eq:edmdef}
\ee
can be generated in the MSSM already at 1-loop level, assuming that
supersymmetric parameters are complex.

\begin{figure}[htbp]
\begin{center}
\begin{tabular}{lr}
\begin{picture}(150,120)(0,0)
\ArrowLine(10,20)(30,20)
\Text(10,10)[]{\mbox{$e^I$}}
\Vertex(30,20){2}
\DashArrowLine(30,20)(90,20){7}
\Text(60,10)[]{\mbox{$\tilde{\nu}_K$}}
\Vertex(90,20){2}
\ArrowLine(90,20)(110,20)
\Text(110,10)[]{\mbox{$e^I$}}
\ArrowLine(30,20)(60,80)
\Text(37,50)[r]{\mbox{$(C_j^+)^c$}}
\ArrowLine(60,80)(90,20)
\Text(85,50)[l]{\mbox{$(C_j^+)^c$}}
\Photon(60,80)(60,110){3}{3}
\Text(50,110)[c]{\mbox{$\gamma$}}
\end{picture}
&
\begin{picture}(150,120)(0,0)
\ArrowLine(10,20)(30,20)
\Text(10,10)[]{\mbox{$e^I$}}
\Vertex(30,20){2}
\ArrowLine(30,20)(90,20)
\Text(60,10)[]{\mbox{$N^0_j$}}
\Vertex(90,20){2}
\ArrowLine(90,20)(110,20)
\Text(110,10)[]{\mbox{$e^I$}}
\DashArrowLine(30,20)(60,80){7}
\Text(37,50)[r]{\mbox{$\tilde{L}_k^-$}}
\DashArrowLine(60,80)(90,20){7}
\Text(85,50)[l]{\mbox{$\tilde{L}_k^-$}}
\Photon(60,80)(60,110){3}{3}
\Text(50,110)[c]{\mbox{$\gamma$}}
\end{picture}
\end{tabular}
\caption{Diagrams contributing to lepton EDM.}
\label{fig:edm_l}
\end{center}
\end{figure}
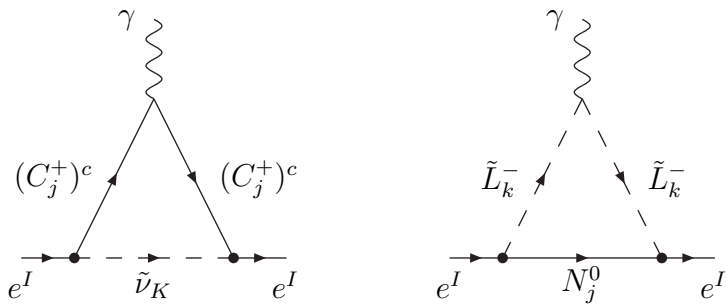

In the mass eigenstate basis for all particles, two diagrams
contribute to the electron electric dipole moment. They are shown in
Fig.~\ref{fig:edm_l} (summation over all charginos, neutralinos,
sleptons and sneutrinos in the loops is understood).
The result for the lepton electric dipole moment reads:

\bea
E_l^I &=& \frac{e m_l^I}{8\pi^2}\sum_{j=1}^2\sum_{K=1}^3
m_{C_j}\mathrm{
Im}\left((V_{l\tilde{\nu}C})_L^{IKj}(V_{l\tilde{\nu}C})_R^{IKj\star}\right)
C_{11}(m_{C_j}^2,m_{\tilde{\nu}_K}^2)\nonumber\\
&-& \frac{e m_l^I}{16\pi^2}\sum_{j=1}^4\sum_{k=1}^6
m_{N_j}\mathrm{
Im}\left((V_{l\tilde{L}N})_L^{Ikj}(V_{l\tilde{L}N})_R^{Ikj\star}\right)
C_{12}(m_{\tilde{L}_k}^2,m_{N_j}^2)
\label{eq:edml_full}
\eea
where $(V_{l\tilde{\nu}C})_L$, $(V_{l\tilde{\nu}C})_R$,
$(V_{l\tilde{L}N})_L$, $(V_{l\tilde{L}N})_R$ are, respectively, the
left- and right- electron-sneutrino-chargino and
electron-selectron-neutralino vertices and $C_{11},C_{12}$ are the
loop integrals. Explicit form of the vertices and integrals can be
found in~\ref{app:lagr}.

The eq.~(\ref{eq:edml_full}) is completely general, but as we discussed
already in the Introduction, in the rest of this paper we assume no
flavour mixing in the slepton sector.  Therefore, in the formulae
below we skip the slepton flavour indices.  

We present now the calculation of the electron EDM in the mass
insertion approximation, for easier understanding of cancellations of
various contributions, and then compare the two results.  We use the
``generalized mass insertion approximation'', i.e. we treat as mass
insertions both the L-R mixing terms in the squark mass mixing
matrices and the off-diagonal terms in the chargino and neutralino
mass matrices. Therefore we assume that the diagonal entries in the
latter: $|\mu|$, $|M_1|$, $|M_2|$ are sufficiently larger than the
off-diagonal entries, which are of the order of $M_Z$.

\begin{figure}[htbp]
\begin{center}
\begin{tabular}{lr}
\begin{picture}(150,120)(0,0)
\ArrowLine(10,20)(30,20)
\Text(10,10)[]{\mbox{$e$}}
\Vertex(30,20){2}
\DashArrowLine(30,20)(90,20){7}
\Text(60,10)[]{\mbox{$\tilde{\nu}_0$}}
\Vertex(90,20){2}
\ArrowLine(90,20)(110,20)
\Text(110,10)[]{\mbox{$e$}}
\ArrowLine(30,20)(45,50)
\ArrowLine(45,50)(60,80)
\Text(35,40)[r]{\mbox{$\tilde{W}^-$}}
\Line(42,47)(48,53)
\Line(48,47)(42,53)
\Text(45,70)[r]{\mbox{$\tilde{h}^-$}}
\ArrowLine(60,80)(90,20)
\Text(85,55)[l]{\mbox{$\tilde{h}^-$}}
\Photon(60,80)(60,110){3}{3}
\Text(50,110)[c]{\mbox{$\gamma$}}
\end{picture}
&
\begin{picture}(150,120)(0,0)
\ArrowLine(10,20)(30,20)
\Text(10,10)[]{\mbox{$e$}}
\Vertex(30,20){2}
\DashArrowLine(30,20)(90,20){7}
\Text(60,10)[]{\mbox{$\tilde{\nu}_0$}}
\Vertex(90,20){2}
\ArrowLine(90,20)(110,20)
\Text(110,10)[]{\mbox{$e$}}
\ArrowLine(30,20)(60,80)
\Text(40,55)[r]{\mbox{$\tilde{W}^-$}}
\ArrowLine(60,80)(75,50)
\ArrowLine(75,50)(90,20)
\Text(75,70)[l]{\mbox{$\tilde{W}^-$}}
\Line(72,47)(78,53)
\Line(78,47)(72,53)
\Text(90,40)[l]{\mbox{$\tilde{h}^-$}}
\Photon(60,80)(60,110){3}{3}
\Text(50,110)[c]{\mbox{$\gamma$}}
\end{picture}
\\
&\\
\begin{picture}(150,120)(0,0)
\ArrowLine(10,20)(30,20)
\Text(10,10)[]{\mbox{$e$}}
\Vertex(30,20){2}
\DashArrowLine(30,20)(90,20){7}
\Text(60,10)[]{\mbox{$\tilde{\nu}_0$}}
\Vertex(90,20){2}
\ArrowLine(90,20)(110,20)
\Text(110,10)[]{\mbox{$e$}}
\ArrowLine(30,20)(45,50)
\ArrowLine(45,50)(60,80)
\Text(30,40)[r]{\mbox{$\tilde{h}^-$}}
\Line(42,47)(48,53)
\Line(48,47)(42,53)
\Text(50,70)[r]{\mbox{$\tilde{W}^-$}}
\ArrowLine(60,80)(90,20)
\Text(85,55)[l]{\mbox{$\tilde{W}^-$}}
\Photon(60,80)(60,110){3}{3}
\Text(50,110)[c]{\mbox{$\gamma$}}
\end{picture}
&
\begin{picture}(150,120)(0,0)
\ArrowLine(10,20)(30,20)
\Text(10,10)[]{\mbox{$e$}}
\Vertex(30,20){2}
\DashArrowLine(30,20)(90,20){7}
\Text(60,10)[]{\mbox{$\tilde{\nu}_0$}}
\Vertex(90,20){2}
\ArrowLine(90,20)(110,20)
\Text(110,10)[]{\mbox{$e$}}
\ArrowLine(30,20)(60,80)
\Text(40,55)[r]{\mbox{$\tilde{h}^-$}}
\ArrowLine(60,80)(75,50)
\ArrowLine(75,50)(90,20)
\Text(75,70)[l]{\mbox{$\tilde{h}^-$}}
\Line(72,47)(78,53)
\Line(78,47)(72,53)
\Text(90,40)[l]{\mbox{$\tilde{W}^-$}}
\Photon(60,80)(60,110){3}{3}
\Text(50,110)[c]{\mbox{$\gamma$}}
\end{picture}
\end{tabular}
\caption{Chargino contribution to lepton EDM in mass insertion expansion.}
\label{fig:edm_c_mi}
\end{center}
\end{figure}
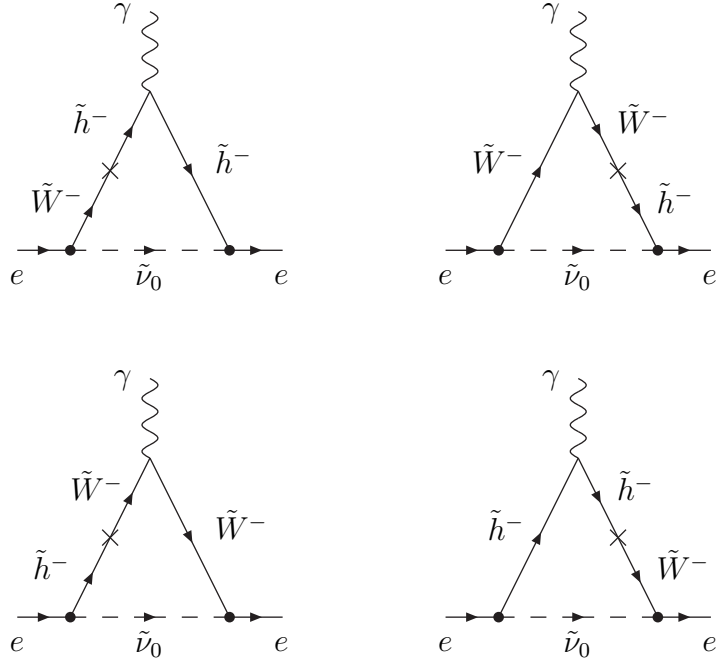

There are four diagrams with wino and charged Higgsino exchange, shown 
in Fig.~\ref{fig:edm_c_mi}. 
Their contribution to the electron EDM is ($E_e
\equiv E_l^1$ for electron):
\bea
(E_e)_C & =& {2eg^2m_e\over (4\pi)^2}\mathrm{Im}(M_2\mu)\tan\beta
{C_{11}(|\mu|^2,m^2_{\tilde{\nu}}) - C_{11}(|M_2|^2,m^2_{\tilde{\nu}})
\over |\mu|^2 - |M_2|^2}
\label{eq:edm_l_c_mi}
\eea
Neutral wino, bino and neutral Higgsino contributions can be split
into two classes: with mass insertion on the fermion or on the
sfermion line. Diagrams belonging to the first class are shown in
Fig.~\ref{fig:edm_nf_mi}. Their contribution has a structure very
similar to that given by eq.~(\ref{eq:edm_l_c_mi}):
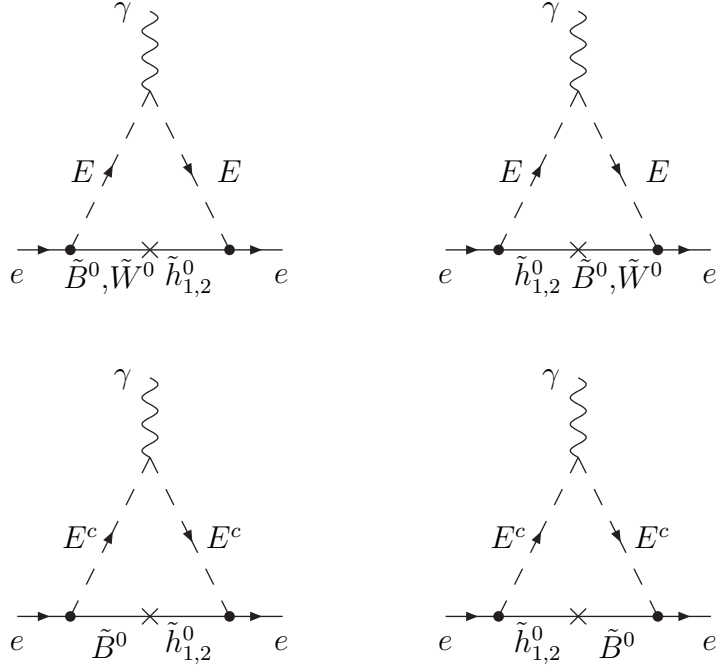
\begin{figure}[htbp]
\begin{center}
\begin{tabular}{lr}
\begin{picture}(150,120)(0,0)
\ArrowLine(10,20)(30,20)
\Text(10,10)[]{\mbox{$e$}}
\Vertex(30,20){2}
\Line(30,20)(60,20)
\Text(45,10)[]{\mbox{$\tilde{B}^0$,$\tilde{W}^0$}}
\Line(57,17)(63,23)
\Line(63,17)(57,23)
\Line(60,20)(90,20)
\Text(75,10)[]{\mbox{$\tilde{h}^0_{1,2}$}}
\Vertex(90,20){2}
\ArrowLine(90,20)(110,20)
\Text(110,10)[]{\mbox{$e$}}
\DashArrowLine(30,20)(60,80){7}
\Text(30,50)[l]{\mbox{$E$}}
\DashArrowLine(60,80)(90,20){7}
\Text(95,50)[r]{\mbox{$E$}}
\Photon(60,80)(60,110){3}{3}
\Text(50,110)[c]{\mbox{$\gamma$}}
\end{picture}
&
\begin{picture}(150,120)(0,0)
\ArrowLine(10,20)(30,20)
\Text(10,10)[]{\mbox{$e$}}
\Vertex(30,20){2}
\Line(30,20)(60,20)
\Text(75,10)[]{\mbox{$\tilde{B}^0$,$\tilde{W}^0$}}
\Line(57,17)(63,23)
\Line(63,17)(57,23)
\Line(60,20)(90,20)
\Text(45,10)[]{\mbox{$\tilde{h}^0_{1,2}$}}
\Vertex(90,20){2}
\ArrowLine(90,20)(110,20)
\Text(110,10)[]{\mbox{$e$}}
\DashArrowLine(30,20)(60,80){7}
\Text(30,50)[l]{\mbox{$E$}}
\DashArrowLine(60,80)(90,20){7}
\Text(95,50)[r]{\mbox{$E$}}
\Photon(60,80)(60,110){3}{3}
\Text(50,110)[c]{\mbox{$\gamma$}}
\end{picture}
\\
&\\
\begin{picture}(150,120)(0,0)
\ArrowLine(10,20)(30,20)
\Text(10,10)[]{\mbox{$e$}}
\Vertex(30,20){2}
\Line(30,20)(60,20)
\Text(75,10)[]{\mbox{$\tilde{h}^0_{1,2}$}}
\Line(57,17)(63,23)
\Line(63,17)(57,23)
\Line(60,20)(90,20)
\Text(45,10)[]{\mbox{$\tilde{B}^0$}}
\Vertex(90,20){2}
\ArrowLine(90,20)(110,20)
\Text(110,10)[]{\mbox{$e$}}
\DashArrowLine(30,20)(60,80){7}
\Text(27,50)[l]{\mbox{$E^c$}}
\DashArrowLine(60,80)(90,20){7}
\Text(95,50)[r]{\mbox{$E^c$}}
\Photon(60,80)(60,110){3}{3}
\Text(50,110)[c]{\mbox{$\gamma$}}
\end{picture}
&
\begin{picture}(150,120)(0,0)
\ArrowLine(10,20)(30,20)
\Text(10,10)[]{\mbox{$e$}}
\Vertex(30,20){2}
\Line(30,20)(60,20)
\Text(75,10)[]{\mbox{$\tilde{B}^0$}}
\Line(57,17)(63,23)
\Line(63,17)(57,23)
\Line(60,20)(90,20)
\Text(45,10)[]{\mbox{$\tilde{h}^0_{1,2}$}}
\Vertex(90,20){2}
\ArrowLine(90,20)(110,20)
\Text(110,10)[]{\mbox{$e$}}
\DashArrowLine(30,20)(60,80){7}
\Text(27,50)[l]{\mbox{$E^c$}}
\DashArrowLine(60,80)(90,20){7}
\Text(95,50)[r]{\mbox{$E^c$}}
\Photon(60,80)(60,110){3}{3}
\Text(50,110)[c]{\mbox{$\gamma$}}
\end{picture}
\\
\end{tabular}
\caption{Neutralino contribution to lepton EDM in mass insertion
expansion: mass insertion on the fermion line.}
\label{fig:edm_nf_mi}
\end{center}
\end{figure}
\begin{figure}[htbp]
\begin{center}
\begin{tabular}{lr}
\begin{picture}(150,120)(0,0)
\ArrowLine(10,20)(30,20)
\Text(10,10)[]{\mbox{$e$}}
\Vertex(30,20){2}
\Line(30,20)(90,20)
\Text(60,10)[c]{\mbox{$\tilde{B}^0$}}
\Vertex(90,20){2}
\ArrowLine(90,20)(110,20)
\Text(110,10)[]{\mbox{$e$}}
\DashArrowLine(30,20)(45,50){3}
\DashArrowLine(45,50)(60,80){3}
\Text(15,40)[]{\mbox{$E(E^c)$}}
\Line(42,47)(48,53)
\Line(48,47)(42,53)
\Text(30,70)[]{\mbox{$E^c(E)$}}
\DashArrowLine(60,80)(90,20){7}
\Text(115,50)[r]{\mbox{$E^c(E)$}}
\Photon(60,80)(60,110){3}{3}
\Text(50,110)[c]{\mbox{$\gamma$}}
\end{picture}
&
\begin{picture}(150,120)(0,0)
\ArrowLine(10,20)(30,20)
\Text(10,10)[]{\mbox{$e$}}
\Vertex(30,20){2}
\Line(30,20)(90,20)
\Text(60,10)[]{\mbox{$\tilde{B}^0$}}
\Vertex(90,20){2}
\ArrowLine(90,20)(110,20)
\Text(110,10)[]{\mbox{$e$}}
\DashArrowLine(30,20)(60,80){7}
\Text(10,50)[l]{\mbox{$E(E^c)$}}
\DashArrowLine(60,80)(75,50){3}
\DashArrowLine(75,50)(90,20){3}
\Text(105,70)[r]{\mbox{$E(E^c)$}}
\Line(72,47)(78,53)
\Line(78,47)(72,53)
\Text(120,40)[r]{\mbox{$E^c(E)$}}
\Photon(60,80)(60,110){3}{3}
\Text(50,110)[c]{\mbox{$\gamma$}}
\end{picture}
\\
&\\
\begin{picture}(150,120)(0,0)
\ArrowLine(10,20)(30,20)
\Text(10,10)[]{\mbox{$e$}}
\Vertex(30,20){2}
\Line(30,20)(90,20)
\Text(60,10)[]{\mbox{$\tilde{h}^0_{1,2}$}}
\Vertex(90,20){2}
\ArrowLine(90,20)(110,20)
\Text(110,10)[]{\mbox{$e$}}
\DashArrowLine(30,20)(45,50){3}
\DashArrowLine(45,50)(60,80){3}
\Text(15,40)[]{\mbox{$E(E^c)$}}
\Line(42,47)(48,53)
\Line(48,47)(42,53)
\Text(30,70)[]{\mbox{$E^c(E)$}}
\DashArrowLine(60,80)(90,20){7}
\Text(115,50)[r]{\mbox{$E^c(E)$}}
\Photon(60,80)(60,110){3}{3}
\Text(50,110)[c]{\mbox{$\gamma$}}
\end{picture}
&
\begin{picture}(150,120)(0,0)
\ArrowLine(10,20)(30,20)
\Text(10,10)[]{\mbox{$e$}}
\Vertex(30,20){2}
\Line(30,20)(90,20)
\Text(60,10)[]{\mbox{$\tilde{h}^0_{1,2}$}}
\Vertex(90,20){2}
\ArrowLine(90,20)(110,20)
\Text(110,10)[]{\mbox{$e$}}
\DashArrowLine(30,20)(60,80){7}
\Text(10,50)[l]{\mbox{$E(E^c)$}}
\DashArrowLine(60,80)(75,50){3}
\DashArrowLine(75,50)(90,20){3}
\Text(105,70)[r]{\mbox{$E(E^c)$}}
\Line(72,47)(78,53)
\Line(78,47)(72,53)
\Text(120,40)[r]{\mbox{$E^c(E)$}}
\Photon(60,80)(60,110){3}{3}
\Text(50,110)[c]{\mbox{$\gamma$}}
\end{picture}
\\
\end{tabular}
\caption{Neutralino contribution to lepton EDM in mass insertion
expansion: mass insertion on the scalar line}
\label{fig:edm_ns_mi}
\end{center}
\end{figure}
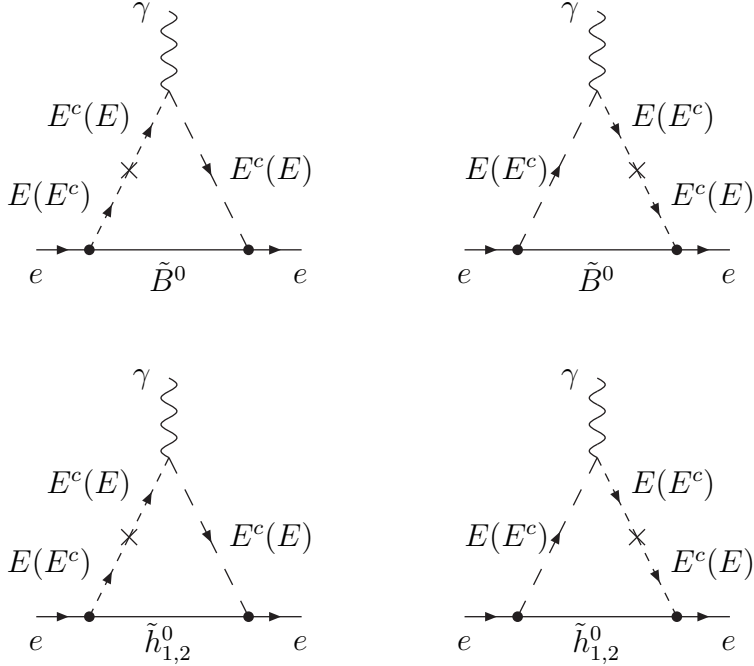
\bea
(E_e)_{Nf} & =& -{eg^2m_e\over 2(4\pi)^2} \mathrm{Im}
(M_2\mu)\tan\beta {C_{12}(m_E^2,|\mu|^2) -
C_{12}(m_E^2,|M_2|^2)\over |\mu|^2 - |M_2|^2}\nonumber\\
&+& {e{g'}^2m_e\over 2(4\pi)^2} \mathrm{Im}(M_1\mu)\tan\beta
{C_{12}(m_E^2,|\mu|^2) -
C_{12}(m_E^2,|M_1|^2)\over |\mu|^2 - |M_1|^2}\nonumber\\
&-& {e{g'}^2m_e\over (4\pi)^2} \mathrm{Im}(M_1\mu)\tan\beta
{C_{12}(m_{E^c}^2,|\mu|^2) -
C_{12}(m_{E^c}^2,|M_1|^2)\over |\mu|^2 - |M_1|^2}
\label{eq:edm_l_nf_mi}
\eea
where we denote by $m_E$, $m_{{E}^c}$ and $m_{\tilde{\nu}}$ the masses
of left- and right- selectron and electron sneutrino, respectively.
Finally, diagrams with mass insertions on the selectron line are shown
in Fig.~\ref{fig:edm_ns_mi}. Only the first two with bino line in the
loop give sizeable contributions. The other two with neutral Higgsino
exchange are suppressed by the additional factor ${\cal O}
\left(m_e^2/M_W^2\right)$ and thus completely negligible. The result
is:
\bea
(E_e)_{Ns} & =& {e{g'}^2m_e\over (4\pi)^2} \mathrm{Im}
\left[M_1(\mu\tan\beta + A_e^{\star})\right]
{C_{12}(m_{E}^2,|M_1|^2) -
C_{12}(m_{E^c}^2,|M_1|^2)\over m_E^2 - m_{E^c}^2}\nonumber\\
&+& \mathrm{terms~suppressed~by~} {\cal
O}\left(\frac{m_e^2}{M_W^2}\right)
\label{eq:edm_l_ns_mi}
\eea 

Eqs.~(\ref{eq:edm_l_c_mi}-\ref{eq:edm_l_ns_mi}) have a simple
structure: they are linear in the CP-invariants~(\ref{eq:phasinv}),
with coefficients that are functions of the real mass
parameters. Thus, the possibility of cancellations depends primarily
on the relative amplitudes and signs of those coefficients. An
immediate conclusion following
from~(\ref{eq:edm_l_c_mi}-\ref{eq:edm_l_ns_mi}) is that limits on the
$M_i\mu$ phases are inversely proportional to $\tan\beta$. Therefore,
in the next Section, we discuss limits on $\sin\phi_{\mu}\tan\beta$
rather than on the $\mu$ phase itself.

\begin{figure}[htbp]
\begin{center}
\mbox{\epsfig{file=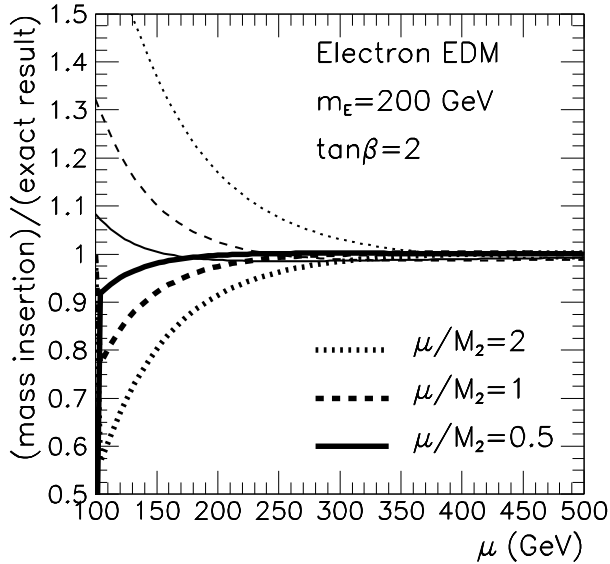,width=0.5\linewidth}}
\vskip -5mm 
\caption{Ratio of the electron EDM calculated in the mass insertion 
approximation to the exact 1-loop result. Thinner lines:
$\phi_{\mu}=0$, thicker lines: $\phi_A=0$. Degenerate left and right
slepton masses and $M_1/\alpha_1 = M_2/\alpha_2$ assumed.
\label{fig:edml_micomp}}
\end{center}
\end{figure}

The approximate formulae~(\ref{eq:edm_l_c_mi}-\ref{eq:edm_l_ns_mi})
work very well already for relatively small $|\mu|$, $|M_1|$ and
$|M_2|$ values, not much above the $M_Z$ scale.
Figure~\ref{fig:edml_micomp} shows the ratio of the electron EDM
calculated in the mass insertion approximation to the exact 1-loop
result given by eq.~(\ref{eq:edml_full}). The accuracy of the mass
insertion expansion may become reasonable already for $|\mu|\geq 150$
GeV (depending on $|M_2/\mu|$ ratio) and becomes very good for
$|\mu|\geq 200-250$ GeV.

Formulae~(\ref{eq:edm_l_c_mi}-\ref{eq:edm_l_ns_mi}) can also be
obtained directly from the exact expression~(\ref{eq:edml_full}) using
the expansion of sfermion and supersymmetric fermion mass matrices
described in~\ref{app:massins} (this gives a very useful cross-check
of the correctness of the calculations). One should note that even
though the expansion~(\ref{eq:zpmexp}) does not work for degenerate
$|M_2|$ and $|\mu|$, the expression~(\ref{eq:edm_l_c_mi}) has already
a well defined limit for $|M_2|=|\mu|$. The same holds for the
$|M_1|=|\mu|$ and the degenerate sfermion masses in
eqs.~(\ref{eq:edm_l_nf_mi},\ref{eq:edm_l_ns_mi}).

\subsection{Limits on $\mu$ and $A_e$ phases}

It is useful to consider two classes of models: one with $M_1/\alpha_1
= M_2/\alpha_2$ for gaugino masses\footnote{Here, by $\alpha_1$,
$\alpha_2$ we understand the gauge couplings in GUT normalization,
i.e.  without a $5/3$ factor for the $U(1)$ coupling.}, that is
universal gaugino masses at the GUT scale (the universal phase can be
set to zero by convention), and the other with non-universal gaugino
masses and arbitrary relative phase between $M_1$ and $M_2$. In the
universal case we choose $\mu$ and $A_e$ phases as the independent
ones, in the second case the $M_1, M_2$ phases are the additional free
parameters.  Constraints from the RGE running and proper electroweak
breaking are insignificant at this point, as we have enough additional
parameters to satisfy them for any chosen set of low energy values for
$\mu$, $M_1$, $M_2$, $m_E$ and $A_e$

\begin{figure}[htbp]
\begin{center}
\begin{tabular}{p{0.48\linewidth}p{0.48\linewidth}}
\mbox{\epsfig{file=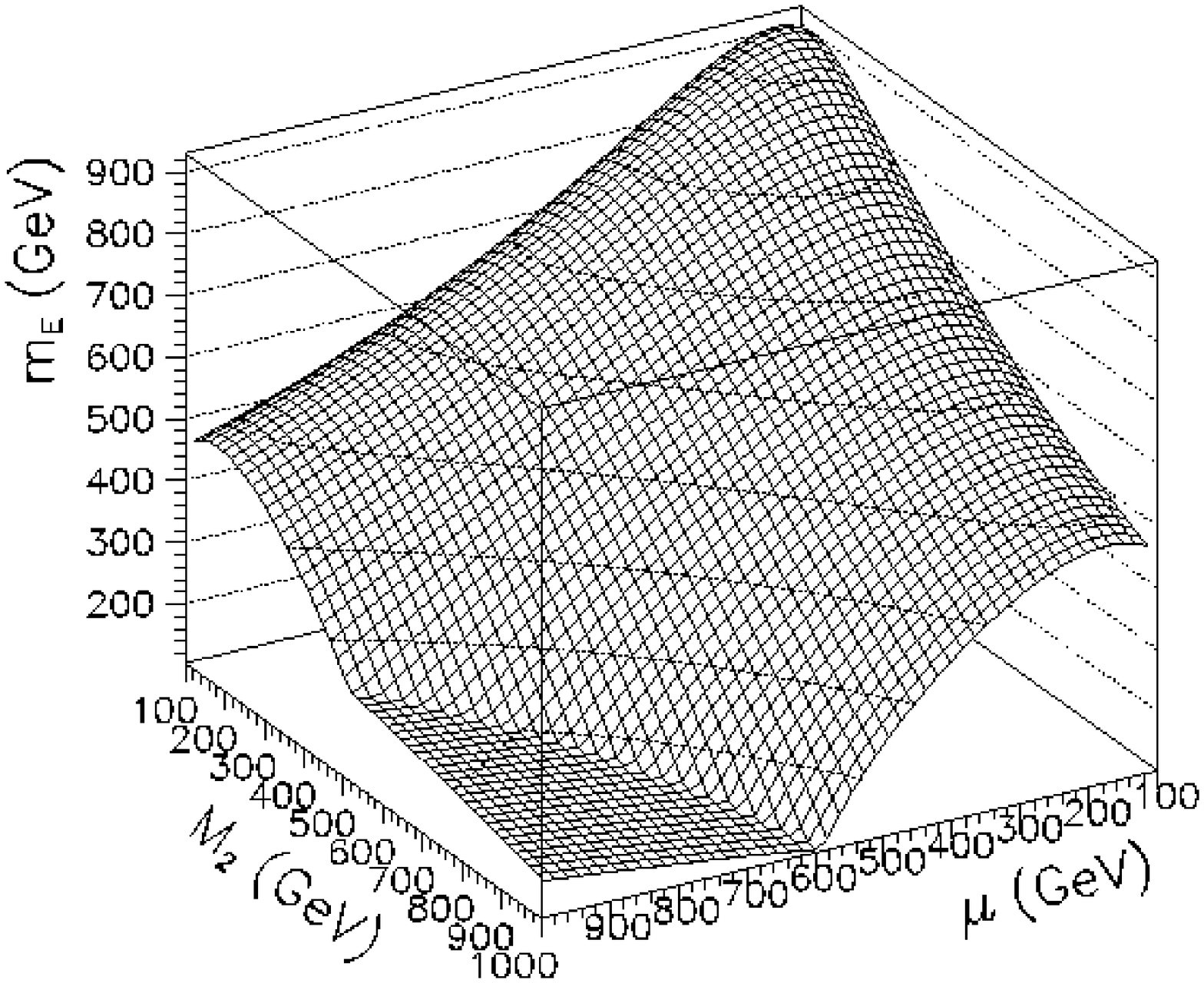,width=\linewidth}}&
\mbox{\epsfig{file=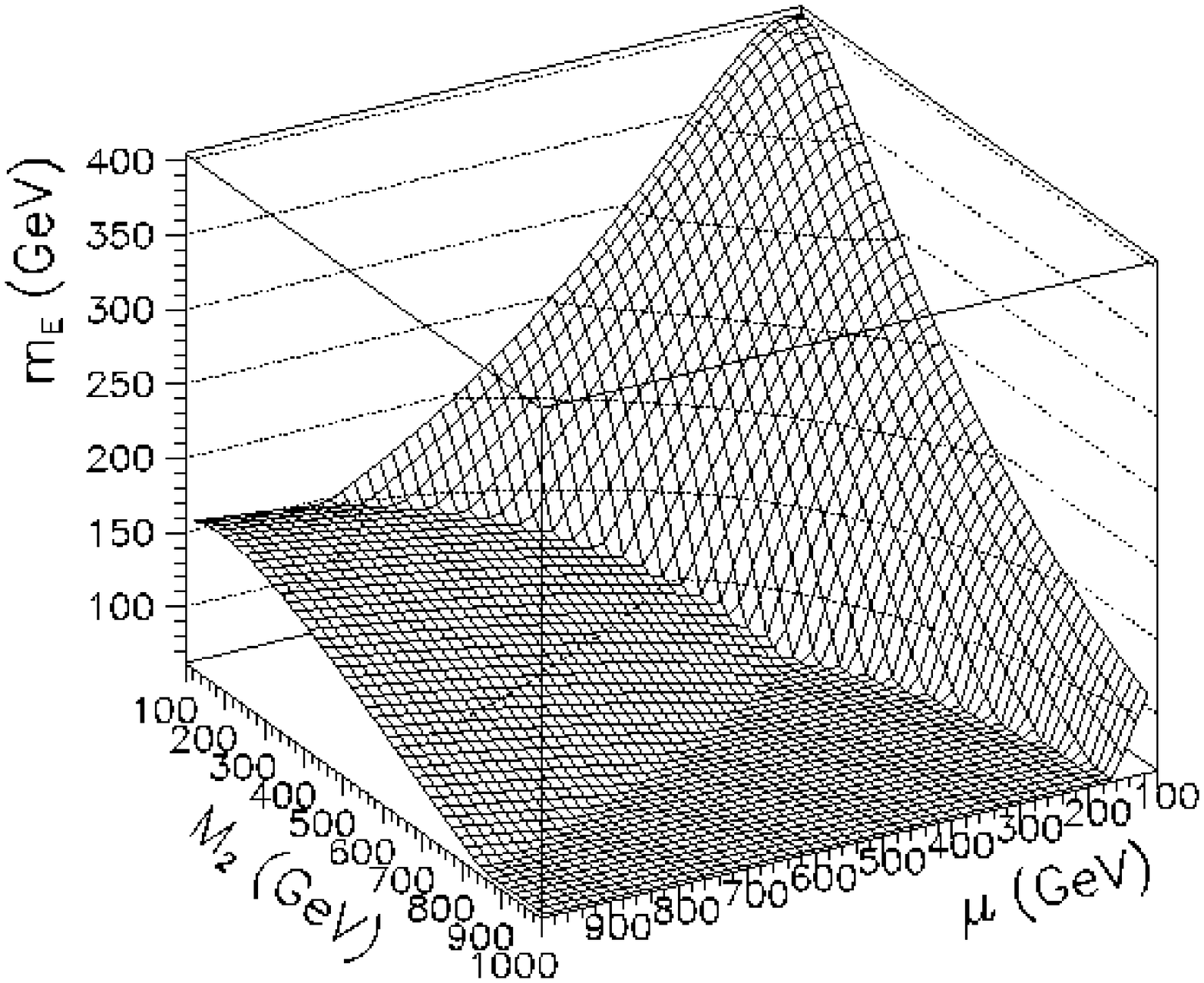,width=\linewidth}}
\end{tabular}
\vskip -5mm 
\caption{Regions (below the dark surface) for which generic limits on
$|\sin\phi_{\mu}|\tan\beta$ are stronger then, respecively, 0.2 (left
plot) and 0.05 (right plot). Degenerate left and right slepton masses
and $M_1/\alpha_1 = M_2/\alpha_2$ assumed.
\label{fig:e_ampl_surf_gut}}
\end{center}
\end{figure}

We shall begin our discussion by presenting the magnitude of each
contribution~(\ref{eq:edm_l_c_mi}),~(\ref{eq:edm_l_nf_mi}) and of the
$\mu$ and $A_e$ terms in eq.~(\ref{eq:edm_l_ns_mi}), separately.  For
the $M_1/\alpha_1= M_2/\alpha_2$ case a sample of results is shown in
Fig.~\ref{fig:e_ampl_surf_gut}. We identify there the parameter region
where at least one of the terms is such that for
$\sin\phi_{\mu}\tan\beta$ fixed at some assumed value, its
contribution to $E_e$ is larger than $E_e^{exp}$. Barring potential
cancellations, the fixed value of $\sin\phi_{\mu}\tan\beta$ is then
the limit on this phase in the identified parameter region. In the
left (right) plot of Fig.~\ref{fig:e_ampl_surf_gut} we show the
regions of masses (below the plotted surface) where the limits on
$|\sin\phi_{\mu}|\tan\beta$ are stronger then 0.2 (0.05),
respectively. We see that this region extends up to 900 (400) GeV in
$m_E$ (physical left selectron mass) for small values of $|\mu|$ and
$|M_1|,|M_2|$ (up to 200-300 GeV, say) and for larger values of
$|\mu|$ and/or $|M_1|,|M_2|$ it gradually shrinks to 100 (0) GeV in
$m_E$ for $|\mu|\sim |M_2|\sim 1$ TeV. We assume left and right
slepton mass parameters equal, $M_L=M_E$, so that the physical masses
of the left and right selectron differ by D-terms only. The regions
below the plotted surfaces are the regions of interest for potential
cancellations. We observe, however, that even without cancellations,
there are interesting regions of small $|\mu|$ and $|M_2|$ and
$m_E>{\cal O}(1 \mathrm{TeV})$ or small $m_E$ and $|\mu|\sim
|M_2|>{\cal O}(500~\mathrm{TeV})$ where the phase of $\mu$ is
weakly constrained.  One should also note that for very large $|\mu|$
and the other masses fixed the limits on the $\mu$ phase get stronger
again. This is due to the term~(\ref{eq:edm_l_ns_mi}), which does not
decouple for large $|\mu|$. The limits on $(|A_e\sin\phi_{A_e}|/m_E$
are typically significantly weaker.

\begin{figure}[htbp]
\begin{center}
\begin{tabular}{p{0.48\linewidth}p{0.48\linewidth}}
\mbox{\epsfig{file=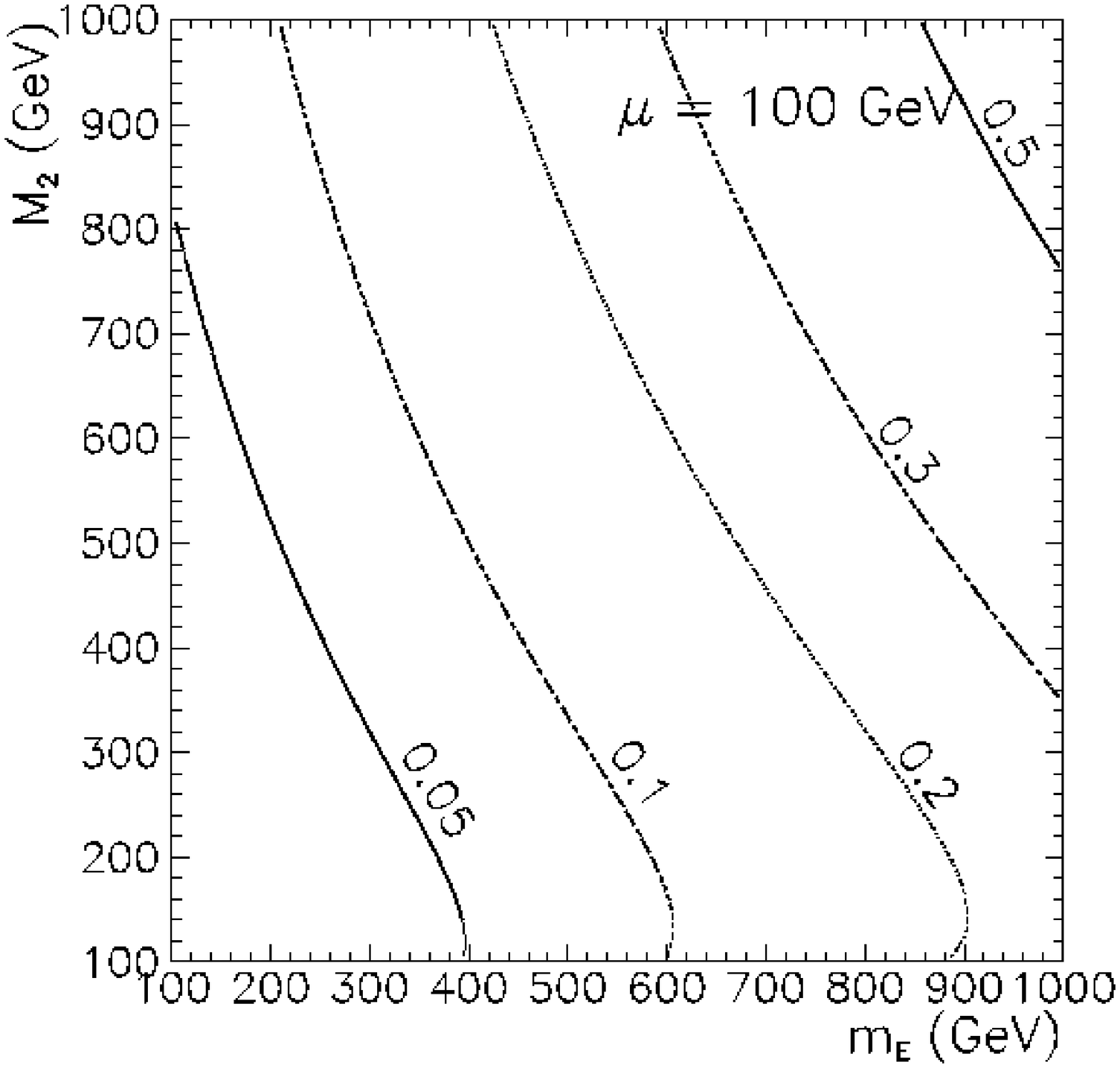,width=\linewidth}}&
\mbox{\epsfig{file=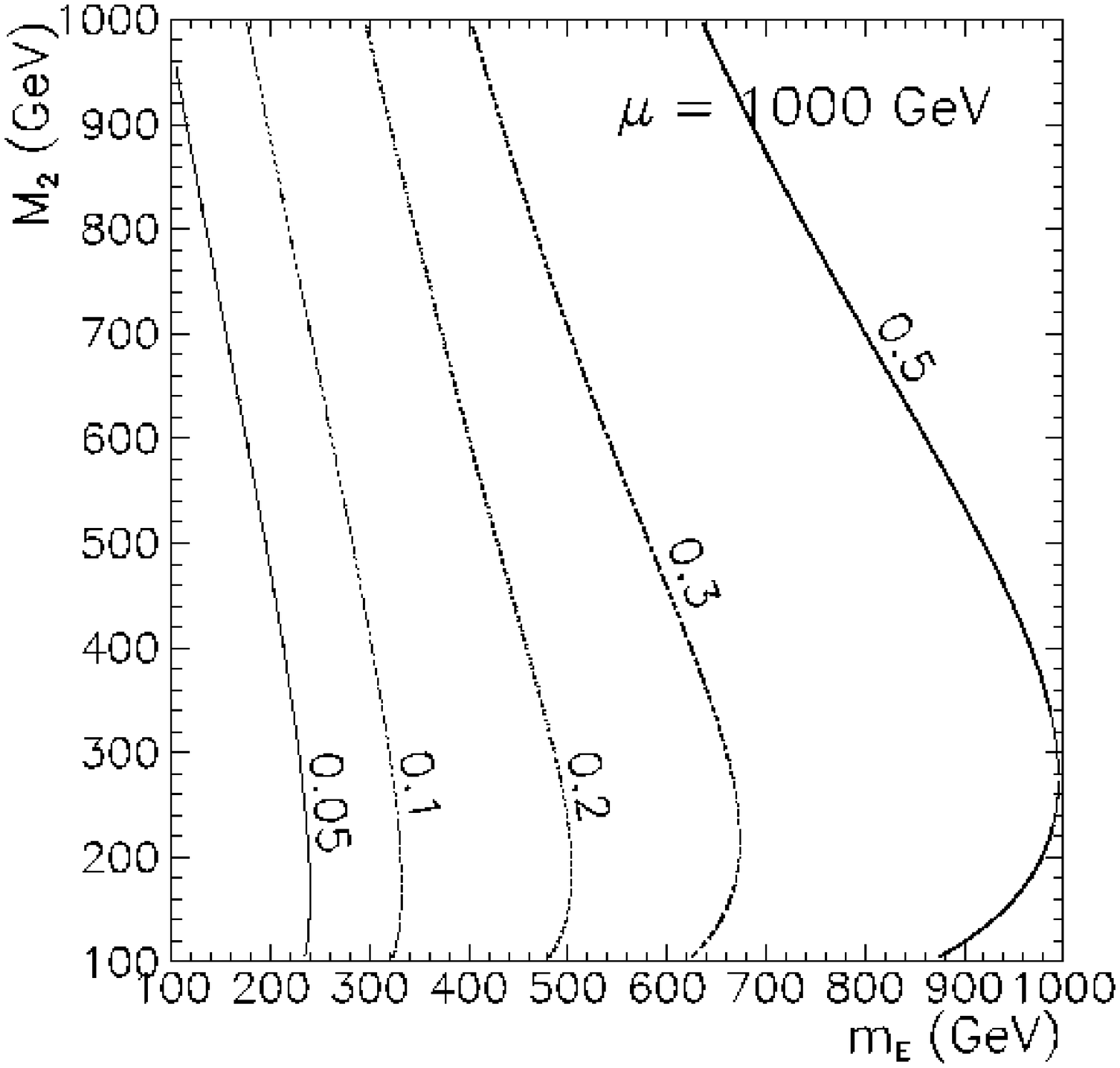,width=\linewidth}}
\end{tabular}
\caption{Limits on $|\sin\phi_{\mu}|\tan\beta$ given by the electron
EDM measurements. $\sin\phi_{A_e}=0$ and $M_1/\alpha_1=M_2/\alpha_2$
assumed.}
\label{fig:e_cont_mu}
\end{center}
\end{figure}

In Fig.~\ref{fig:e_cont_mu} we show again the limits on $\mu$ phase,
this time as a two-dimensional plot in the $(m_E,|M_2|)$ plane,
assuming $M_1/\alpha_1=M_2/\alpha_2$ and $\phi_{A_e}=0$. The limits
plotted in Fig.~\ref{fig:e_cont_mu} are given by the sum of all
terms~(\ref{eq:edm_l_c_mi}-\ref{eq:edm_l_ns_mi}),
not by the largest of them like in Fig.~\ref{fig:e_ampl_surf_gut}.

\begin{figure}[htbp]
\begin{center}
\mbox{\epsfig{file=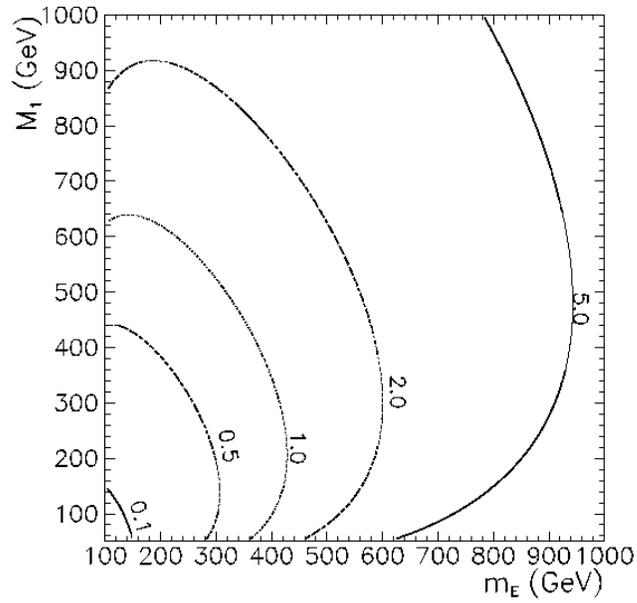,width=0.5\linewidth}}
\caption{Limits on $|A_e/m_E \sin\phi_{A_e}|$ given by the electron
EDM measurements ($|\mu|=200$ GeV, $\sin\phi_{\mu}=0$ and
$M_1/\alpha_1=M_2/\alpha_2$ assumed).\label{fig:e_cont_a}}
\end{center}
\end{figure}

In Fig.~\ref{fig:e_cont_a} we show similar limits on the $A_e$ phase
on $(m_E$,$|M_1|)$ plane, assuming vanishing $\mu$ phase. The limits
on the $A_e$ parameter phase are significantly weaker and decrease
more quickly with increasing particle masses\footnote{One should note
that the off diagonal entry in the slepton mass matrices describing
LR-mixing is proportional to $A_e$ and to the electron mass $m_e$ (see
eq.~(\ref{eq:sfmass})). Therefore, even if imaginary part of $A_e$
parameter is weakly constrained, $\mathrm{Im}A_e/m_E\sim 1$, the full
LR-mixing in the selectron sector can have only very small imaginary
part, of the order of $m_e m_E$. For the electron, Gabbiani
et. al~\cite{MASIERO} give the limit $\mathrm{Im}A_e/m_E\leq 10^{-6}$,
but their definition of $A_e$ contains $m_e$ in it. After extracting
it, their limit is similar to ours.}. They are almost independent of
$M_2$ and $\mu$, as can be also seen immediately from
eq.~(\ref{eq:edm_l_ns_mi}).

\begin{figure}[htbp]
\begin{center}
\mbox{\epsfig{file=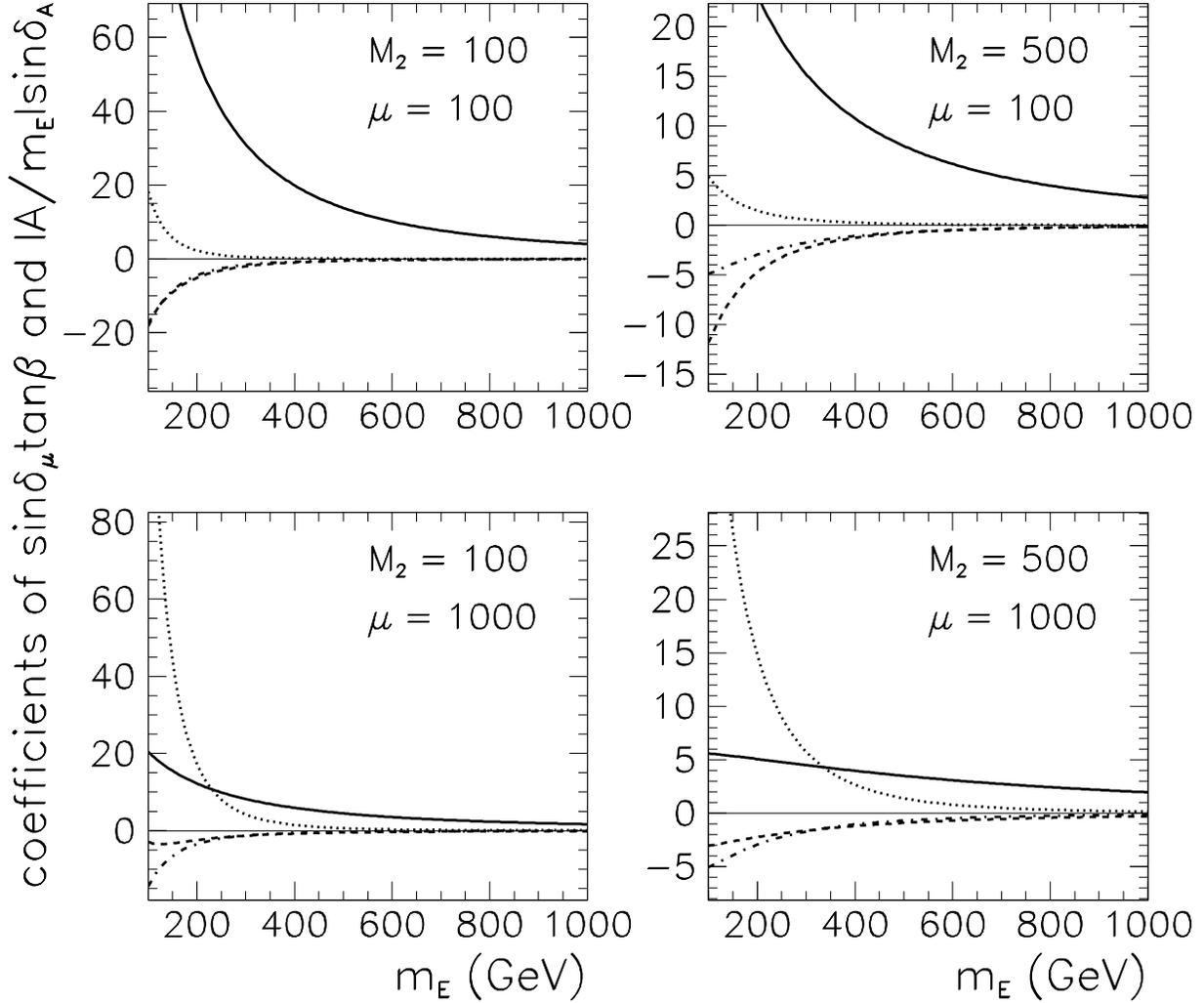,width=\linewidth}}
\vskip -5mm 
\caption{Relative signs and amplitudes of various contributions to the 
electron EDM, normalized to (divided by) the experimental
limit. Solid, dashed, dotted lines: coefficients of
$\sin\phi_{\mu}\tan\beta$ given by chargino
(eq.~(\ref{eq:edm_l_c_mi})) and neutralino contributions
(eqs.~(\ref{eq:edm_l_nf_mi}) and~(\ref{eq:edm_l_ns_mi}))
respectively. Dotted-dashed line: coefficient of
$|A_e\sin\phi_{A_e}|/m_E$.  Degenerate left and right slepton masses
and $M_1/\alpha_1 = M_2/\alpha_2$ assumed.
\label{fig:e_canc}}
\end{center}
\end{figure}

\begin{figure}[htbp]
\begin{center}
\begin{tabular}{p{0.48\linewidth}p{0.48\linewidth}}
\mbox{\epsfig{file=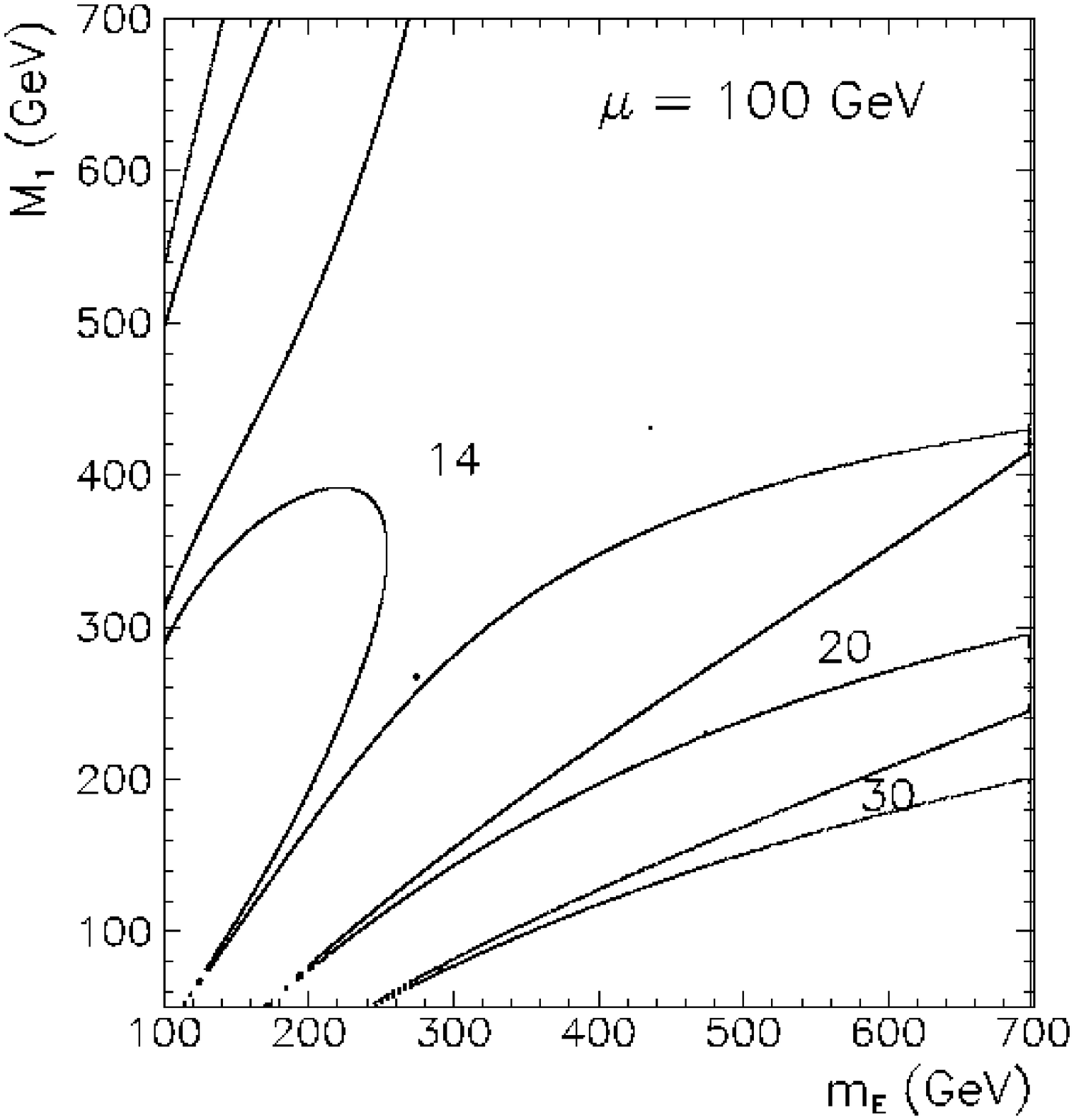,width=\linewidth}}&
\mbox{\epsfig{file=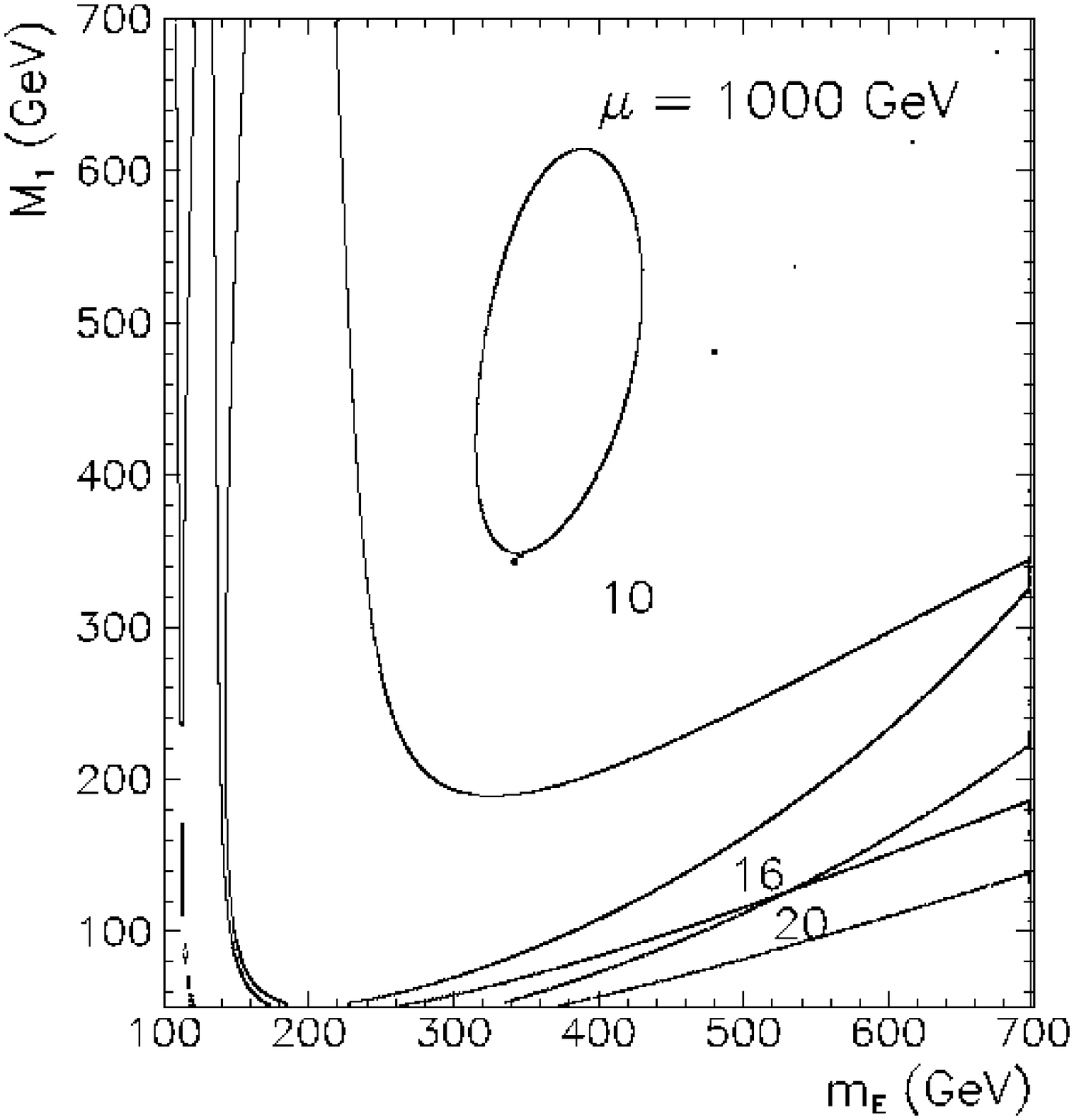,width=\linewidth}}
\end{tabular}
\vskip -5mm 
\caption{Regions of $m_E-M_1$ plane allowed by the electron EDM
measurement assuming ``maximal'' CP violation
$\phi_{\mu}=\phi_{A_e}=\pi/2$ and the corresponding values of
$A_e/m_E$ parameter (marked on the plots), necessary for the $\mu-A$
cancellation.
\label{fig:e_cont_a_mu}}
\end{center}
\end{figure}

The magnitude and signs of individual contributions as a function of
$m_E$ are illustrated in Fig.~\ref{fig:e_canc}. We plot there the
coefficients of $\mu$ and $A_e$ phases obtained from the exact 1-loop
result and normalized by dividing them by the experimental limit on
the electron EDM. Their shape depends mostly on the $m_E/|\mu|$ ratio,
much less on the $|\mu/M_2|$ ratio and scales like $1/m_E^2$. We see
that either the chargino contribution to the term proportional to the
$\mu$ phase dominates (for small $|\mu|$), or, if they become
comparable (possible only for larger values of $|\mu|> 700$ GeV), the
chargino and the dominant neutralino contribution, given by
eq.~(\ref{eq:edm_l_ns_mi}), to the $\mu$ phase coefficient are of the
same sign. The neutralino contribution given by
eq.~(\ref{eq:edm_l_nf_mi}) has opposite sign than that of
eq.~(\ref{eq:edm_l_ns_mi}), but their sum is positive. The only
exception is the case of $|\mu|,|M_1|\leq 100$ GeV, where both
neutralino contributions are much smaller than the chargino one. Thus,
the full coefficient of the $\mu$ phase cannot vanish and the only
possible cancellations are between the $A_e$ and $\mu$ phases.
Since the $A_e$ phase coefficient is in the interesting region much
smaller such cancellations always require large $A_e$ in the selectron
sector, $A_e/m_E\gg 1$. This is shown in Fig.~\ref{fig:e_cont_a_mu},
where we assume ``maximal'' CP violation
$\phi_{\mu}=\phi_{A_e}=\pi/2$.

Better understanding of the $\mu$--$A_e$ cancellation can be achieved
after some approximations.  For light supersymmetric fermions,
significantly lighter than sleptons, chargino exchanges dominate
(Fig.~\ref{fig:edm_c_mi}), whereas in the opposite limit the biggest
contribution is given by the diagrams with bino exchanges
(Fig.~\ref{fig:edm_ns_mi}). Eqs.~(\ref{eq:edm_l_c_mi}-\ref{eq:edm_l_ns_mi})
can be greatly simplified in both cases, giving for degenerate slepton
masses $m_E\approx m_{E^c}\approx m_{\tilde{\nu}}$:

\noindent 1) $|M_{1,2}|,|\mu|\ll m_E$.
\bea
E_e&\approx& {eg^2m_e\over (4\pi)^2}{\mathrm{Im}
(M_2\mu)\tan\beta\over m_E^2(|\mu^2|-
|M_2|^2)}\log\frac{|\mu|^2}{|M_2|^2} +
{e{g'}^2m_e\over2(4\pi)^2}{\mathrm{Im}(M_1A_e^{\star})\over m_E^4}
\label{eq:edm_l_mulight}
\eea
2) $|M_{1,2}|,|\mu|\gg m_E$. 
\bea
E_e&\approx& {eg^2m_e\over 4(4\pi)^2}{\mathrm{Im}
(M_2\mu)\tan\beta \over |\mu|^2|M_2|^2} - {e{g'}^2m_e\over
4(4\pi)^2}{\mathrm{Im}(M_1\mu)\tan\beta \over
|\mu|^2|M_1|^2}\nonumber\\
&-&{e{g'}^2m_e\over 2(4\pi)^2}{\mathrm{Im}\left[M_1(\mu\tan\beta +
A_e^{\star})\right] \over |M_1|^4}
\left(5  + 2\log{m_E^2\over|M_1|^2}\right)
\label{eq:edm_l_muheavy}
\eea

The behaviour of the lepton EDM is different in both limits. For heavy
sleptons, $|M_{1,2}|,|\mu|\ll m_E$ the coefficient of the $\mu$ phase
decreases with the increasing slepton mass as $1/m_E^2$. The
coefficient of the $A_e$ phase decreases faster, as
$1/m_E^4$. Therefore, in this limit the exact cancellation between
$A_e$ and $\mu$ phases requires large $A_e$ value, growing with
increasing $m_E$. However, because all contributions simultaneously
decrease with increasing $m_E$, partial cancellation between $\mu$ and
$A_e$ phase is already sufficient to push the electron EDM below the
experimental limits, what may be observed in
Fig.~\ref{fig:e_cont_a_mu} as a widening of the allowed regions for
large $m_E$.

For sufficiently small slepton masses the full cancellation between
$\mu$ and $A_e$ terms occurs approximately for
\bea
\sin\phi_{A_e}|A_e| = \sin\phi_{\mu}|\mu|\tan\beta
\left(1 + {|M_1|^2\over |\mu|^2} {3\over 5 +
2\log\frac{m_E^2}{|M_1^2|}}\right)\nonumber\\
\label{eq:ma_canc}
\eea
where we assumed $M_1/\alpha_1=M_2/\alpha_2$ and redefined the $M_1$
phase to zero. Since this result is valid for $|\mu|,|M_1|,|M_2|\gg
m_E$ we see that for comparable $\phi_{\mu}$ and $\phi_{A_e}$ the
cancellation is again possible only for large $A_e/m_E\gg 1$. For
large $|\mu|\gg |M_1|$, when one can neglect the second term in the
parenthesis in eq.~(\ref{eq:ma_canc}), the $A_e$ giving maximal
cancellation is almost independent of $|M_1|$, what can be observed in
the right plot of Fig.~\ref{fig:e_cont_a_mu}. The allowed regions also
widen with increasing $|M_1|$, but slower then for large $m_E$ because
the $\mu$ and $A_e$ phases are in this case suppressed by lower powers
of $|M_1|$: $1/|M_1|$ and $1/|M_1|^3$ respectively, instead of
$1/m_E^2$ and $1/m_E^4$.

It is worthwhile to note that in the most interesting region of light
SUSY masses, where the limits on phases are strongest, the
cancellation between (fixed) $\mu$ and $A_e$ phases may occur only for
very precisely correlated mass parameters, i.e. it requires strong
fine tuning between $|\mu|$, $|M_1|$, $|M_2|$ and $|A_e|$.

\begin{figure}[htbp]
\begin{center}
\begin{tabular}{p{0.48\linewidth}p{0.48\linewidth}}
\mbox{\epsfig{file=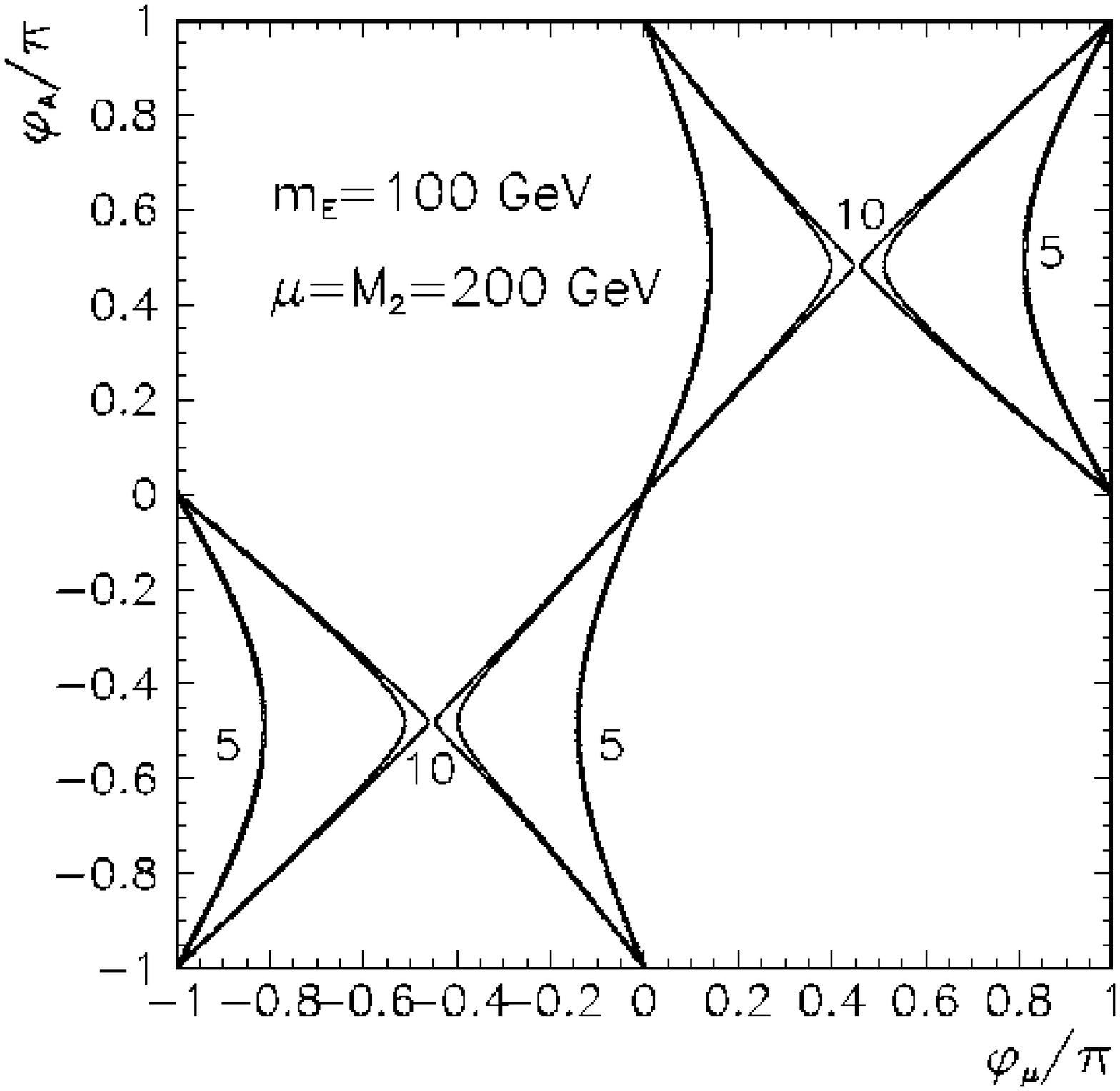,width=\linewidth}}&
\mbox{\epsfig{file=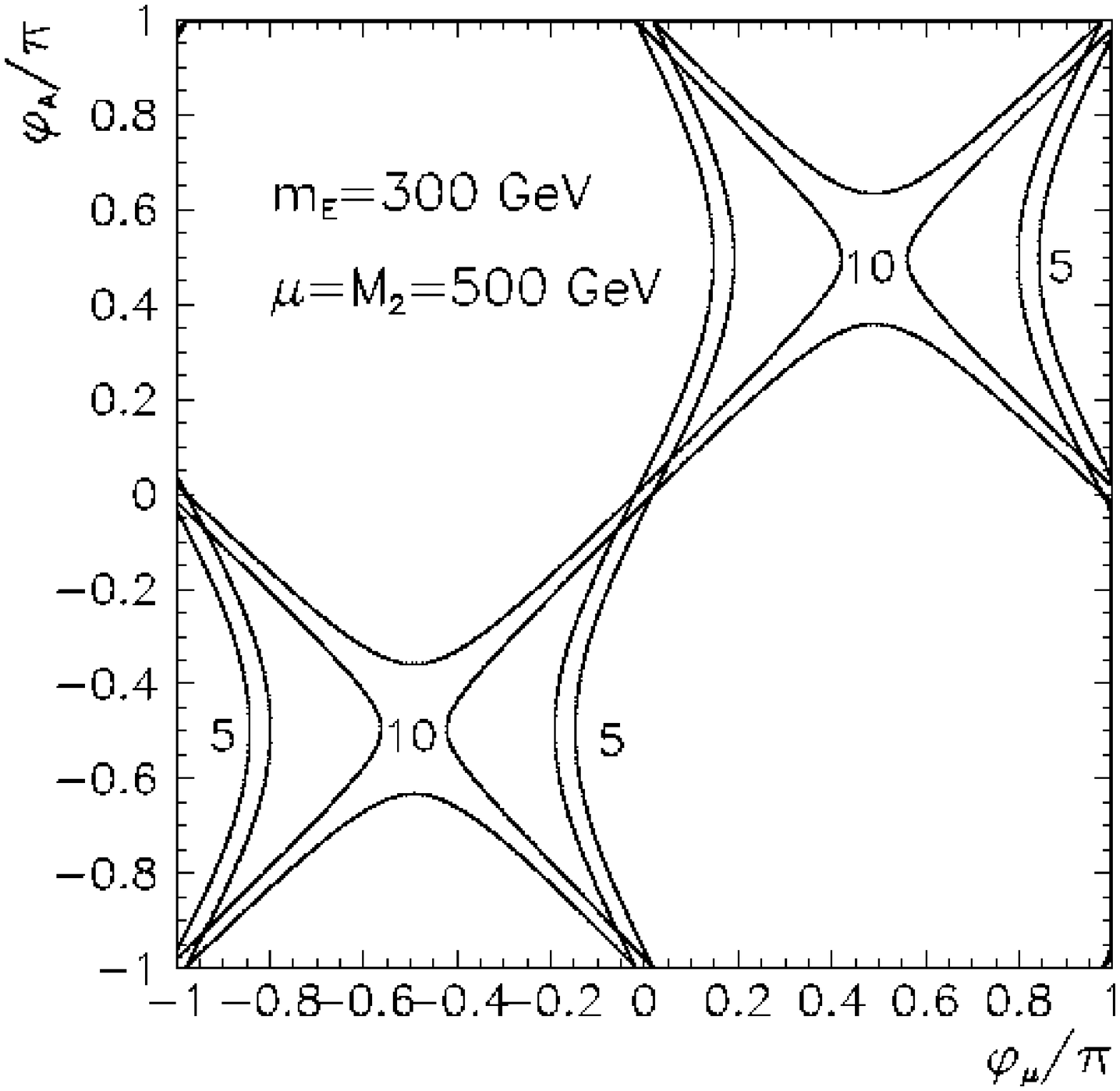,width=\linewidth}}
\end{tabular}
\vskip -5mm 
\caption{Illustration of the fine tuning between the phases for fixed mass 
parameters (listed in the plots), $\tan\beta=2$ and two values of
$A_e/m_E=5,~10$ (marked near the respective allowed
regions). Degenerate left and right slepton masses and $M_1/\alpha_1
= M_2/\alpha_2$ assumed.
\label{fig:e_cont_a_mu_ph}}
\end{center}
\end{figure}

In Fig.~\ref{fig:e_cont_a_mu_ph} we plot the allowed regions of the
$\phi_{A_e}-\phi_{\mu}$ plane for chosen fixed values of mass
parameters. For light SUSY masses they are very narrow. This means
that for fixed light mass parameters one needs strong fine tuning
between the phases in order to fulfill experimental limits.

\begin{figure}[htbp]
\begin{center}
\begin{tabular}{p{0.48\linewidth}p{0.48\linewidth}}
\mbox{\epsfig{file=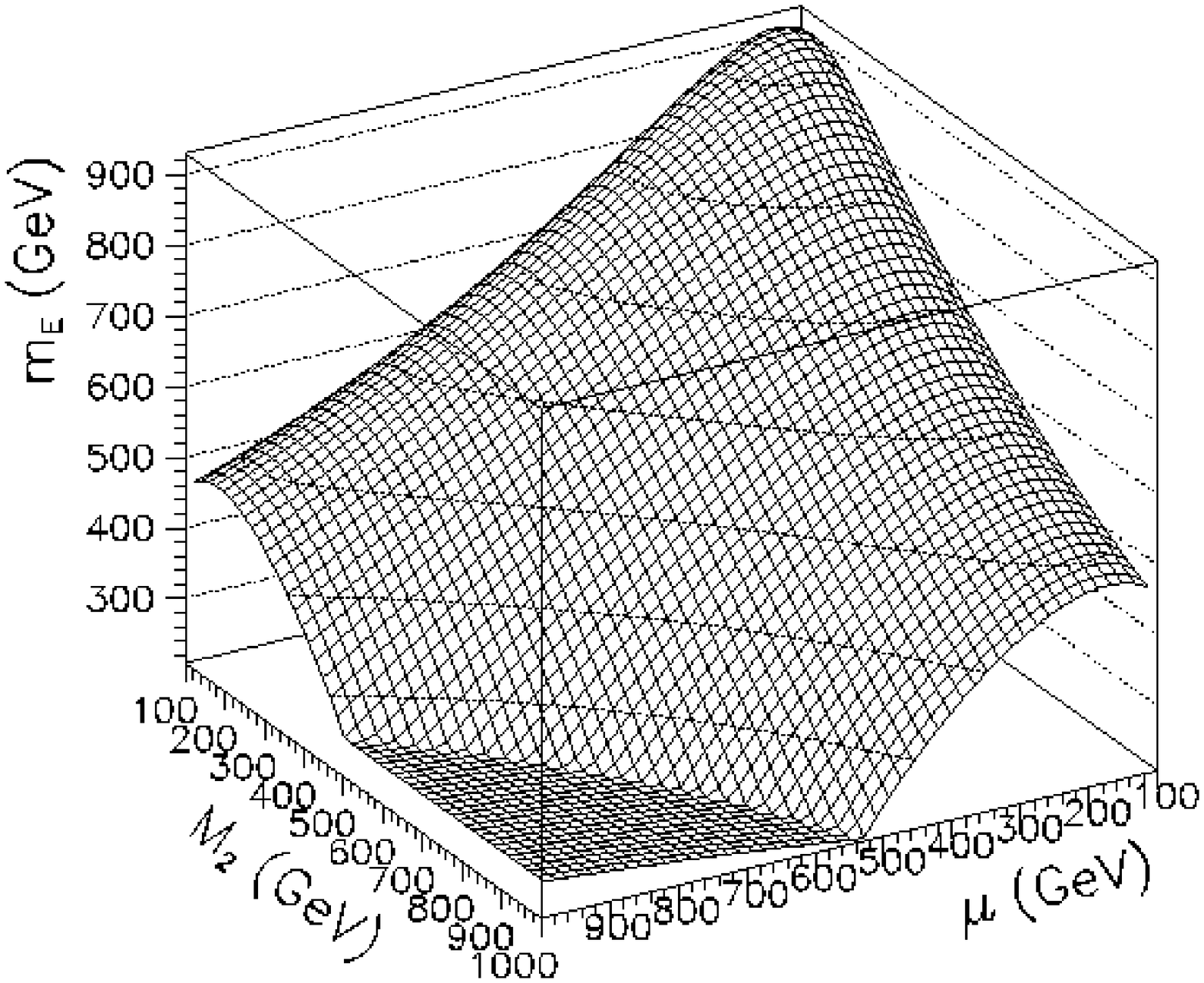,width=\linewidth}}&
\mbox{\epsfig{file=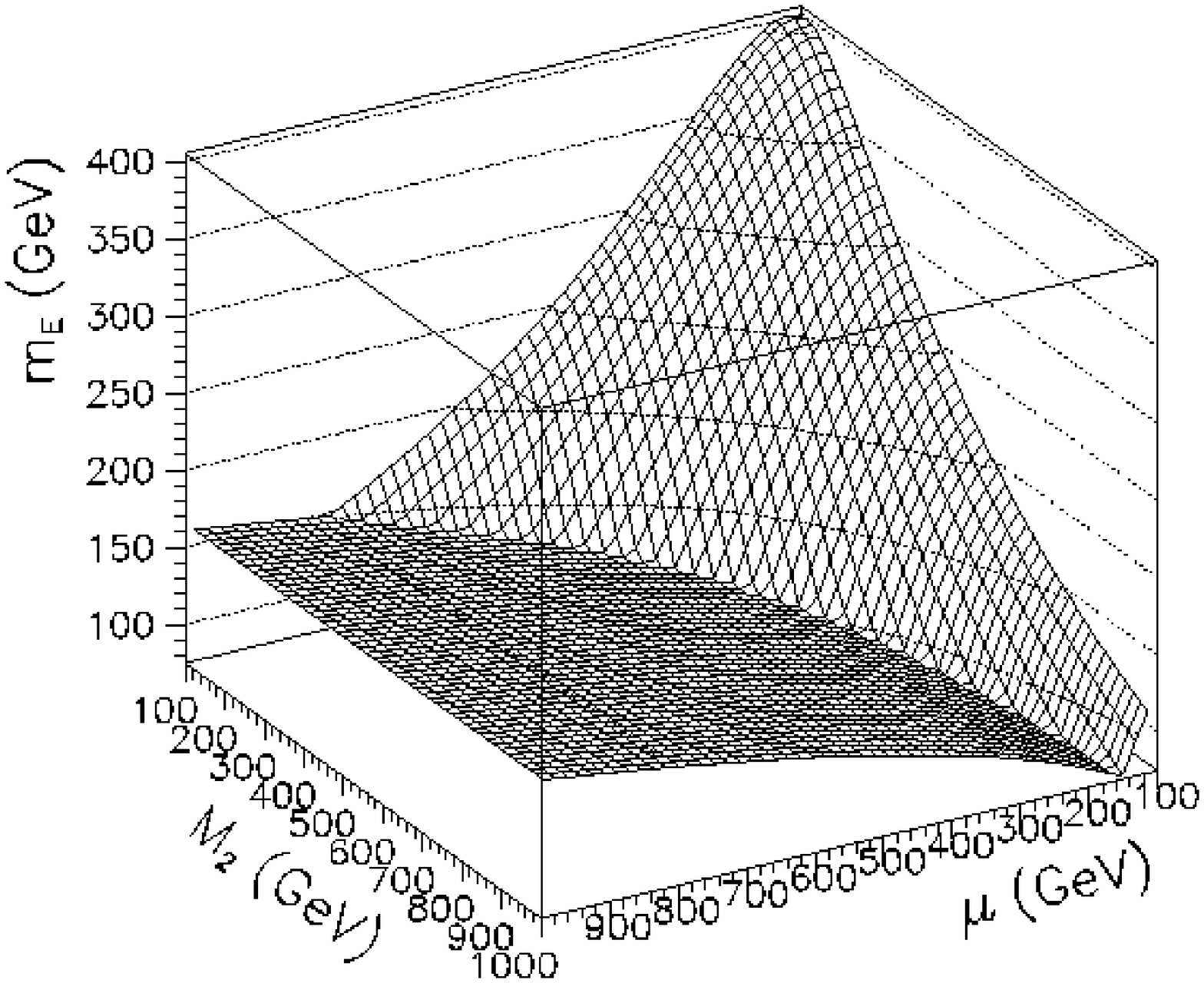,width=\linewidth}}\\
\end{tabular}
\vskip -5mm 
\caption{Regions for which generic limits on
$|\sin\phi_{\mu}|\tan\beta$ are stronger then, respectively, 0.2 (left
plots) and 0.05 (right plots). The plots are done for $M_1=100$ GeV,
degenerate left and right slepton masses are assumed.
\label{fig:e_ampl_surf_nogut}}
\end{center}
\end{figure}

\begin{figure}[htbp]
\begin{center}
\begin{tabular}{p{0.48\linewidth}p{0.48\linewidth}}
\mbox{\epsfig{file=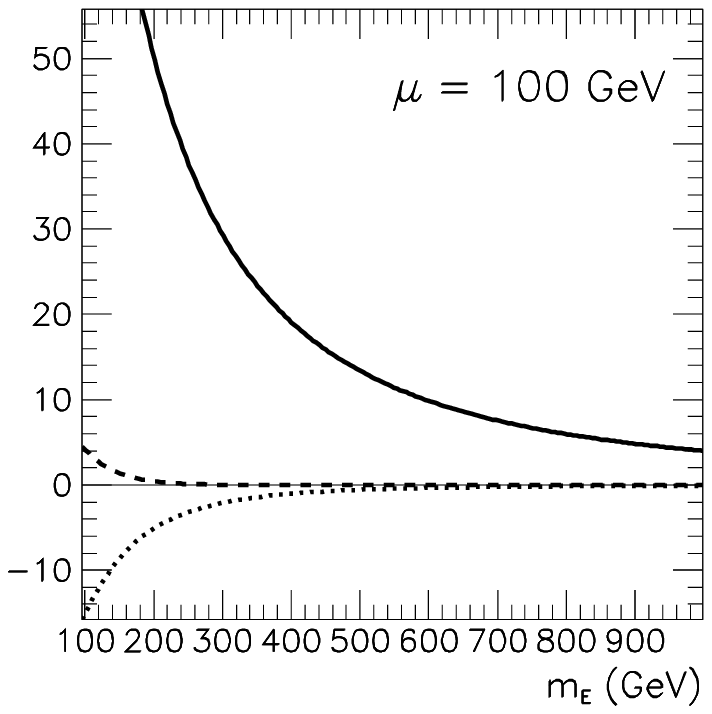,width=\linewidth}}&
\mbox{\epsfig{file=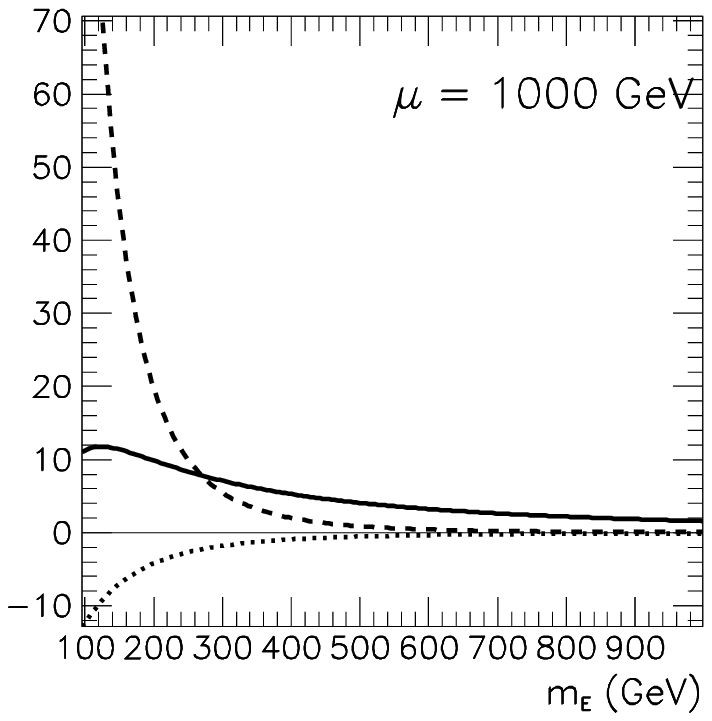,width=\linewidth}}
\end{tabular}
\vskip -5mm 
\caption{Relative signs and amplitudes of various contributions to the 
electron EDM, normalized to (divided by) the experimental
limit. Solid, dashed and dotted lines: coefficients of
$\sin(\phi_{\mu}+\phi_2)\tan\beta$, $\sin(\phi_{\mu}+\phi_1)\tan\beta$
and $|A_e|/m_E\sin(\phi_{A_e}-\phi_1)$ respectively. $|M_1|=|M_2|=100$
GeV and degenerate left and right slepton masses assumed.
\label{fig:e_canc_terms}}
\end{center}
\end{figure}

We shall discuss now the general case, with non-universal gaugino
masses. The results for the magnitude of individual terms remain
qualitatively similar. This is shown in
Fig.~\ref{fig:e_ampl_surf_nogut} for $M_1=100$ GeV. The region of
strong constraints on the $\mu$ phase shrinks in $m_E$ with increasing
$M_1$.  One can also see again some subtle effects like the expansion
in $m_E$ of this region, for fixed $M_1$ and $M_2$ and increasing
$|\mu|$. This behaviour can be easily understood from the analytic
results of eq.~(\ref{eq:edm_l_ns_mi}), where one can identify the term
increasing with $|\mu|$.
The magnitude of individual contributions as a function of $m_E$ has
very similar behaviour as in the universal case -- again, for small
$|\mu|$ chargino contribution dominates for all values of $|M_1|$ and
$|M_2|$. The only possible cancellations for this $|\mu|$ range are
between $\mu$ and $A_e$ phases. For larger values of $|\mu|> 700$
GeV the magnitude of individual terms may become comparable. For
instance the $\mu$ phase coefficients in $E_c$ and $E_{Ns}$ terms
(eqs.~(\ref{eq:edm_l_c_mi},\ref{eq:edm_l_ns_mi})) becomes comparable
for $|\mu|/m_E$ fixed by the ratio $|M_1/M_2|$. With arbitrary
relative phase of $M_1$ and $M_2$ it is possible to cancel the terms
proportional to the $\mu$ phase.  To study this possibility it is more
convenient to plot the contributions proportional to $\mathrm{Im}(\mu
M_1)$ and $\mathrm{Im}(\mu M_2)$ (Fig.~\ref{fig:e_canc_terms}).
They are comparable for $m_E/|\mu|\sim 1/5-1/3$, depending on
$|M_1/M_2|$ ratio. It is clear that choosing $\phi_1$ and $\phi_2$
phases such that $\sin(\phi_{\mu}+\phi_2)$ and
$\sin(\phi_{\mu}+\phi_1)$ have opposite signs, e.g. $\phi_1 - \phi_2
\sim \pi$, would give cancellation at these points.

\begin{figure}[htbp]
\begin{center}
\begin{tabular}{p{0.48\linewidth}p{0.48\linewidth}}
\mbox{\epsfig{file=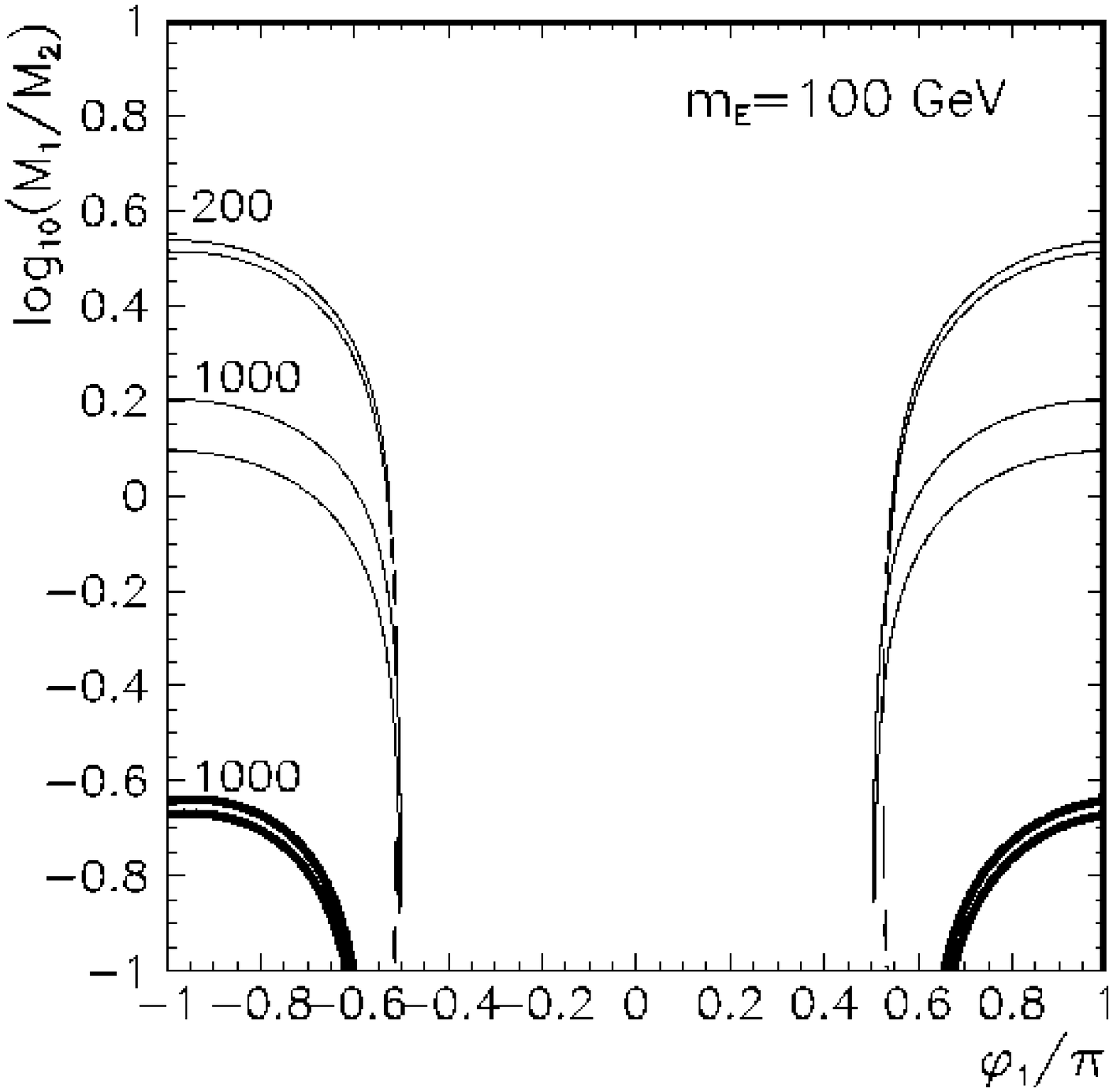,width=\linewidth}}&
\mbox{\epsfig{file=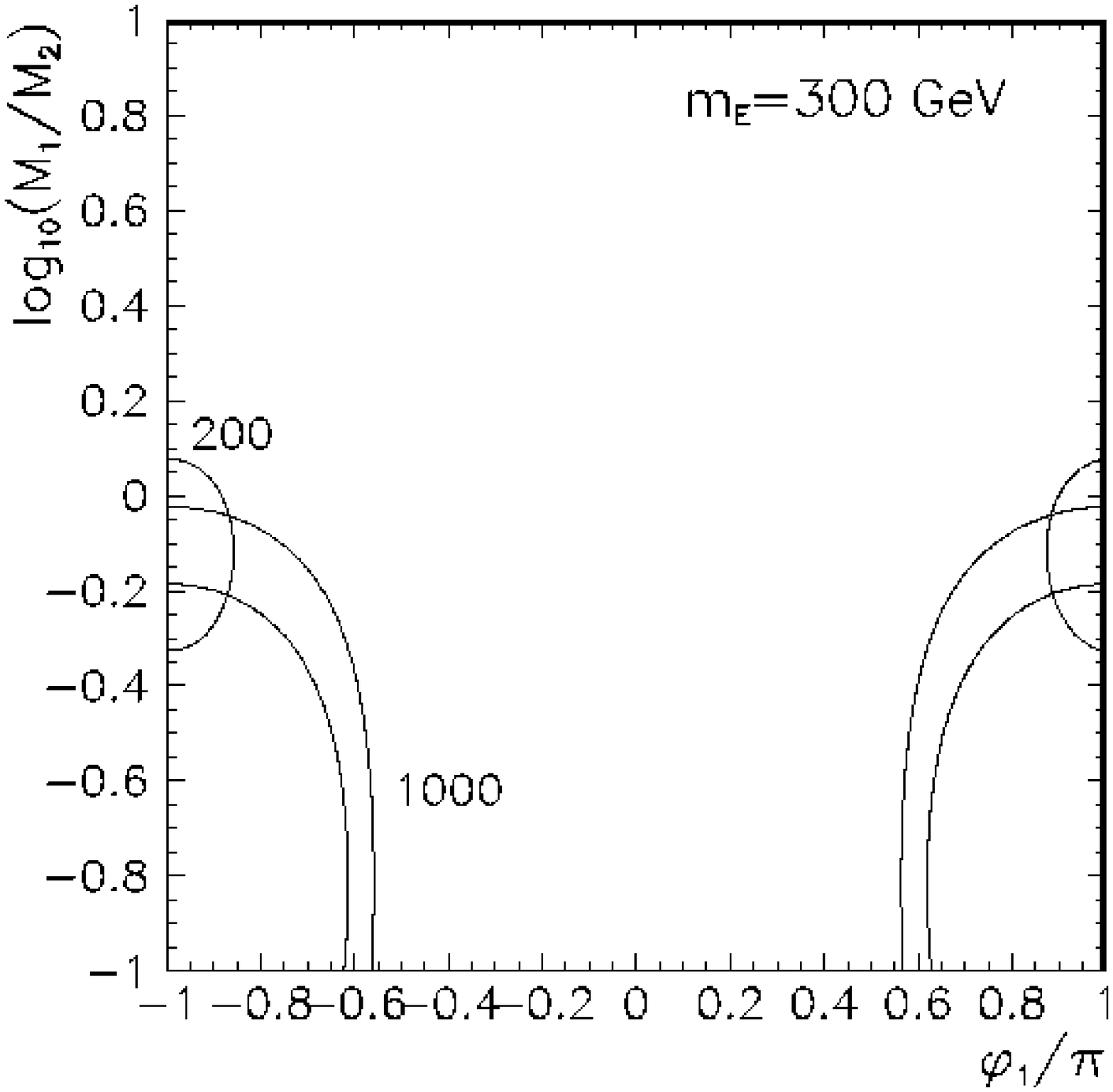,width=\linewidth}}
\end{tabular}
\vskip -5mm 
\caption{Allowed regions for $M_1$ phase as a function
of $|M_1/M_2|$ for some choices of mass parameters, $\tan\beta=2$ and
$\phi_{\mu}=\pi/2$. Thicker lines: $|\mu|=200$ GeV, thinner lines:
$|\mu|=1000$ GeV. $|M_2|=200$ or $1000$ GeV (marked on the plots),
$\phi_2=\phi_{A_e}=0$ and degenerate left and right slepton masses
assumed.
\label{fig:e_cont_mph}}
\end{center}
\end{figure}

To give a more specific example, lets assume $\phi_2=0$ and maximal
$\phi_{\mu}=\pi/2$ (one can always achieve $\phi_2=0$ by field
redefinition). In such a case, this new possibility of cancellation
applies for $|\phi_1-\phi_2|\equiv|\phi_1|>\pi/2$.  For very light
$|M_1|$ and $m_E$, $|\phi_1|\sim\pi/2$ is required, in order to
suppress the term proportional to $\mathrm{Im}(\mu M_1)$. For heavier
$M_1$ and $m_E$ to avoid limits on $\phi_{\mu}$ one needs
$|\phi_1|\sim\pi$ -- otherwise the $\mathrm{Im}(\mu M_1)$ term is
suppressed too strongly. This behaviour is illustrated in
Fig.~\ref{fig:e_cont_mph}.

\section{EDM of the neutron}
\label{sec:neut}

\subsection{Formulae for the neutron EDM}

\begin{figure}[htbp]
\begin{center}
\begin{tabular}{lcr}
\begin{picture}(150,120)(0,0)
\ArrowLine(10,20)(30,20)
\Text(10,10)[]{\mbox{$d^I$}}
\Vertex(30,20){2}
\DashArrowLine(30,20)(90,20){7}
\Text(60,10)[]{\mbox{$U^+_k$}}
\Vertex(90,20){2}
\ArrowLine(90,20)(110,20)
\Text(110,10)[]{\mbox{$d^I$}}
\ArrowLine(30,20)(60,80)
\Text(37,50)[r]{\mbox{$(C_j^+)^C$}}
\ArrowLine(60,80)(90,20)
\Text(85,50)[l]{\mbox{$(C_j^+)^C$}}
\Photon(60,80)(60,110){3}{3}
\Text(50,110)[c]{\mbox{$\gamma$}}
\end{picture}
&
\begin{picture}(150,120)(0,0)
\ArrowLine(10,20)(30,20)
\Text(10,10)[]{\mbox{$d^I$}}
\Vertex(30,20){2}
\ArrowLine(30,20)(90,20)
\Text(60,10)[]{\mbox{$N^0_j$}}
\Vertex(90,20){2}
\ArrowLine(90,20)(110,20)
\Text(110,10)[]{\mbox{$d^I$}}
\DashArrowLine(30,20)(60,80){7}
\Text(23,50)[l]{\mbox{$D_k^-$}}
\DashArrowLine(60,80)(90,20){7}
\Text(100,50)[r]{\mbox{$D_k^-$}}
\Photon(60,80)(60,110){3}{3}
\Text(50,110)[c]{\mbox{$\gamma$}}
\end{picture}
&
\begin{picture}(150,120)(0,0)
\ArrowLine(10,20)(30,20)
\Text(10,10)[]{\mbox{$d^I$}}
\Vertex(30,20){2}
\ArrowLine(30,20)(90,20)
\Text(60,10)[]{\mbox{$\tilde{g}^a$}}
\Vertex(90,20){2}
\ArrowLine(90,20)(110,20)
\Text(110,10)[]{\mbox{$d^I$}}
\DashArrowLine(30,20)(60,80){7}
\Text(23,50)[l]{\mbox{$D_k^-$}}
\DashArrowLine(60,80)(90,20){7}
\Text(100,50)[r]{\mbox{$D_k^-$}}
\Photon(60,80)(60,110){3}{3}
\Text(50,110)[c]{\mbox{$\gamma$}}
\end{picture}
\end{tabular}
\caption{Diagrams contributing to $d$-quark EDM.}
\label{fig:edm_d}
\end{center}
\end{figure}
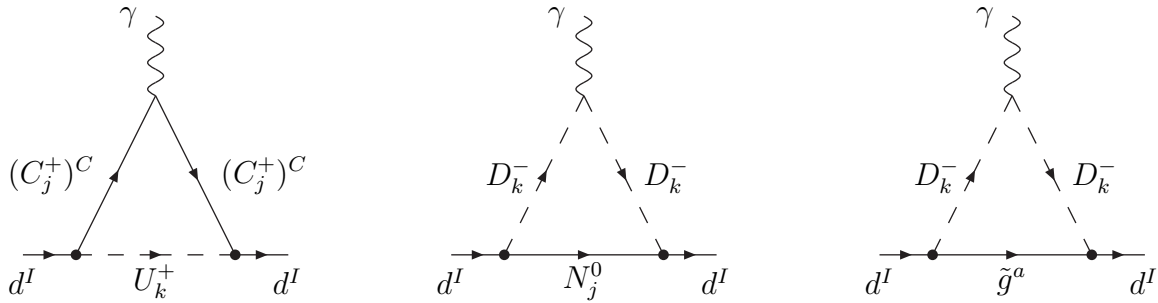

The structure of the neutron EDM is more complicated then in the
electron case. It can be approximately calculated as the sum of the
electric dipole moments of the constituent $d$ and $u$ quarks plus
additional contributions coming from the chromoelectric dipole moments
of quarks and gluons.  The chromoelectric dipole moment $C_q$ of a
quark is defined as:
\bea
{\cal L}_C = - \frac{i}{2} C_q \bar{q} \sigma_{\mu\nu} \gamma_5 T^a q
G^{\mu\nu a}
\label{eq:cdmdef}
\eea
The gluonic dipole moment $C_g$ is defined as:
\bea
{\cal L}_g = - \frac{1}{6} C_g f_{abc} G^a_{\mu\rho} G^{b\rho}_{\nu}
G^c_{\lambda\sigma}\epsilon^{\mu\nu\lambda\sigma}
\label{eq:gdmdef}
\eea
As an example, in Fig.~\ref{fig:edm_d} we list the diagrams
contributing to the $d$-quark electric dipole moment.

Exact calculation of the neutron EDM requires the full knowledge of
its wave function. We use the ``naive'' chiral quark model
approximation~\cite{QUARKMODEL}, which gives the following expression:
\bea
E_n = \frac{\eta_e}{3}(4E_d - E_u) + \frac{e\eta_c}{4\pi}(4C_d - C_u) 
+\frac{e\eta_g\Lambda_X}{4\pi}C_g
\label{eq:fullneut}
\eea
where $\eta_i$ and $\Lambda_X$ are the QCD correction factors and
chiral symmetry breaking scale, respectively, $\eta_e\approx 1.53$,
$\eta_c\approx\eta_g\approx 3.4$~\cite{ETAQCD}, $\Lambda_X=1.19$
GeV~\cite{QUARKMODEL}. We use also following values of light quark
masses: $m_d(\Lambda_X)=10$ MeV, $m_u(\Lambda_X)=7$
MeV~\cite{WEINBERG}.

Eq.~(\ref{eq:fullneut}) contains sizeable theoretical uncertainities
due to non-perturbative strong interactions. However, as we show in
the next section, for most parameter choices $E_d$ alone gives the
leading contribution to the neutron EDM. Therefore, one may hope that
those uncertainities affect mainly the overall normalization of the
neutron EDM.  They do not affect significantly the possible
cancellations between the phases (or in their coefficients), as long
as such cancellations must occur predominantly inside the $E_d$. For
instance, if one uses relativistic quark-parton
model~\cite{FLORES,BARTL} instead of the naive chiral quark model, the
contribution of quarks to the neutron EDM is weighted by~\cite{FLORES}
their contributions $\Delta_q$ to the spin of proton, measured in
polarized lepton-nucleon scattering: $E_n = \eta_e(\Delta_u E_d +
\Delta_d E_u + \Delta_s E_s)$.  For the neutron, applying the isospin
relations, we have $(\Delta_u)_n=\Delta_d$, $(\Delta_d)_n=\Delta_u$,
$(\Delta_s)_n=\Delta_s$. With the assumption of degenerate masses and
phases of the $A$ parameters of the first two generations of squarks,
$E_d$ and $E_s$ have exactly the same structure and differ only by the
quark mass multiplying the appropriate loop integrals:
$E_d/m_d=E_s/m_s$ (see eq.~(\ref{eq:edm_d_exp})). Therefore, accepting
the values given in ref.~\cite{KARLINER}: $\Delta_u=0.82$,
$\Delta_d=-0.44$, $\Delta_s=-0.11$, one can estimate that, in the
quark-parton model, the contribution from the $s$ sea-quark, not
present at all in the naive chiral model, is predicted to dominate.
The leading contributions to the neutron EDM in the considered models
are (we assume $m_s\approx 150$ MeV):
\bea
\mathrm{Chiral~quark~model:}&\hskip 1cm& E_n\approx \frac{4}{3}\eta_e
E_d\approx 1.33\eta_e E_d\nonumber\\
\mathrm{Quark~parton~model:}&\hskip 1cm& 
E_n\approx (\Delta_u+\frac{m_s}{m_d}\Delta_s)\eta_e E_d\approx -0.83 
\eta_e E_d
\label{eq:neut_appr}
\eea
and the $u$ quark contribution is smaller. The coefficients of $E_d$
for the two models differ in magnitude and even in sign but in both
models the limits on the phases can be avoided if the $E_d$ itself
vanishes due to some internal cancellations. Of course, one should
remember that the estimate~(\ref{eq:neut_appr}) is subject to some
uncertainity in $\Delta_s$ and a larger cancellation between $d$ and
$s$ quark contributions, leading (at least in the low $\tan\beta$
region~\cite{FLORES}) to the dominance of the $u$ quark contribution,
is not excluded. Thus, there are uncertainities related to the choice
of a particular quark model and uncertainities within the chosen model
(like e.g. $\mathrm{Im}A_d = \mathrm{Im}A_s$ assumption). Any
theoretical calculation of the overall normalization of the neutron
EDM should be considered as a qualitative one. The pattern of
cancellations we discuss in the naive chiral quark model can be more
generally trusted to the extend the smallness of the $u$ quark
contribution compared to the $d$ and $s$ quark contributions can be
trusted as a model independent feature. At present the limits on the
phases given by the electron EDM are more precise and better
established.

It was recently pointed out~\cite{PILAFTSIS} that 2-loop contributions
to the neutron EDM may be numerically significant, especially for
large $\tan\beta$ regime. Unlike most of the terms in
eq.~(\ref{eq:fullneut}), they depend mainly on the masses and mixing
parameters of the third generation of squarks. Therefore, they are
especially important in the case of the third generation of squarks
significantly lighter than the first two generation, so that the
1-loop contributions are suppressed.  We do not include such
corrections in the present analysis.

The formulae for the up- and down quark electric dipole moment are the
following:
\bea
E_d^I&=&\frac{e}{8\pi^2}\sum_{j=1}^2\sum_{k=1}^6 m_{C_j}\mathrm{Im}
\left((V_{d\tilde{U}C})_L^{Ikj}(V_{d\tilde{U}C})_R^{Ikj\star}\right)
\left(C_{11}(m_{C_j}^2,m_{\tilde{U}_k}^2)
+\frac{1}{3}C_{12}(m_{\tilde{U}_k}^2,m_{C_j}^2)\right)\nonumber\\
&-& \frac{e}{48\pi^2}\sum_{j=1}^4\sum_{k=1}^6 m_{N^0_j}\mathrm{Im}
\left((V_{d\tilde{D}N})_L^{Ikj}(V_{d\tilde{D}N})_R^{Ikj\star}\right)
C_{12}(m_{\tilde{D}_k}^2,m_{N^0_j}^2)
\nonumber\\
&-& \frac{2e\alpha_s}{9\pi}|M_3|\sum_{k=1}^6\mathrm{Im}
(Z_{DL}^{Ik}Z_{DR}^{Ik\star})C_{12}(m_{\tilde{D}_k}^2,|M_3|^2)
\eea
\bea
E_u^I&=&-\frac{e}{8\pi^2}\sum_{j=1}^2\sum_{k=1}^6m_{C_j}\mathrm{Im}
\left((V_{u\tilde{D}C})_L^{Ikj}(V_{u\tilde{D}C})_R^{Ikj\star}\right)
\left(C_{11}(m_{C_j}^2,m_{\tilde{D}_k}^2)
+\frac{1}{6}C_{12}(m_{\tilde{D}_k}^2,m_{C_j}^2)\right)\nonumber\\
&+&\frac{e}{24\pi^2}\sum_{j=1}^4\sum_{k=1}^6m_{N^0_j}\mathrm{Im}
\left((V_{u\tilde{U}N})_L^{Ikj}(V_{u\tilde{U}N})_R^{Ikj\star})
C_{12}(m_{\tilde{U}_k}^2,m_{N^0_j}^2\right)
\nonumber\\
&+& \frac{4e\alpha_s}{9\pi}|M_3|\sum_{k=1}^6\mathrm{Im}
(Z_{UL}^{Ik}Z_{UR}^{Ik\star})C_{12}(m_{\tilde{U}_k}^2,|M_3|^2)
\eea
Chromoelectric dipole moments of the quarks are given by:
\bea
C_d^I&=&\frac{g_s}{16\pi^2}\sum_{j=1}^2\sum_{k=1}^6m_{C_j}\mathrm{Im}
\left((V_{d\tilde{U}C})_L^{Ikj}(V_{d\tilde{U}C})_R^{Ikj\star}\right)
C_{12}(m_{\tilde{U}_k}^2,m_{C_j}^2)\nonumber\\ &+&
\frac{g_s}{16\pi^2}\sum_{j=1}^4\sum_{k=1}^6m_{N^0_j}\mathrm{Im}
\left((V_{d\tilde{D}N})_L^{Ikj}(V_{d\tilde{D}N})_R^{Ikj\star}\right)
C_{12}(m_{\tilde{D}_k}^2,m_{N^0_j}^2)\nonumber\\ &-&
\frac{g_s^3}{8\pi^2}|M_3|\sum_{k=1}^6\mathrm{Im}
(Z_{DL}^{Ik}Z_{DR}^{Ik\star})
\left(3C_{11}(|M_3|^2,m_{\tilde{D}_k}^2) +\frac{1}{6}
C_{12}(m_{\tilde{D}_k}^2,|M_3|^2)\right)
\eea
\bea
C_u^I&=&\frac{g_s}{16\pi^2}\sum_{j=1}^2\sum_{k=1}^6m_{C_j}\mathrm{Im}
\left((V_{u\tilde{D}C})_L^{Ikj}(V_{u\tilde{D}C})_R^{Ikj\star}\right)
C_{12}(m_{\tilde{D}_k}^2,m_{C_j}^2)\nonumber\\
&+&\frac{g_s}{16\pi^2}\sum_{j=1}^4\sum_{k=1}^6m_{N^0_j}\mathrm{Im}
\left((V_{u\tilde{U}N})_L^{Ikj}(V_{u\tilde{U}N})_R^{Ikj\star}\right)
C_{12}(m_{\tilde{U}_k}^2,m_{N^0_j}^2)\nonumber\\ &-&
\frac{g_s^3}{8\pi^2}|M_3|\sum_{k=1}^6\mathrm{Im}
(Z_{UL}^{Ik}Z_{UR}^{Ik\star})
\left(3C_{11}(|M_3|^2,m_{\tilde{U}_k}^2)
+\frac{1}{6}C_{12}(m_{\tilde{U}_k}^2,|M_3|^2)\right)
\eea
Finally, the gluon chromoelectric dipole moment is given by:
\bea
C_g &=&{3 \alpha_s^2 g_s m_t\over
16\pi^2}\mathrm{Im}(Z_{UL}^{36\star}Z_{UR}^{36}) {m_{\tilde{t}_1}^2 -
m_{\tilde{t}_2}^2\over |M_3|^5} H\left({m_{\tilde{t}_1}^2 \over
|M_3|^2}, {m_{\tilde{t}_2}^2 \over |M_3|^2}, {m_t^2 \over |M_3|^2}
\right)
\eea
where the definition of the 2-loop function $H$ can be found
in~\cite{CG_HDEF}.

The full form of all necessary fermion-sfermion-chargino/neutralino
vertices can be found in \ref{app:lagr}. In \ref{app:massins} we give
also mass insertion expressions for the quark electric and
chromoelectric dipole moments. Although somewhat more complicated than
in the case of leptons, they are very useful for qualitative
understanding of the neutron EDM behaviour.

\subsection{Limits on phases}

The neutron EDM depends on more phases than the electron EDM. All
electric and chromoelectric dipole moments depend on the common $\mu$
phase, but some of them are proportional to $\mu\tan\beta$ and others
to $\mu\cot\beta$, hence the limit on $\mu$ phase does not scale
simply like $1/\tan\beta$, as it was in the electron case. In
addition, the quark moments depend on the phases of the two LR mixing
parameters of the first generation of squarks, $A_d$ and $A_u$. The
gluonic chromoelectric dipole moment depends in principle on all $A$
parameters and squark masses, but contributions from different squark
generation are weighted by the respective fermion mass, so we take
into account only the dominant stop contribution, dependent on the
$A_t$ parameter.

In practice, the analysis of the dependence of the neutron EDM on SUSY
parameters appears less complicated than suggested by the above list,
as some of the parameters have small numerical importance. As
discussed below, the result is most sensitive to the squark masses of
the first generation (left and right), gaugino masses, $|\mu|$, $A_u$, 
$A_d$ and $\tan\beta$. 

The number of free parameters can be further reduced by assuming GUT
unification with universal boundary conditions. Such a variant was
thoroughly discussed in~\cite{NATH,BARTL}, so we do not repeat the
full RGE analysis here, however its results can be qualitatively read
also from the figures presented in this section. This can be done with
the use of the following observations:
\begin{itemize}
\item[{\sl i)}] As mentioned above, the neutron EDM is sensitive mostly to 
the masses of the first generation of squarks.  Assuming universal
sfermion masses at the GUT scale one can to a good approximation keep
them degenerate also at $M_Z$ scale. The remnant of the GUT evolution
is their relation to the gaugino masses: $m_Q^2\approx m_D^2\approx
m_U^2\approx m_0^2 + 6.5M_{1/2}^2\approx m_0^2 + 10|M_2|^2$, which
leads to the relation $m_Q\approx m_U\approx m_D\geq 3 M_2$.
\item[{\sl ii)}] The $\mu$ phase does not run. The $\mu$ itself runs weakly,
it means that $\mathrm{Im}\mu$ also runs weakly. It is a free
parameter anyway.
\item[{\sl iii)}] The imaginary parts of the first generation $A$
parameters, $\mathrm{Im}A_u$ and $\mathrm{Im}A_d$, do not run, apart
from the small corrections proportional to the Yukawa couplings of
light fermions. Real parts of $A_u$ and $A_d$ run approximately in the
same way. Therefore universal boundary conditions at the GUT scale
lead simply to $\phi_{A_u}=\phi_{A_d}$ at the $M_Z$ scale. 
\item[{\sl iv)}] RGE running suppresses the $A_t$ phase (present in
the chromoelectric dipole moment of gluons $C_g$).  Therefore, the low
energy constraints are easy to satisfy even with large $\phi_{A_t}$ at
the GUT scale. The limits on $\phi_{A_t}$ at the electroweak scale
appear themeselves to be rather weak.
\item[{\sl v)}] With universal gaugino masses and phases,
$M_1/\alpha_1=M_2/\alpha_2=M_3/\alpha_3$, the common gaugino phase can
be completely rotated away.
\end{itemize}
Using {\sl i)-v)} one can use our plots for the universal GUT case,
just assuming common $A$ phase, neglecting $\phi_{A_t}$ and looking at
the part of plots for which $m_Q\geq 3 M_2$. Actually, in all Figures
of this Section we keep degenerate squark mass parameters
$M_Q=M_D=M_U$, so that the physical masses differ by D-terms only. We
plot the results in terms of the physical mass of the $D$-squark
$m_D$.

\begin{figure}[htbp]
\begin{center}
\begin{tabular}{p{0.48\linewidth}p{0.48\linewidth}}
\mbox{\epsfig{file=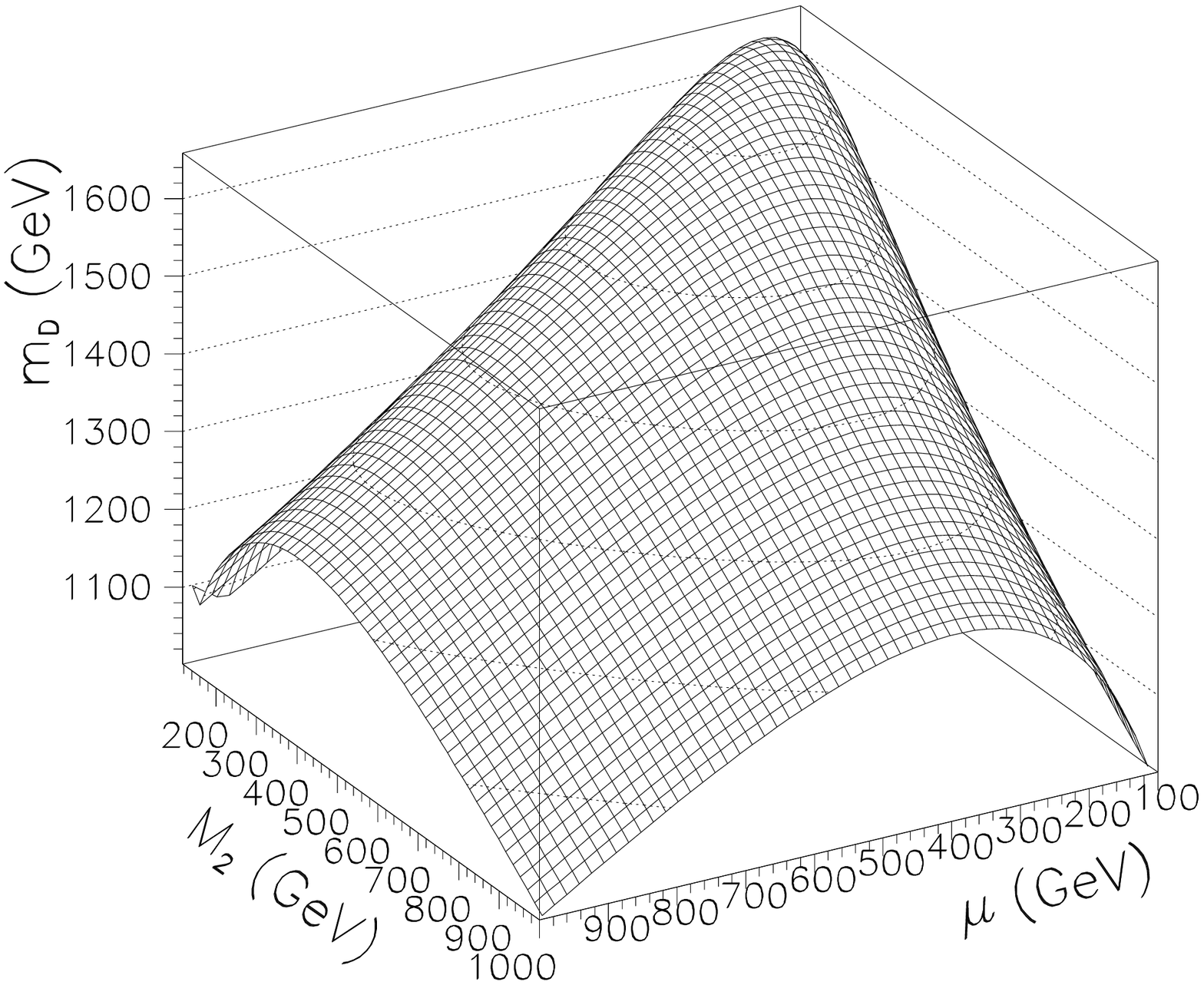,width=\linewidth}}&
\mbox{\epsfig{file=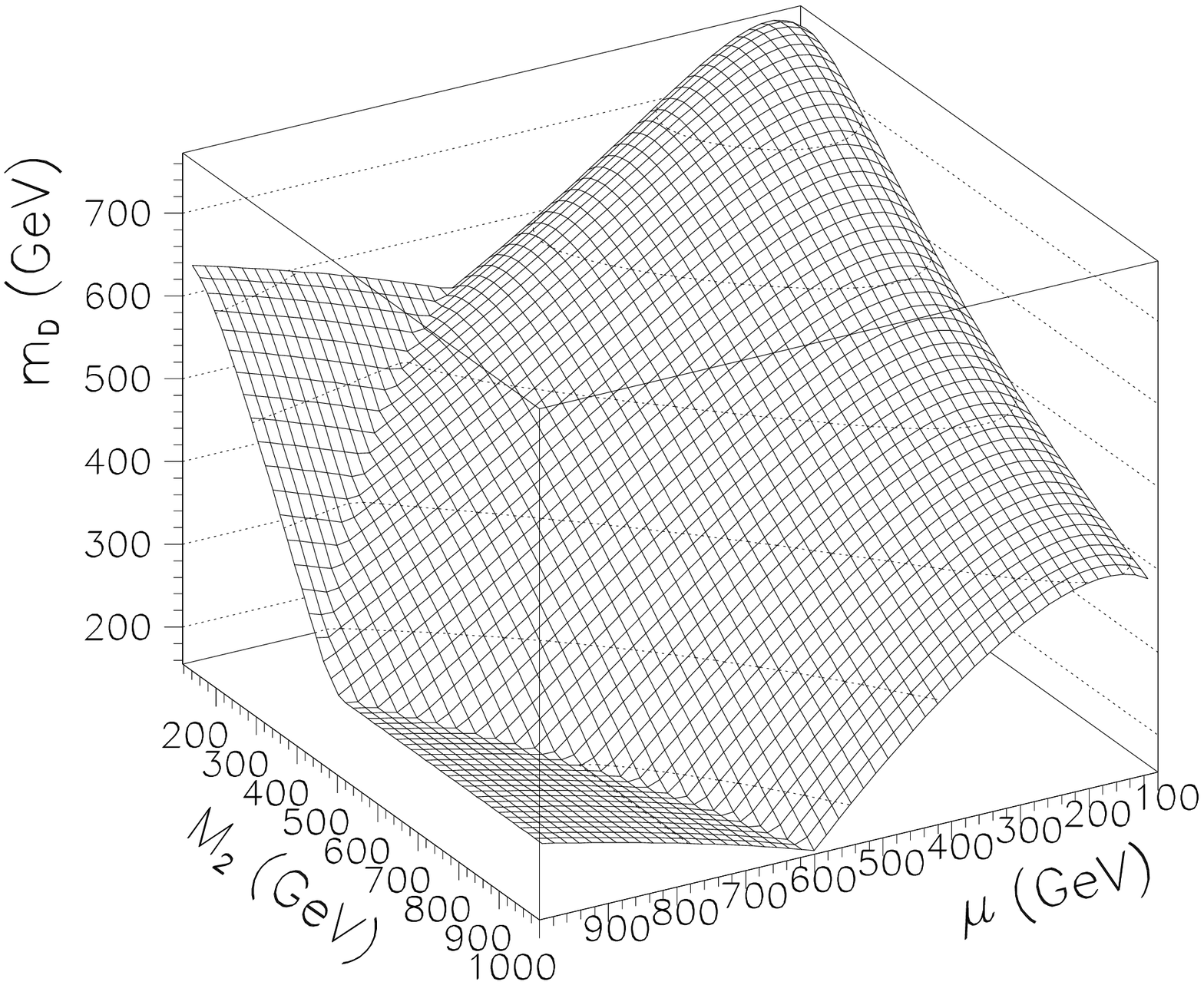,width=\linewidth}}
\end{tabular}
\vskip -5mm 
\caption{Regions for which generic limits on
$|\sin\phi_{\mu}|\tan\beta$ given by neutron EDM are stronger then,
respecively, 0.2 (left plot) and 0.05 (right plot). Degenerate squark
masses and GUT related $M_1, M_2, M_3$ assumed.
\label{fig:n_ampl_surf_gut}}
\end{center}
\end{figure}

We consider first the limits on the $\mu$ phase, neglecting the
possibility of $\mu-A$ cancellations. In
Fig.~\ref{fig:n_ampl_surf_gut} (analogous to
Fig.~\ref{fig:e_ampl_surf_gut}) we show where the generic limits for
the $\mu$ phase given by the neutron EDM are strong. We plot there the
area where the limit on $|\sin\phi_{\mu}|\tan\beta$ given separately
by each of the contributions present in~eq.(\ref{eq:fullneut}) is
stronger then 0.2 or 0.05 (in addition we consider separately the
chargino, neutralino and gluino contributions to $E_d$). For small
$|\mu|$, $|M_2|$, squark masses $m_Q\sim m_D\sim m_U>1600(750)$ GeV are
required to avoid the assumed limits. 
%
%

\begin{figure}[htbp]
\begin{center}
\begin{tabular}{p{0.48\linewidth}p{0.48\linewidth}}
\epsfig{file=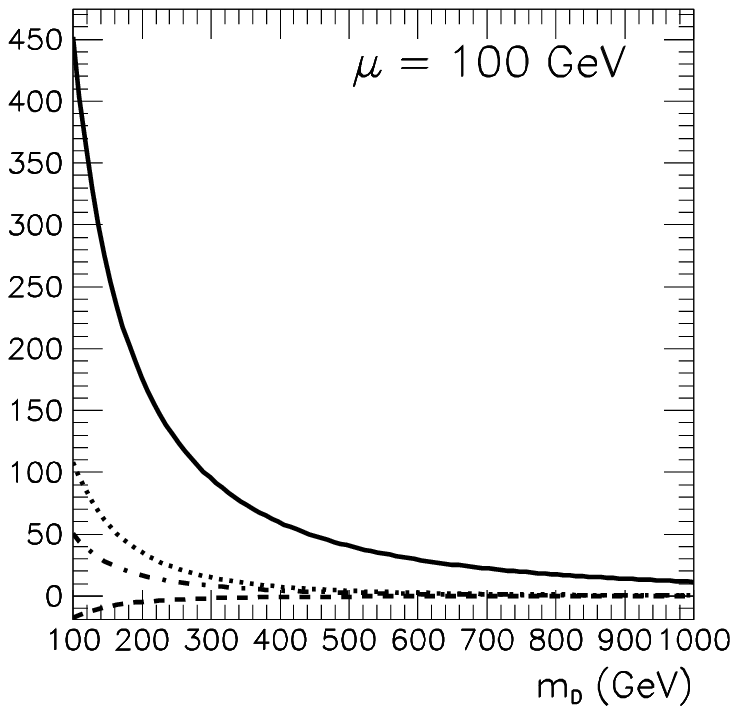,width=\linewidth}&
\epsfig{file=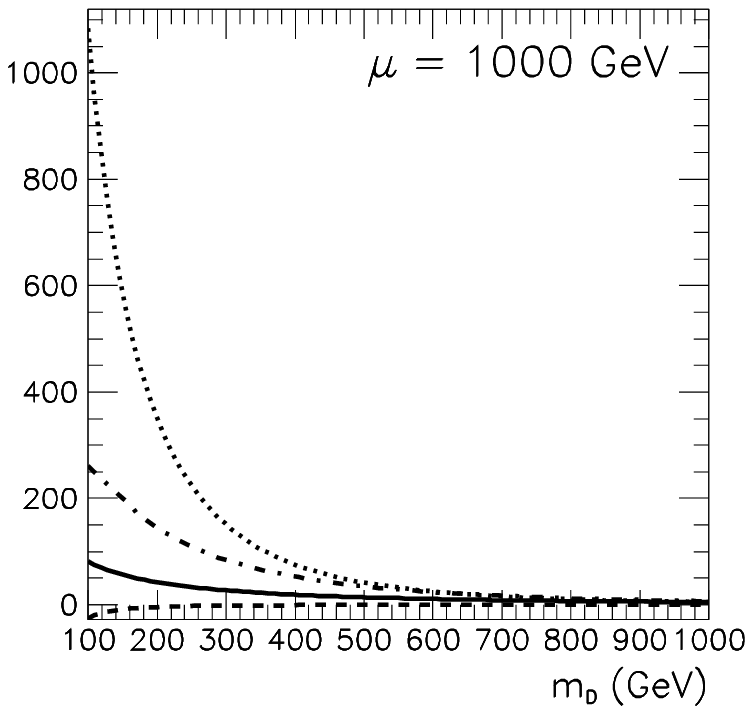,width=\linewidth}
\end{tabular}
\vskip -5mm 
\caption{Contributions to $\sin\phi_{\mu}\tan\beta$
coefficient in the neutron EDM, normalized to (divided by) the
experimental limit. Solid, dashed, dotted line: chargino, neutralino
and gluino contributions to $E_d$; dotted-dashed line:
$E_u+C_d+C_u+C_g$ summed up. $|M_2|=100$ GeV, $M_1,M_2,M_3$
GUT-related, $\phi_{A_i}=0$ and degenerate squark masses assumed.
\label{fig:n_ampl}}
\end{center}
\end{figure}

\begin{figure}[htbp]
\begin{center}
\begin{tabular}{p{0.48\linewidth}p{0.48\linewidth}}
\mbox{\epsfig{file=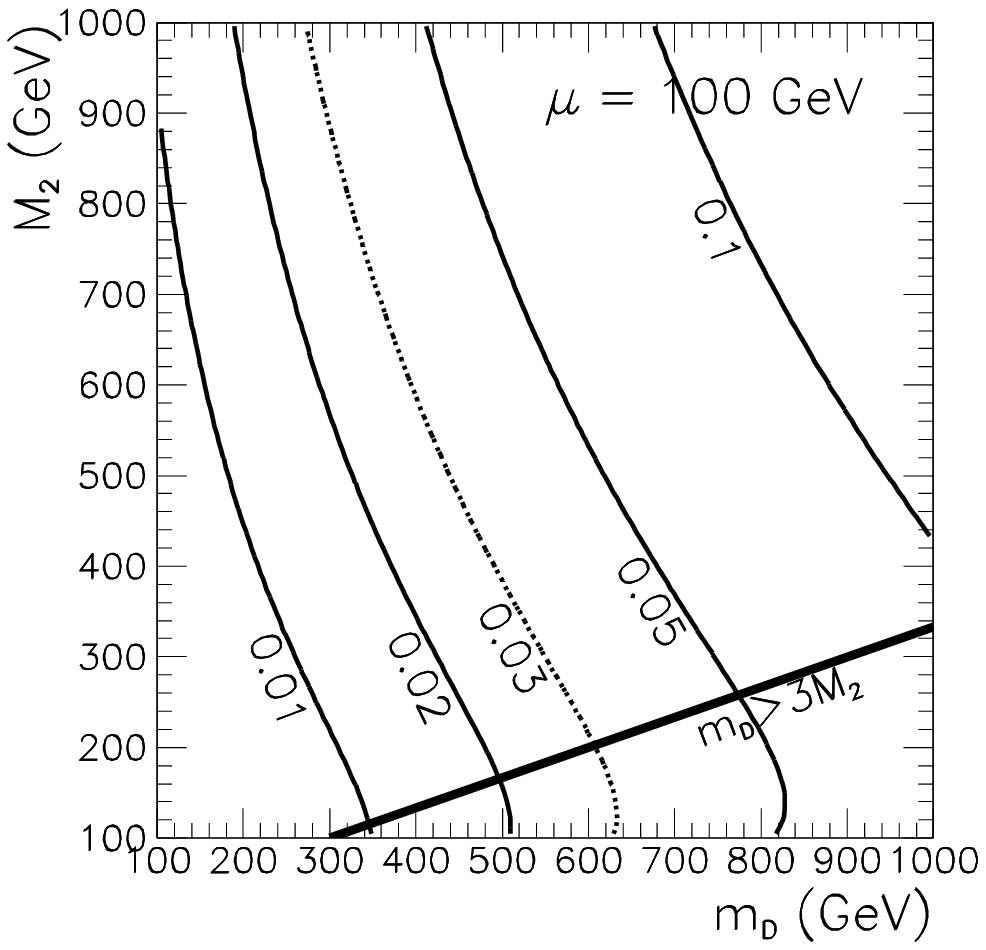,width=\linewidth}}&
\mbox{\epsfig{file=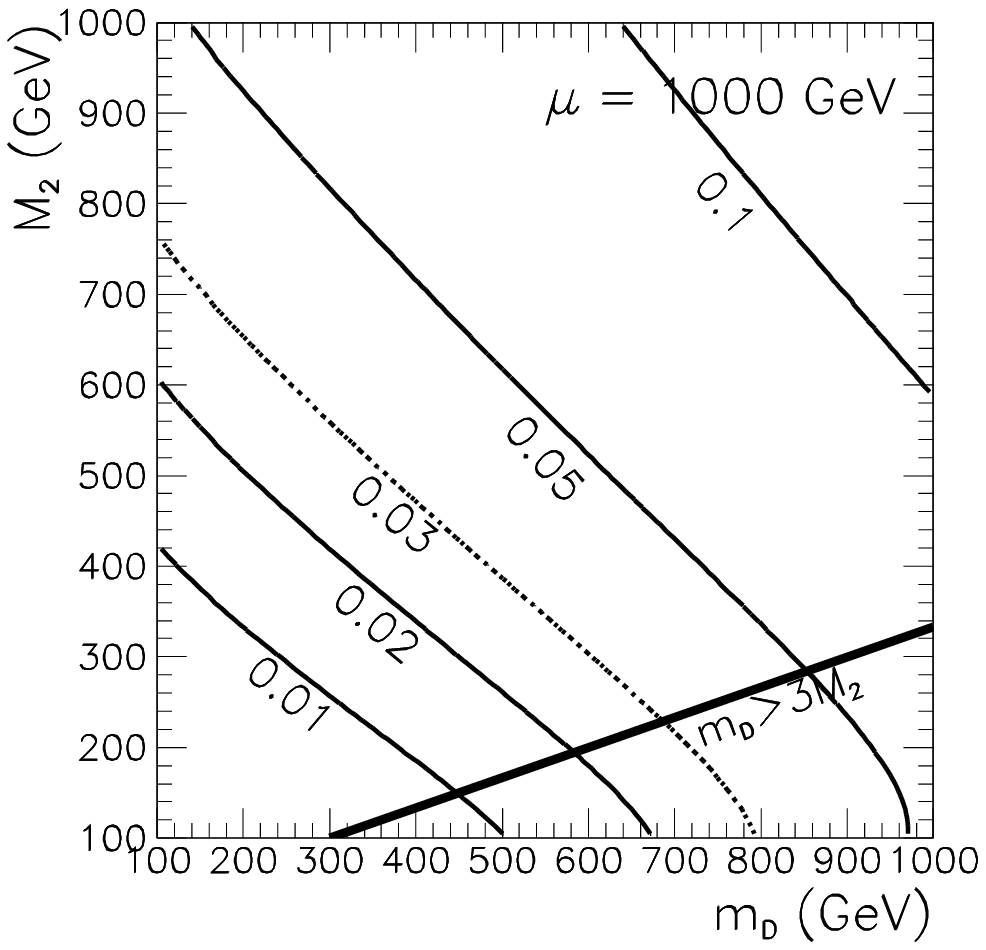,width=\linewidth}}
\end{tabular}
\caption{Limits on $|\sin\phi_{\mu}|\tan\beta$ given by the neutron
EDM measurements. $M_1,M_2,M_3$ GUT-related,
$\phi_{A_u}=\phi_{A_d}=\phi_{A_t}=0$ and degenerate squark masses
assumed.}
\label{fig:n_cont_mu}
\end{center}
\end{figure}

The dominant contributions to the coefficient multiplying
$\sin\phi_{\mu}$ come from the first term of eq.~(\ref{eq:fullneut}),
i.e. from the d-quark electric dipole moment, as illustrated in
Fig.~\ref{fig:n_ampl}. The only exception is large $|\mu|$ and light
gauginos case, where also $C_d$ becomes comparable to the other
term. In principle, relative importance of various contributions
changes with $\tan\beta$, as $E_d,C_d$ are proportional to
$\mu\tan\beta$ and $E_u,C_u,C_g$ to $\mu\cot\beta$. However, $E_d$ and
eventually $C_d$ always dominate and the $\mu$ phase coefficient
scales again, like in the electron case, as $\tan\beta$
(Fig.~\ref{fig:n_ampl} has been done for $\tan\beta=2$).
The largest contributions to $\mu$ phase coefficient are given by the
chargino and gluino (for small and large $|\mu|$, respectively)
diagrams. They have the same sign, so again, like in the electron
case, the total $\mu$ phase coefficient may disappear only if one
allows the non-universal gaugino phases. The limits on
$|\sin\phi_{\mu}|\tan\beta$ on $m_D-|M_2|$ plane are plotted in
Fig.~\ref{fig:n_cont_mu}.

\begin{figure}[htbp]
\begin{center}
\mbox{\epsfig{file=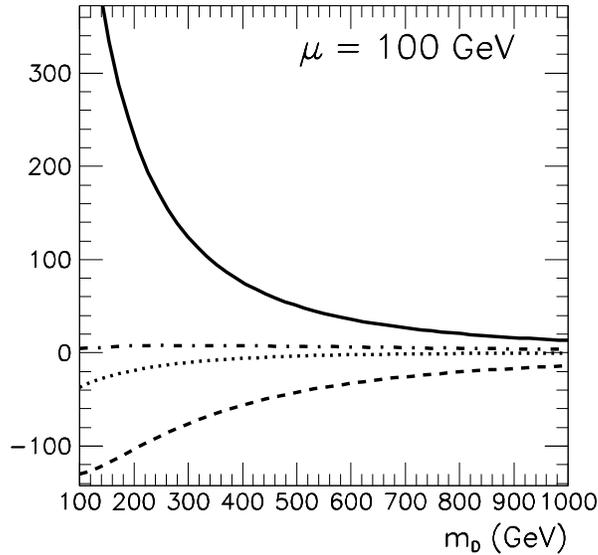,width=0.5\linewidth}}
\vskip -5mm 
\caption{Coefficients of $\sin\phi_{\mu}\tan\beta$, $|A_d|/m_D\sin\phi_{A_d}$,
$|A_u|/m_U\sin\phi_{A_u}$ and $|A_t|/m_T\sin\phi_{A_t}$ terms in the
neutron EDM (solid, dashed, dotted and dashed-dotted lines
respectively). $M_1,M_2,M_3$ GUT-related and degenerate squark masses
assumed.
\label{fig:n_ampl_a}}
\end{center}
\end{figure}

\begin{figure}[htbp]
\begin{center}
\begin{tabular}{p{0.48\linewidth}p{0.48\linewidth}}
\mbox{\epsfig{file=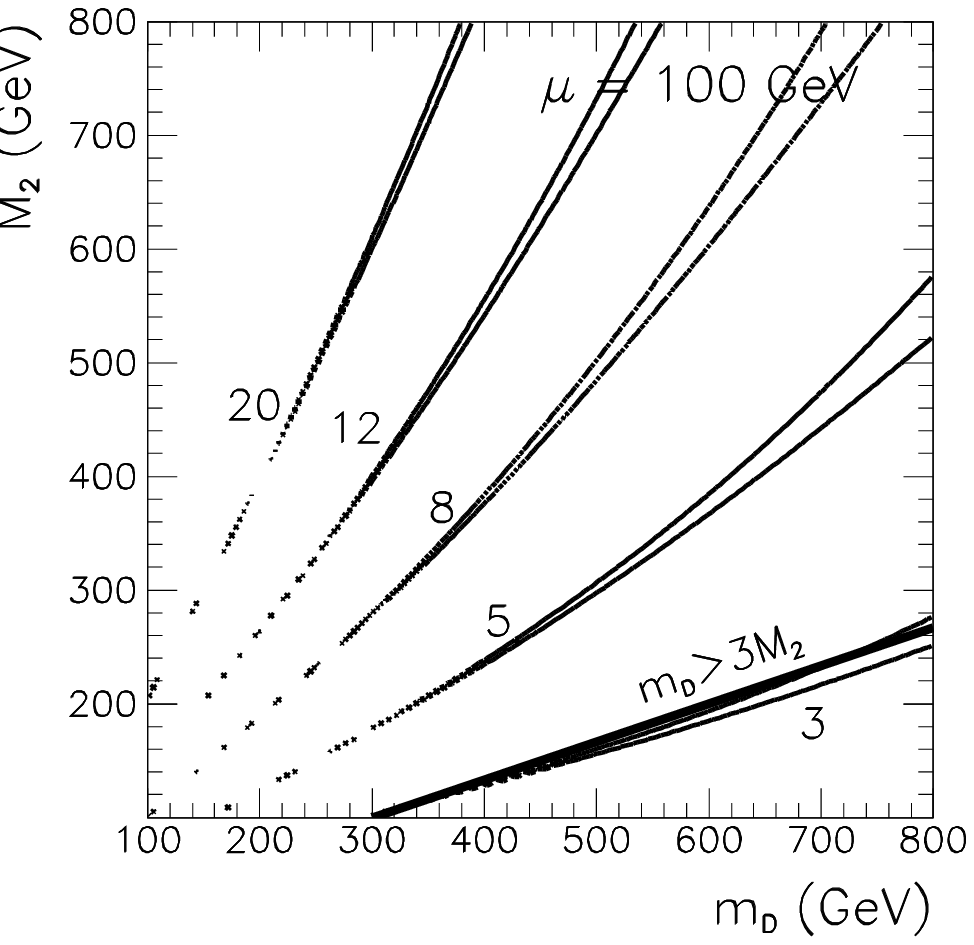,width=\linewidth}}&
\mbox{\epsfig{file=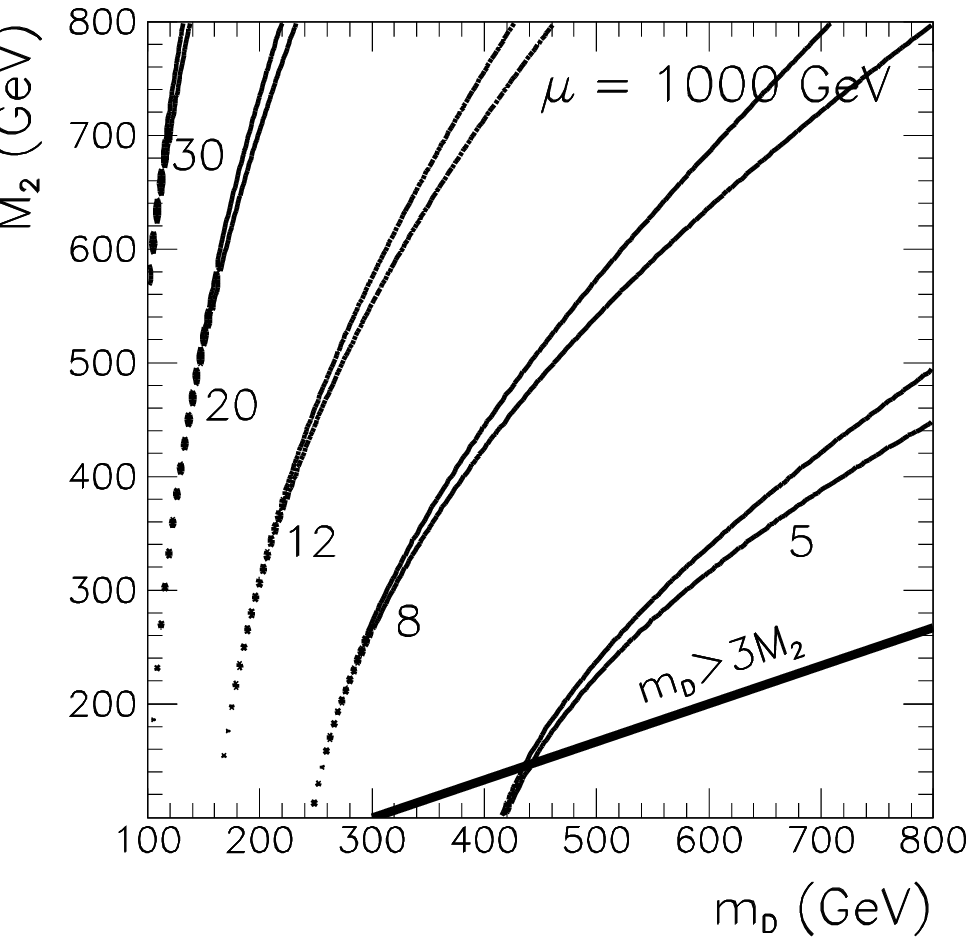,width=\linewidth}}
\end{tabular}
\vskip -5mm 
\caption{Regions of $m_D-|M_1|$ plane allowed by the neutron EDM
measurement assuming ``maximal'' CP violation
$\phi_{\mu}=\phi_{A_d}=\phi_{A_u}=\pi/2$, various values of
$A_d/m_D=A_u/m_U$ (marked on the plots), $A_t=0$, $m_Q=m_U=m_D$
and $M_1/\alpha_1=M_2/\alpha_2=M_3/\alpha_3$.
\label{fig:n_cont_a_mu}}
\end{center}
\end{figure}

Some differences with the electron case may be observed in the
structure of possible $\mu-A$ cancellations. For $E_e$ the term
proportional to $A_e$ originates from the neutralino exchange diagram.
For the neutron, additional contributions proportional to $A_u$, $A_d$
and $A_t$ are given by the diagrams with gluino exchange and they have
larger magnitude than those induced by neutralino loops, as
illustrated in Fig.~\ref{fig:n_ampl_a} (this efect is particularly
strong for large $|\mu|$ and light gauginos).
This means that, on the one hand constraints on the $A_I$ phases are
somewhat stronger than in the electron case but, on the other hand,
smaller $A_I$ values are necessary for cancellations. For small
$|\mu|\sim 100$ GeV one needs $A_e/m_E\geq 14$ but only $A_d/m_D\sim
A_u/m_U\geq 3$.  Furthermore, in unconstrained MSSM we have bigger
freedom because of several different $A_i$ parameters present in the
formulae for $E_n$.  Contributions proportional to $A_u,A_t$, coming
from $E_u,C_u$ and $C_g$, are more important comparing to those given
by the $E_d, C_d$ as they are not suppressed by the relative factor
$\cot^2\beta$ (like it is for the $\mu$ phase). Therefore, one has to
take into account all $A_i$ phases.  In Fig.~\ref{fig:n_cont_a_mu}
(compare Fig.~\ref{fig:e_cont_a_mu}) we plot the regions of
$m_D-|M_1|$ plane allowed by the neutron EDM measurement assuming
$\phi_{\mu}=\phi_{A_d}=\phi_{A_u}=\pi/2$ and various values of
$A_u/m_U=A_d/m_D$.

The overall conlusion is that eventual cancellations in neutron EDM
are more likely than in the electron case.  They require somewhat
smaller values of $A$ parameters when one considers $\mu-A$ phases
cancellations. Furthermore, assuming non-universal $A_i$ parameters it
is possible to suppress simultaneously both $E_e$ and $E_n$ values
below the experimental constraints, at the cost of rather strong
fine-tuning if the SUSY mass parameters are light.

\section{$\mu$ phase dependence of \epsk, $\Delta m_B$ and \bsg}

Analysing the dependence of \kk and \bb mixing on the SUSY phases, we
assume that there is no flavour violation in the squark mass matrices,
so that only chargino and charged Higgs contributions to the matrix
element do not vanish (gluino and neutralino contributions are always
proportional to the flavour off-diagonal entries in the squark mass
matrices). Furthermore, only chargino exchange contribution depends on
the $\mu$, $M_2$ and $A$ phases and is interesting for our analysis.
The leading chargino contribution is proportional to the
$\bar{d}_L^{I\alpha}\gamma_{\mu}d_L^{J\alpha}\cdot
\bar{d}_L^{I\beta}\gamma_{\mu}d_L^{J\beta}$ matrix element and has the
form:
\bea
(M_C)_{LLLL} = \frac{1}{8}\sum_{i,j,k,l}
(V_{d\tilde{U}C})_L^{Iki}
(V_{d\tilde{U}C})_L^{Ilj}(V_{d\tilde{U}C})_L^{Jli\star}
(V_{d\tilde{U}C})_L^{Jkj\star}
D_2(m^2_{C_i}, m^2_{C_j},m^2_{\tilde{U}_k}, m^2_{\tilde{U}_l})
\label{eq:mc}
\eea
where one should put $I=2,J=1$ for \kk mixing, $I=3,J=1$ for \bbd
mixing and $I=3,J=2$ for \bbs mixing (see~\ref{app:lagr} for the
expression for loop function $D_2$).

\begin{figure}[htbp]
\begin{center}
\mbox{\epsfig{file=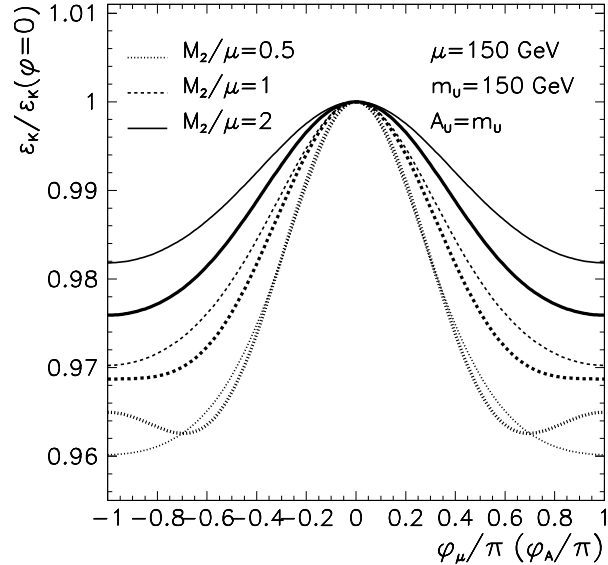,width=0.5\linewidth}}
\vskip -5mm
\caption{Dependence of \epsk on $\mu$ and $A_u$ phases, normalized to
$\phi_{\mu}=\phi_{A_u}=0$ case. Thin lines: dependence on $\phi_{\mu}$
for $\phi_{A_u}=0$, thick lines: dependence on $\phi_{A_u}$ for
$\phi_{\mu}=0$.
\label{fig:kk_mu}
}
\end{center}
\end{figure}

In order to analyze the dependence of the matrix element~(\ref{eq:mc})
on the phases, we consider the simplest case of flavour-diagonal and
degenerate up-squark mass and L-R mixing matrices and $|\mu|,|M_2|\geq
2M_Z$. In this case we can expand the matrix element in the mass
insertion approximation, both in the sfermion and chargino sectors, as
described in~\ref{app:massins}. The eq.~(\ref{eq:mc}) gives in such
approximation:
\bea
(M_C)_{LLLL} &\approx& \frac{1}{8}\left(K^{\dagger}Y_u^2 K\right)^2_{JI}
\left[D_2(|\mu|^2,|\mu|^2,m^2_U,m^2_U)\right.\nonumber\\
&+&8M_W^2\mathrm{Re}\left[(\mu^{\star}\cos\beta +
M_2\sin\beta)(\mu\cos\beta + A_U^{\star}\sin\beta)\right]\nonumber\\
&\times&\left. {\partial\over \partial m^2_U} {D_2(|\mu|^2,
|\mu|^2,m^2_U, m^2_U) - D_2(|\mu|^2, |M_2|^2,m^2_U, m^2_U) \over
|\mu|^2 - |M_2|^2}\right]
\label{eq:mcexp}
\eea
\epsk and $\Delta m_B$ are proportional, respectively, to the
imaginary and real part of the matrix element.  One can see
immediately from the equation above that in the leading order it is
sensitive only to $|\mu|$ and to the real parts of the $M_2\mu$,
$A_U\mu$ and $M_2A_U^{\star}$ products, i.e. to cosins of the
appropriate phase combinations, not sins of them like it was in the
EDM case. Eventual effects of the phases can be thus visible only for
large phase values. Even then, they are suppressed by
$M_W^2/m_{\tilde{U}}^2$ ratio and small numerical coefficient
mutiplying them. An example of the \epsk dependence on the $\mu$ and
$A_U$ phases (based on exact expression~(\ref{eq:mc}) and assuming
$M_2$ to be real) is presented in Fig.~\ref{fig:kk_mu}. As can be seen
from the Figure, even for light SUSY particle masses the change of the
\epsk value with variation of $\mu$ and $A$ phases is smaller than
5\%.
\begin{figure}[htbp]
\begin{center}
\begin{tabular}{p{0.48\linewidth}p{0.48\linewidth}}
\mbox{\epsfig{file=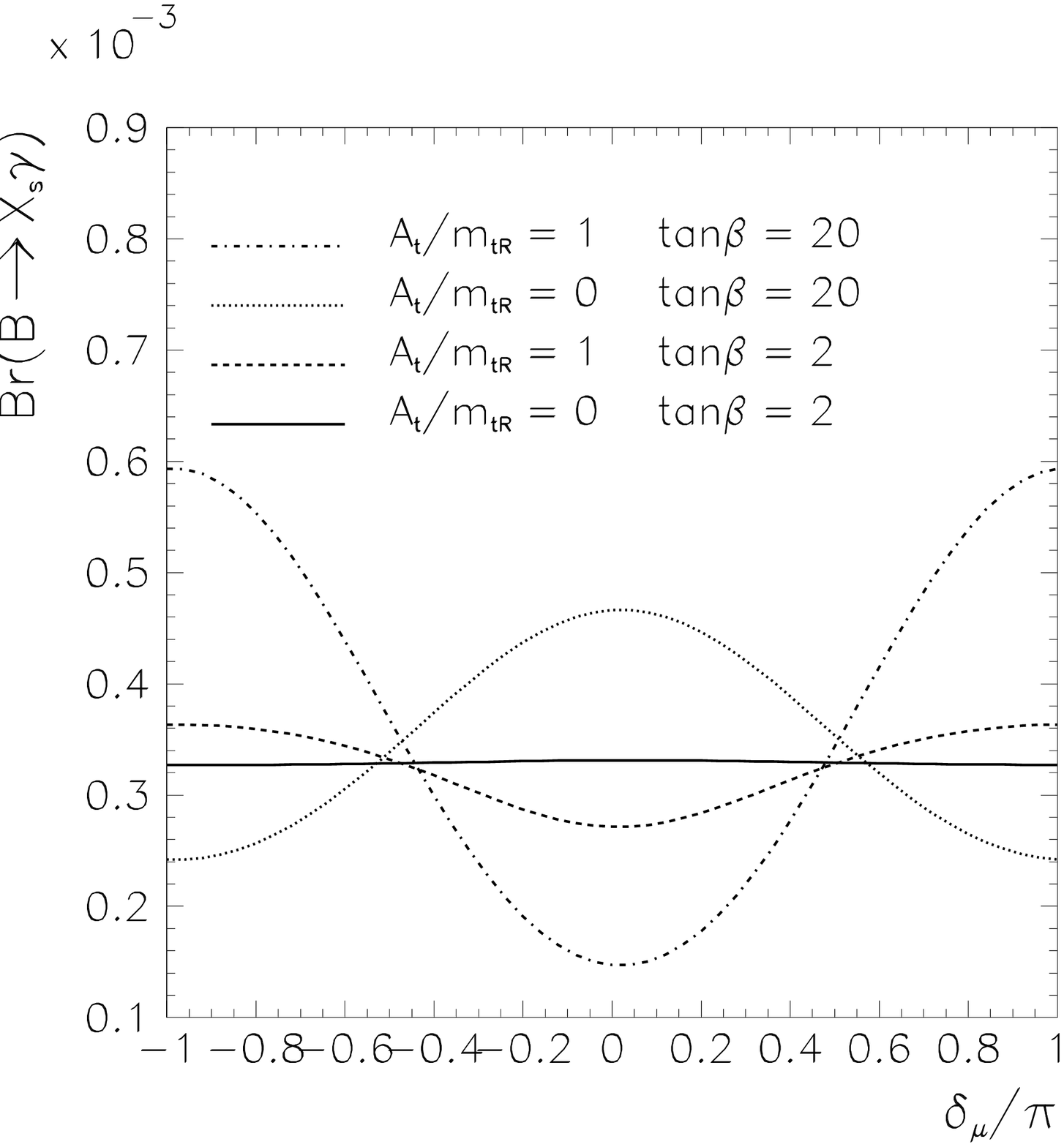,width=\linewidth}}&
\mbox{\epsfig{file=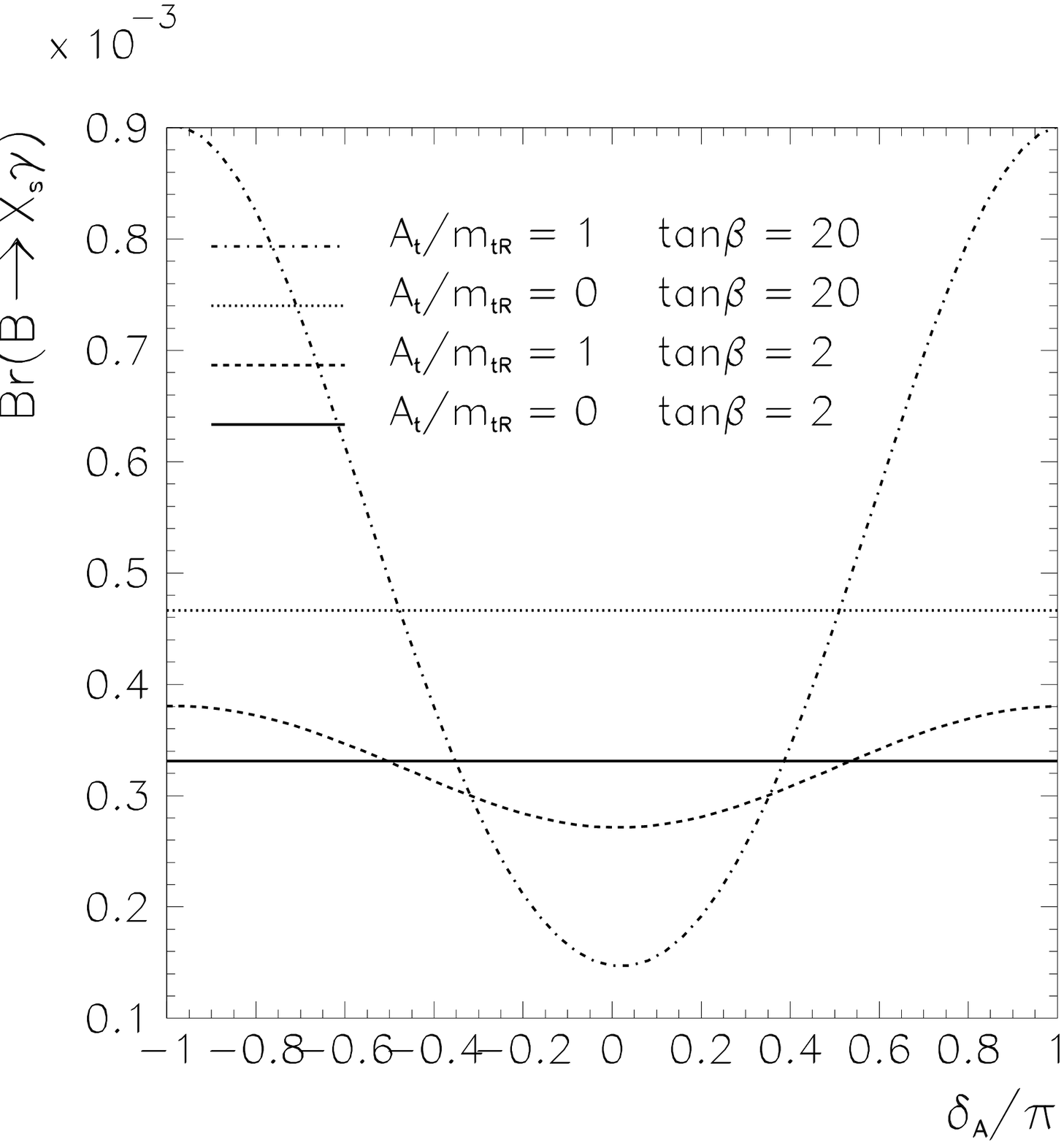,width=\linewidth}}
\end{tabular}
\vskip -5mm
\caption{Dependence of $\mathrm{Br}(B\rightarrow X_s\gamma)$ on $\mu$
and $A_t$ phases for $m_A=500$ GeV,
$|\mu|=|M_2|=\alpha_2/\alpha_1|M_1|=200$ GeV, $m_{\tilde{t}_L}=300$
GeV, $m_{\tilde{t}_R}=100$ GeV.
\label{fig:bsg_mu}
}
\end{center}
\end{figure}

Recently the potential importance of non-leading chargino
contributions to \epsk and \bb mi\-xing (proportional to LLRR and RRRR
matrix elements) was discussed in the literature~\cite{MASDEMIR}. Such
contributions are suppressed by the factor
$m_s^2(m_b^2)\tan^2\beta/M_W^2$ for \kk (\bb) mixing. Therefore they
can be significant only for large $\tan\beta$ values. Demir {\sl et
  al.}~\cite{MASDEMIR} estimate that $\epsk\sim 3\cdot 10^{-3}$ can be
obtained solely due to the supersymmetric phases of $\mu$ and $A_t$,
assuming vanishing Kobayashi-Maskawa phase $\delta_{KM}=0$. It
requires however large $\tan\beta=60$, large phases
$\phi_{\mu}\sim\phi_{A_t}\sim -\pi/2$ and light stop and chargino: in
the example they give $|\mu|=|M_2|=125$ GeV,
$m_{\tilde{t}_L}=m_{\tilde{t}_R}=150$ GeV, $|A_t|=250$ GeV, what gives
physical masses $m_{\tilde{t}_1}\approx 83$ GeV, $m_{C_1}\approx 80$
GeV. As follows from our discussion in sections~~\ref{sec:electron}
and~\ref{sec:neut}, it is rather unlikely, although not completely
impossible, to avoid limits on $\mu$ phase for light SUSY spectrum and
simultaneously large $\tan\beta$ (limits on $\mu$ phase are inversely
proportional to $\tan\beta$). The generic limits are in such a case
very tight, so in order to avoid them one needs strong fine tuning and
large cancellations between various contributions. For example,
assuming the above set of parameters, universal gaugino masses and
phases, $m_E=m_U=m_D=200$ GeV, $A_u=A_d$, $\phi_{A_i}=\pi/2$ one needs
very large $|A_e|/m_E\sim 750$ and $|A_u|/m_U=|A_d|/m_D\sim 170$ in
order to avoid the limits from the electron and neutron EDM
measurements. Alternatively, one can keep smaller $A_i$ but give up
the assumption of gaugino mass universality and adjust precisely their
masses and relative phases.  In each case, a very peculiar choice of
parameters is required to fulfill simultaneously experimental
measurements of \epsk and EDM's if one choose to keep
$\delta_{KM}=0$\footnote{Bounds on the $\mu$ phase cannot be avoided
  even by assuming heavy first two generations of sfermions. In this
  case, for large $\tan\beta$ considered here, 2-loop contributions
  from the third generation discussed in ref.~\cite{PILAFTSIS} are
  themselves sufficient to exclude the large $\mu$ phase necessary to
  reproduce the result for \epsk - see the discussion
  in~\cite{BAEK}.}.

Because of the weak dependence of \epsk and $\Delta m_B$ on flavour
conserving supersymmetric CP phases (excluding possibly very large
$\tan\beta$ case), the determination of the KM matrix phase
$\delta_{KM}$ (see e.g.~\cite{FCNC97}) is basically unaffected by
their eventual presence. However, one should remeber that the KM phase
detemination in the MSSM depends on the chargino and charged Higgs
masses and mixing angles, even if not on their phases.

In contrast to \epsk and $\Delta m_B$, \bsg decay appears to depend
strongly on the $\mu$ and $A_t$ phases. This is illustrated in
Fig.~\ref{fig:bsg_mu}. Exact formulae for the matrix element for \bsg
decay can be found e.g. in~\cite{FCNC97}. Expansion in the sfermion
and chargino/neutralino mass insertions (assuming degenerate stop
masses) gives the following approximate result for the so-called
$C_7^{(0)}$ coefficient, which primarily defines the size of $Br(B\ra
X_s\gamma)$:
\bea
C_7^{(0)}(M_W)&\approx&- {1\over 12 |\mu|^2 \sin^2\beta}
f_1({m_{\tilde{t}}^2\over|\mu|^2}) -{M_W^2\over 6 |M_2|^4}
f_1'({m_{\tilde{t}}^2\over|M_2|^2})
-{(|\mu|^2\cot\beta + A_t\mu)\over 3|\mu|^4\sin 2\beta}
f'_2({m_{\tilde{t}}^2\over |\mu|^2})\nonumber\\
&-&{M_W^2\mathrm{Re}\left[(\mu\cot\beta+A_t^{\star})
(\mu^{\star}\cot\beta+M_2)\right]\over 3(|M_2|^2 - |\mu|^2)}
\left({f'_1(\frac{m_{\tilde{t}}^2}{|M_2|^2})\over |M_2|^4} - 
{f'_1(\frac{m_{\tilde{t}}^2}{|\mu|^2})\over |\mu|^4}\right)\\ 
&-&{M_W^2\over 3(|M_2|^2-|\mu|^2)}\left[
\left({f'_2({m_{\tilde{t}}^2\over |M_2|^2})\over |M_2|^2} 
- {f'_2({m_{\tilde{t}}^2\over |\mu|^2}) \over |\mu|^2} \right)\tan\beta
+ \mu M_2 \left({f'_2({m_{\tilde{t}}^2\over |M_2|^2})  \over|M_2|^4}
- {f'_2({m_{\tilde{t}}^2\over |\mu|^2})\over |\mu|^4} \right) 
\right]\nonumber
\label{eq:bsg_exp}
\eea
where
\bea
f_1(x) &=& {x(3x-2)\over (1-x)^4}\log x + {8x^2+5x-7\over (1-x)^3}
\nonumber\\
f_2(x) &=& {-x(3x-2)\over (1-x)^3}\log x - {x(5x-3)\over 2(1-x)^2}
\label{eq:bsg_fun}
\eea
Expression~(\ref{eq:bsg_exp}) contains terms proportional to both real
and imaginary parts of the $\mu$ and $A_t$ parameters. However, the
branching ratio $Br(B\ra X_s\gamma)$ depends on $|C_7^{(0)}(M_W)|^2$
which, like in the \epsk case, depends mainly on the real parts of the
$\mu$ and $A_t$ parameters, i.e. on cosins of the phases. This is
clearly visible in Fig.~\ref{fig:bsg_mu}. However, contrary to the
\epsk case, this dependence is quite strong and growing with increase
of $\tan\beta$ and of the stop LR-mixing $A_t$ parameter. Also, as
follows from the discussion in the previous Section, the limits on
$A_t$ phase are rather weak, independently on $\tan\beta$, so one can
expect large effects of this phase in $Br(B\ra X_s\gamma)$ decay.

\section{Conclusions}
\label{sec:conclusions}

In this paper we have reanalyzed the constraints on the phases of
flavour conserving supersymmetric couplings that follow from the
electron and neutron EDM measurements. Also, we have discussed the
dependence on those phases of \epsk, $\Delta m_B$ and the branching
ratio for \bsg. We find that the constraints on the phases
(particularly on the phase of $\mu$ and of the gaugino masses) are
generically strong $\phi\leq 10^{-2}$ if all relevant supersymmetric
masses are light, say $\leq{\cal O}(500~\GeV)$.  However, we also find
that the constraints disappear or are substantially relaxed if just
one of those masses, e.g. slepton mass, is large, $m_E> {\cal O}
(1~\TeV)$. Thus, the phases can be large even if some masses, e.g. the
chargino masses, are small.

In the parameter range where the constraints are generically strong,
there exist fine-tuned regions where cancellations between different
contributions to the EDM can occur even for large phases. However,
such cancellations have no obvious underlying symmetry principle. From
the low energy point of view they look purely accidental and require
not only $\mu-A$, $\mu-M_{gaugino}$ or $M_1-M_2$ phase adjustment but
also strongly correlated with the phases and among themselves values
of soft mass parameters. Therefore, with all soft masses, say,
$\leq{\cal O}(1~\TeV)$ models with small phases look like the easiest
solution to the experimental EDM constraints. This conclusion becomes
stronger the higher is the value of $\tan\beta$, as the constraints on
$\mu$ phase scale as $1/\tan\beta$, and will be substantially stronger
also for low $\tan\beta$ after order of magnitude improvements in the
experimental limits on EDM's. Nevertheless, since the notion of fine
tuning is not precise, particularly from the point of view of GUT
models, it is not totally inconceivable that the rationale for large
cancellations exists in the large energy scale physics (in a very
recent paper~\cite{KANE2} it is pointed out that non-universal gaugino
phases necessary (but not sufficient) for large cancellations can be
obtained in some String I type models). Therefore all experimental
bounds on the supersymmetric parameters, and particularly on the Higgs
boson masses~\cite{WAGNER}, should include the possibility of large
phases even if with large cancellations, to claim full model
independence.

The dependence of \epsk and $\Delta m_B$ on the supersymmetric phases
is weak and gives no clue about their values. Hence, the $\delta_{KM}$
determination remains essentially unaffected. Large effects may be
observed in \bsg decay, but, apart from the $\phi_{\mu}$ and
$\phi_{A_t}$ phases, \bsg amplitude depends on many free mass
parameters, including the Higgs mass and the masses of squarks of the
third generation.

\renewcommand{\thesection}{Appendix~\Alph{section}}
\renewcommand{\theequation}{\Alph{section}.\arabic{equation}}

\setcounter{equation}{0}
\setcounter{section}{0}

\section{Conventions and Feynman rules}
\label{app:lagr}

For easy comparison with other references we spell out our
conventions.  They are similar to the ones used in
ref.~\cite{FEYRUL}. The MSSM matter fields form chiral left-handed
superfields in the following representations of the $SU(3) \times
SU(2) \times U(1)$ gauge group (the generation index is suppressed):

\begin{eqnarray*}
\begin{array}{lll}
\mathrm{Scalar~field}&  \mathrm{Weyl~Fermion~field} & 
SU(3)\times SU(2) \times U(1) ~\mathrm{representation}\\
L=\left(\begin{array}{c}\tilde\nu_0\\E\end{array}\right)& 
\mbox{\hskip 1cm} l=\left(\begin{array}{c}{\nu}\\
e\end{array}\right)&
\mbox{\hskip 2cm}(0,2,-1)\\
E^{c} &\mbox{\hskip 1cm}e^c&\mbox{\hskip 2cm} (0,0,2)\\
Q=\left(\begin{array}{c} U\\D\end{array}\right)& 
\mbox{\hskip 1cm}q=\left(\begin{array}{c}u\\
d\end{array}\right) &
\mbox{\hskip 2cm}(3,2,\frac{1}{3})\\
D^{c} &\mbox{\hskip 1cm} d^c &\mbox{\hskip 2cm}
(\bar{3},0,\frac{2}{3})\\
U^{c} &\mbox{\hskip 1cm} u^c &\mbox{\hskip 2cm}
(\bar{3},0,-\frac{4}{3})\\
H^{1}\left(\begin{array}{c} H_1^1\\H_2^1\end{array}\right)&
\mbox{\hskip 1cm}\tilde{h}^1\left(\begin{array}{c} \tilde{h}_1^1\\
\tilde{h}_2^1\end{array}\right)& 
\mbox{\hskip 2cm}(0,2,-1) \\
H^{2}\left(\begin{array}{c} H_1^2\\H_2^2\end{array}\right)&
\mbox{\hskip 1cm}\tilde{h}^2\left(\begin{array}{c} \tilde{h}_1^2\\
\tilde{h}_2^2\end{array}\right)& 
\mbox{\hskip 2cm}(0,2,1)
\label{eq:superkm}
\end{array}
\end{eqnarray*}
Two $SU(2)$-doublets can be contracted into an $SU(2)$-singlet,
e.g. $H^1 H^2 = \epsilon_{ij}H^1_i H^2_j = - H^1_1 H^2_2 + H^1_2
H^2_1$ (we choose $\epsilon_{12}=-1$; lower indices (when present)
will label components of $SU(2)$-doublets). The superpotential and the
soft terms are defined as:
\bea
W = W_0 + W_{CP}
\label{eq:superpot}
\eea
\bea
W_0 = Y_e H^1 L E + Y_d H^1 Q D + Y_u H^2 Q U
\label{eq:superpot-0}
\eea
\bea
W_{CP} = \mu H^1 H^2
\label{eq:superpot-cp}
\eea
\bea
{\cal L}_{soft} &=& {\cal L}_{soft-0} +{\cal L}_{soft-CP}
\label{eq:lsoft}
\eea 
\bea
{\cal L}_{soft-0} &=& 
- M_{H^1}^2 H^{1\dagger} H^1 - M_{H^2}^2 H^{2\dagger} H^2
- L^{\dagger} M_L^2  L 
- E^{c\dagger} M_E^2 E^c
\nonumber \\ &&
- Q^{ \dagger} M_Q^2 Q 
- D^{c\dagger} M_D^2 D^c 
- U^{c\dagger}M_U^2 U^c
\label{eq:lsoft-0}
\eea 
\bea
{\cal L}_{soft-CP} &=& 
\f{1}{2} \left( M_3 \tilde{G}^a \tilde{G}^a
              + M_2 \tilde{W}^i \tilde{W}^i
              + M_1 \tilde{B}\tilde{B} \right)
              + m^2_{12} H^1 H^2
\nonumber \\ &&
         + Y_e A_e H^1 L E^c 
         + Y_d A_d H^1 Q D^c 
         + Y_u A_u H^2 Q U^c
         + \mathrm{H.c.}
\label{eq:lsoft-cp}
\eea 
where we divided all terms into two subclasses, collecting in $W_{CP}$
and ${\cal L}_{soft-CP}$ those of them which may contain
flavour-diagonal CP breaking phases. We also extracted Yukawa coupling
matrices from the definition of the $A_I$ coefficients.

In general, the Yukawa couplings and the masses are matrices in the
flavour space. Rotating the fermion fields one can diagonalize the
Yukawa couplings (and simultaneously fermion mass matrices). This
procedure is well known from the Standard Model: it leads to the
appearance of the Kobayashi-Maskawa matrix $K$ in the charged current
vertices. In the MSSM, simultaneous parallel rotations of the fermion
and sfermion fields from the same supermultiplets lead to so-called
``super-KM'' basis, with flavour diagonal Yukawa couplings and neutral
current fermion and sfermion vertices. As we neglect flavour violation
effects in this paper, we give all the expressions already in the
super-KM basis (see e.g.~\cite{FCNC97} for a more detailed
discussion).

The diagonal fermion mass matrices are connected with Yukawa matrices
by the formulae:
\bea
m_l = -\f{v_1}{\sqrt{2}} Y_l
\hspace{0.5cm}
m_u =  \f{v_2}{\sqrt{2}} Y_u
\hspace{0.5cm}
m_d = -\f{v_1}{\sqrt{2}} Y_d
\nonumber 
\eea

The physical Dirac chargino and Majorana neutralino eigenstates are
linear combinations of left-handed Winos, Binos and Higgsinos
\bea
C^+_i = \left(\begin{array}{c}
-iZ_{+}^{i1\star}\tilde{W}^+ + Z_{+}^{i2\star}\tilde{h}^1_2 \\
iZ_{-}^{i1}\overline{\tilde{W}^-} + Z_{-}^{i2}\overline{\tilde{h}^2_1}
\end{array}\right)
\label{eq:chargmass}
\eea
where $\tilde{W}^{\pm} = (\tilde{W}^1\mp\tilde{W}^2)/\sqrt{2}$.
\bea
N^0_i = \left(\begin{array}{c}
-iZ_N^{i1\star}\tilde{B}  -iZ_N^{i2\star}\tilde{W}^3 
+ Z_N^{i3\star}\tilde{h}^1_1 + Z_N^{i4\star}\tilde{h}^2_2 \\
iZ_N^{i1}\overline{\tilde{B}} + iZ_N^{i2}\overline{\tilde{W}^3} 
+ Z_N^{i3} \overline{\tilde{h}^1_1} + Z_N^{i4}\overline{\tilde{h}^2_2}
\end{array}\right)
\eea
The unitary transformations $Z^+$, $Z^-$ and $Z_N$ diagonalize the mass
matrices of these fields
\bea
{\cal M}_C = Z_{-}^T \left( \begin{array}{cc} M_2 &
\frac{gv_2}{\sqrt{2}}  \\
\frac{gv_1}{\sqrt{2}} & \mu \end{array} \right) Z_{+}
\eea
and
\bea
{\cal M}_N = Z_N^T \left( \begin{array}{cccc} 
M_1 & 0 & -\frac{g'v_1}{2} &  \frac{g'v_2}{2}\\
0 & M_2 &  \frac{gv_1}{2}  & -\frac{gv_2}{2} \\
-\frac{g'v_1}{2} & \frac{gv_1}{2} & 0    & -\mu \\ 
\frac{g'v_2}{2}& -\frac{gv_2}{2}  & -\mu & 0
\end{array} \right) Z_N
\eea
4-component gluino field $\tilde{g}$ is defined as
\bea
\tilde{g}^a = \left(\begin{array}{c}-i\tilde{G}^a\\i\overline{\tilde{G}^a}
\end{array}\right)
\eea

The sfermion mass matrices in the super-KM basis have the following
form:
\bea
{\cal M}^2_{\tilde{U}} &=&
\left( \begin{array}{cc}
K M^2_Q K^{\dagger} + m_u^2 - \frac{\cos 2\beta}{6}(M_Z^2 - 4 M_W^2)\hat{\mbox{\large 1}} & 
- m_u(\cot\beta \mu\hat{\mbox{\large 1}} + A_u^{\star})\\ 
- m_u(\cot\beta \mu^{\star}\hat{\mbox{\large 1}} + A_u)& 
M^2_U+m_u^2+\frac{2\cos 2\beta}{3} (M_Z^2 - M_W^2)\hat{\mbox{\large 1}}\\
\end{array}\right)\nonumber\\ 
\nonumber\\
{\cal M}^2_{\tilde{D}} &=& 
\left( \begin{array}{cc}
M^2_Q + m_d^2 - \frac{\cos 2\beta}{6}(M_Z^2 + 2M_W^2)\hat{\mbox{\large 1}} & 
-m_d(\tan\beta \mu\hat{\mbox{\large 1}} +A_d^{\star}) \\
-m_d(\tan\beta \mu^{\star}\hat{\mbox{\large 1}} +A_d) &
M^2_D+m_d^2-\frac{\cos 2\beta}{3} (M_Z^2 - M_W^2)\hat{\mbox{\large 1}}\\
\end{array}\right) \nonumber\\
\nonumber\\
{\cal M}^2_{\tilde{L}} &=&  
\left( \begin{array}{cc}
M^2_L + m_e^2 + \frac{\cos 2\beta}{2}(M_Z^2 - 2M_W^2)\hat{\mbox{\large 1}} & 
-m_e(\tan\beta \mu\hat{\mbox{\large 1}} + A_e^{\star})\\
-m_e(\tan\beta \mu^{\star}\hat{\mbox{\large 1}} + A_e) & 
M^2_E + m_e^2 -\cos 2\beta (M_Z^2 - M_W^2)\hat{\mbox{\large 1}}\\
\end{array}\right)\nonumber\\ 
\nonumber\\
{\cal M}^2_{\tilde{\nu}}& =&
M^2_L +  \frac{\cos 2\beta}{2} M_Z^2\hat{\mbox{\large 1}}
\label{eq:sfmass}
\eea

\noindent where $\theta_W$ is the Weinberg angle and
$\hat{\mbox{\large 1}}$ stands for the $3 \times 3$ unit matrix.

Throughout this paper we assume that there is no flavour and CP
violation due to the flavour mixing in the sfermion mass
matrices. i.e. matrices $M^2_I, A_I$ are diagonal in the super-KM
basis. However, one should remember that in this basis mass matrices
of the left up and down squarks are connected due to gauge
invariance~\cite{FCNC97}. This means that it is impossible to set all
the $(M^2_{IJ})_{LL}$ to zero simultaneously, unless $M^2_Q \sim
\hat{\mbox{\large 1}}$.

The matrices ${\cal M}^2_{\tilde{\nu}}$, ${\cal M}^2_{\tilde{L}}$,
${\cal M}^2_{\tilde{U}}$ and ${\cal M}^2_{\tilde{D}}$ can be
diagonalized by additional unitary matrices $Z_{\nu}$ ($3\times 3$)
and $Z_L$, $Z_U$, $Z_D$ ($6\times 6$), respectively
\bea 
\left({\cal M}^2_{\tilde{\nu}}\right)^{diag} = 
Z_{\nu}^{\dagger} {\cal M}^2_{\tilde{\nu}} Z_{\nu}
&\hskip 3cm&
\left({\cal M}^2_{\tilde{U}}\right)^{diag} = 
Z_U^{\dagger} {\cal M}^2_{\tilde{U}} Z_U
\label{eq:zdef}
\nonumber\\
\left({\cal M}^2_{\tilde{L}}\right)^{diag}  =  
Z_L^{\dagger} {\cal M}^2_{\tilde{L}} Z_L 
&\hskip 3cm& 
\left({\cal M}^2_{\tilde{D}}\right)^{diag} = 
Z_D^{\dagger} {\cal M}^2_{\tilde{D}} Z_D 
\eea

The physical (mass eigenstates) sfermions are then defined in terms of
super-KM basis fields~(\ref{eq:superkm}) as:
\bea
\tilde{\nu} = Z_{\nu}^{\dagger} \tilde\nu_0
\hspace{0.5cm}
\tilde{L} = Z_L^{\dagger} \left( \begin{array}{c} E \\ E^{c\star} \end{array} 
\right)
\hspace{0.5cm}
\tilde{U} = Z_U^{\dagger} \left( \begin{array}{c} U \\ U^{c\star} \end{array} 
\right)
\hspace{0.5cm}
\tilde{D} =Z_D^{\dagger} \left( \begin{array}{c} D\\ D^{c\star}  \end{array} 
\right)
\label{eq:sferdef}
\eea
In order to compactify notation, it is convenient to split $6\times6$
matrices $Z_L,Z_U,Z_D$ into $3\times 6$ sub-blocks:
\bea
Z_X^{Ii}\equiv Z_{XL}^{Ii}\hskip 2cm Z_X^{I+3,i}\equiv
Z_{XR}^{Ii}\hskip 2cm I=1..3,i=1..6
\label{eq:zsplit}
\eea
(the index $i$ numbers the physical states).  Formally $Z_{XL}$ and
$Z_{XR}$ are projecting respectively left and right sfermion fields in
the super-KM basis into mass eigenstate fields.

Using the notation of this Appendix, one gets for the Feynman
rules in the mass eigenstate basis:

1) Charged current vertices:\\[5mm]
\begin{tabular}{ll}
\begin{picture}(115,60)(0,0)
\DashArrowLine(50,10)(10,10){6}
\Vertex(50,10){2}
\ArrowLine(50,10)(90,10)
\ArrowLine(50,50)(50,10)
\Text(0,10)[]{$\tilde{\nu}^J$}
\Text(110,10)[]{$(C_j^+)^c$}
\Text(50,60)[]{$e^I$}
\end{picture}&
\raisebox{10\unitlength}{\begin{tabular}{l}
$ -i\left[g_2 Z_{1j}^+ P_L + Y_l^{I} Z_{2j}^{-*} P_R \right] 
Z_{\tilde{\nu}}^{IJ\star}$\\
\end{tabular}}\\
\begin{picture}(115,60)(0,0)
\DashArrowLine(50,10)(10,10){6}
\Vertex(50,10){2}
\ArrowLine(50,10)(90,10)
\ArrowLine(50,50)(50,10)
\Text(0,10)[]{$U_i^+$}
\Text(110,10)[]{$(C_j^+)^c$}
\Text(50,60)[]{$d^I$}
\end{picture}&
\raisebox{10\unitlength}{\begin{tabular}{l}
$ i\left[\left(-g_2
Z_{UL}^{Ji*} Z_{1j}^+ + Y_u^{J} Z_{UR}^{Ji*} Z_{2j}^+\right) P_L 
- Y_d^{I} Z_{UL}^{Ji*} Z_{2j}^{-*} P_R \right] K^{JI}$\\
\end{tabular}}\\
\begin{picture}(115,60)(0,0)
\DashArrowLine(50,10)(10,10){6}
\Vertex(50,10){2}
\ArrowLine(50,10)(90,10)
\ArrowLine(50,50)(50,10)
\Text(0,10)[]{$D_i^-$}
\Text(100,10)[]{$C_j^+$}
\Text(50,60)[]{$u^J$}
\end{picture}&
\raisebox{10\unitlength}{\begin{tabular}{l}
$ i\left[-\left(g_2
Z_{DL}^{Ii\star} Z_{1j}^+ + Y_d^{I} Z_{DR}^{Ii\star} Z_{2j}^-\right) P_L 
+ Y_u^{J} Z_{DL}^{Ii\star} Z_{2j}^{+*} P_R \right] K^{JI\star}$\\
\end{tabular}}\\
\end{tabular}

2) Neutral current vertices:\\[5mm]
\begin{tabular}{ll}
\begin{picture}(100,60)(0,0)
\DashArrowLine(50,10)(10,10){6}
\Vertex(50,10){2}
\ArrowLine(50,10)(90,10)
\ArrowLine(50,50)(50,10)
\Text(0,10)[]{$L_i^-$}
\Text(100,10)[]{$N_j^0$}
\Text(50,60)[]{$e^I$}
\end{picture}&
\raisebox{30\unitlength}{\begin{tabular}{l}
$i\left[\left({1 \over \sqrt{2}} Z_{LL}^{Ii\star}\left(g_1Z_N^{1j}
+g_2 Z_N^{2j}\right) + Y_l^{I} Z_{LR}^{Ii\star} Z_N^{3j}\right)P_L\right.$\\
$+ \left.\left(-g_1 \sqrt{2}Z_{LR}^{Ii\star}Z_N^{1j\star}
+ Y_l^{I} Z_{LL}^{Ii\star} Z_N^{3j\star}\right) P_R \right]$\\
\end{tabular}}\\
\begin{picture}(100,60)(0,0)
\DashArrowLine(50,10)(10,10){6}
\Vertex(50,10){2}
\ArrowLine(50,10)(90,10)
\ArrowLine(50,50)(50,10)
\Text(0,10)[]{$U_i^+$}
\Text(100,10)[]{$N_j^0$}
\Text(50,60)[]{$u^I$}
\end{picture}&
\raisebox{30\unitlength}{\begin{tabular}{l}
$i\left[\left({-1 \over \sqrt{2}} Z_{UL}^{Ii\star}\left(\frac{g_1}{3}Z_N^{1j}
+g_2 Z_N^{2j}\right) - Y_u^{I} Z_{UR}^{Ii\star} Z_N^{4j}\right)P_L\right.$\\
$+ \left.\left({2g_1 \sqrt{2} \over 3}Z_{UR}^{Ii\star}Z_N^{1j\star}
- Y_u^{I} Z_{UL}^{Ii\star} Z_N^{4j\star}\right) P_R \right]$\\
\end{tabular}}\\
\begin{picture}(100,60)(0,0)
\DashArrowLine(50,10)(10,10){6}
\Vertex(50,10){2}
\ArrowLine(50,10)(90,10)
\ArrowLine(50,50)(50,10)
\Text(0,10)[]{$D_i^-$}
\Text(100,10)[]{$N_j^0$}
\Text(50,60)[]{$d^I$}
\end{picture}&
\raisebox{30\unitlength}{\begin{tabular}{l}
$i\left[\left({-1 \over \sqrt{2}} Z_{DL}^{Ii\star}\left(\frac{g_1}{3}Z_N^{1j}
-g_2 Z_N^{2j}\right)+ Y_d^{I} Z_{DR}^{Ii\star} Z_N^{3j}\right) P_L\right.$\\
$+ \left.\left({-g_1 \sqrt{2} \over 3}Z_{DR}^{Ii\star}Z_N^{1j\star}
+ Y_d^{I} Z_{DL}^{Ii\star} Z_N^{3j\star}\right) P_R \right]$\\
\end{tabular}}\\
\end{tabular}

3) Neutral current vertices with strong coupling constant:\\[5mm]
\begin{tabular}{ll}
\begin{picture}(100,60)(0,0)
\DashArrowLine(50,10)(10,10){6}
\Vertex(50,10){2}
\ArrowLine(50,10)(90,10)
\ArrowLine(50,50)(50,10)
\Text(0,10)[]{$U_{i\alpha}^+$}
\Text(100,10)[]{$\tilde{g}_a$}
\Text(50,60)[]{$u^I_{\beta}$}
\end{picture}&
\raisebox{30\unitlength}{\begin{tabular}{l}
$-ig_s\sqrt{2}T^a_{\alpha\beta}\left(Z_{UL}^{Ii\star}P_L -
Z_{UR}^{Ii\star}P_R\right)$\\
\end{tabular}}\\
\begin{picture}(100,60)(0,0)
\DashArrowLine(50,10)(10,10){6}
\Vertex(50,10){2}
\ArrowLine(50,10)(90,10)
\ArrowLine(50,50)(50,10)
\Text(0,10)[]{$D_{i\alpha}^-$}
\Text(100,10)[]{$\tilde{g}_a$}
\Text(50,60)[]{$d^I_{\beta}$}
\end{picture}&
\raisebox{30\unitlength}{\begin{tabular}{l}
$-ig_s\sqrt{2}T^a_{\alpha\beta}\left(Z_{DL}^{Ii\star}P_L -
Z_{DR}^{Ii\star}P_R\right)$\\
\end{tabular}}\\
\end{tabular}

Finally, we give here explicit formulae for the loop functions. Three
point functions $C_{11},C_{12}$ are defined as:
\bea
\left.\int{d^4 k\over (2\pi)^4} {k^{\mu}\over [k^2 -
m_1^2][(p+k)^2-m_1^2][(k+p+q)^2 - m_2^2]}\right|_{p,q\rightarrow 0}=
~~~~~~~~~~~~~~~~\nonumber\\ -{i\over (4\pi)^2}
\left(p^{\mu}C_{11}(m_1^2,m_2^2) + q^{\mu}C_{12}(m_1^2,m_2^2)\right)
\label{eq:cijdef}
\eea
\bea
C_{11}(x, y) & = & {-x + 3y\over 4(x - y)^2} 
+ {y^2\over 2(x-y)^3} \log \frac{y}{x}
\label{eq:cp11}
\\
C_{12}(x, y) & = & -{x + y\over 2(x - y)^2} 
- {xy\over (x-y)^3} \log \frac{y}{x}
\label{eq:cp12}
\eea
The four point loop function $D_2$ has the form
\bea
D_2(v,x,y,z) &=& - {x^2\over (x - v)(x - y)(x - z)}\log\frac{x}{v} -
{y^2\over (y - v)(y - x)(y - z)}\log\frac{y}{v}\nonumber\\ 
&-& {z^2\over (z - v)(z - x)(z - y)}\log\frac{z}{v}
\label{eq:dfun2}
\eea

\section{Feynman rules for mass insertion calculations}
\label{app:massins}

We list below the Feynman rules necessary to calculate contributions
to lepton EDM in the mass insertion approximation.  We treat now
off-diagonal terms in chargino and neutralino mass matrices
(proportional to $v_1,v_2$) as the mass insertions. In order to reduce
remaining charged and neutral SUSY fermion mass terms to their
canonical forms:
\bea
-|\mu|\overline{\tilde{h}^-}\tilde{h}^-
-\frac{1}{2}|\mu|\overline{\tilde{h}_i^0}\tilde{h}_i^0
-|M_2|\overline{\tilde{W}^-}\tilde{W}^-
-\frac{1}{2}|M_2|\overline{\tilde{W}^0}\tilde{W}^0
-\frac{1}{2}|M_1|\overline{\tilde{B}^0}\tilde{B}^0
\eea
we define the 4-component spinor fields in terms of the initial
2-component spinors as:
\bea
\tilde{W}^- = \frac{i}{\sqrt{2}}
\left(\begin{array}{c}
e^{i\phi_2}(\tilde{W}^1 + i\tilde{W}^2)\\ (\overline{\tilde{W}^1} +
i\overline{\tilde{W}^2})
\end{array}\right)
\hskip 2cm
\tilde{h}^- = \left(\begin{array}{c}
\tilde{h}^1_2\\
e^{-i\phi_{\mu}}\overline{\tilde{h}^2_1}
\end{array}\right)
\eea
\bea
\tilde{W}^0 = i\left(\begin{array}{c}
e^{i\phi_2/2}\tilde{W}^3\\ -e^{-i\phi_2/2}\overline{\tilde{W}^3}\\
\end{array}\right)
\hskip 2cm
\tilde{B}^0 = i\left(\begin{array}{c}
e^{i\phi_1/2}\tilde{B}\\
-e^{-i\phi_1/2}\overline{\tilde{B}}\\
\end{array}\right)
\eea
\bea
\tilde{h}_1^0 = \frac{i}{\sqrt{2}}
\left(\begin{array}{c}
-e^{i\phi_{\mu}}(\tilde{h}^1_1 + \tilde{h}^2_2)\\
e^{-i\phi_{\mu}}(\overline{\tilde{h}^1_1} + \overline{\tilde{h}^2_2})
\end{array}\right)
\hskip 2cm
\tilde{h}_2^0 = \frac{1}{\sqrt{2}}
\left(\begin{array}{c}
e^{i\phi_{\mu}}(\tilde{h}^1_1 - \tilde{h}^2_2)\\
e^{-i\phi_{\mu}}(\overline{\tilde{h}^1_1} - \overline{\tilde{h}^2_2})
\end{array}\right)
\eea

Then the necessary Feynman rules are:

\begin{tabular}{ll}
\begin{minipage}[c]{60mm}
\begin{picture}(80,90)(0,0)
\Photon(0,20)(40,20){3}{4}
\Text(0,10)[]{\mbox{$\gamma$}}
\Vertex(40,20){2}
\DashArrowLine(40,20)(80,20){4}
\Text(60,10)[]{\mbox{$k$}}
\Text(80,10)[l]{\mbox{$E(E^c)$}}
\DashArrowLine(40,60)(40,20){4}
\Text(30,40)[]{\mbox{$p$}}
\Text(50,60)[l]{\mbox{$E(E^c)$}}
\end{picture}
\end{minipage}
&$\pm ie(p+k)^{\mu}$
\end{tabular}

\begin{tabular}{ll}
\begin{minipage}[c]{60mm}
\begin{picture}(80,90)(0,0)
\Photon(0,20)(40,20){3}{4}
\Text(0,10)[]{\mbox{$\gamma$}}
\Vertex(40,20){2}
\ArrowLine(40,20)(80,20)
\Text(80,10)[c]{\mbox{$\tilde{W}^-,\tilde{h}^-$}}
\ArrowLine(40,60)(40,20)
\Text(50,60)[l]{\mbox{$\tilde{W}^-,\tilde{h}^-$}}
\end{picture}
\end{minipage}
&$ie\gamma^{\mu}$
\end{tabular}

\begin{tabular}{ll}
\begin{minipage}[c]{60mm}
\begin{picture}(80,90)(0,0)
\Line(0,20)(40,20)
\Text(0,10)[l]{\mbox{$\tilde{h}_{1,2}^0$}}
\Vertex(40,20){2}
\ArrowLine(40,20)(80,20)
\Text(80,10)[l]{\mbox{$e$}}
\DashArrowLine(40,60)(40,20){4}
\Text(50,60)[]{\mbox{$E$}}
\end{picture}
\end{minipage}
& $\frac{i}{\sqrt{2}}Y_e e^{-i\phi_{\mu}/2}
\left\{\begin{array}{c} i \\ 1 \end{array}\right\} P_L$
\end{tabular}

\begin{tabular}{ll}
\begin{minipage}[c]{60mm}
\begin{picture}(80,90)(0,0)
\ArrowLine(0,20)(40,20)
\Text(0,10)[]{\mbox{$e$}}
\Vertex(40,20){2}
\Line(40,20)(80,20)
\Text(80,10)[l]{\mbox{$\tilde{h}_{1,2}^0$}}
\DashArrowLine(40,60)(40,20){4}
\Text(50,60)[l]{\mbox{$E^c$}}
\end{picture}
\end{minipage}
& $\frac{i}{\sqrt{2}}Y_e e^{-i\phi_{\mu}/2}
\left\{\begin{array}{c} i \\ 1 \end{array}\right\} P_L$
\end{tabular}

\begin{tabular}{ll}
\begin{minipage}[c]{60mm}
\begin{picture}(80,90)(0,0)
\ArrowLine(0,20)(40,20)
\Text(0,10)[]{\mbox{$e$}}
\Vertex(40,20){2}
\ArrowLine(40,20)(80,20)
\Text(80,10)[l]{\mbox{$\tilde{W}^-$}}
\DashArrowLine(40,20)(40,60){4}
\Text(50,60)[l]{\mbox{$\tilde{\nu}_0$}}
\end{picture}
\end{minipage}
& $igP_L$
\end{tabular}

\begin{tabular}{ll}
\begin{minipage}[c]{60mm}
\begin{picture}(80,90)(0,0)
\ArrowLine(0,20)(40,20)
\Text(0,10)[]{\mbox{$e$}}
\Vertex(40,20){2}
\Line(40,20)(80,20)
\Text(80,10)[l]{\mbox{$\tilde{W}^0,\tilde{B}^0$}}
\DashArrowLine(40,20)(40,60){4}
\Text(50,60)[l]{\mbox{$E$}}
\end{picture}
\end{minipage}
& -$\frac{i\sqrt{2}}{2}
\left\{\begin{array}{c} ge^{-i\phi_2/2} \\ g'e^{-i\phi_1/2} 
\end{array}\right\} P_L$
\end{tabular}

\begin{tabular}{ll}
\begin{minipage}[c]{60mm}
\begin{picture}(80,90)(0,0)
\Line(0,20)(40,20)
\Text(0,10)[]{\mbox{$\tilde{B}^0$}}
\Vertex(40,20){2}
\ArrowLine(40,20)(80,20)
\Text(80,10)[l]{\mbox{$e$}}
\DashArrowLine(40,20)(40,60){4}
\Text(50,60)[l]{\mbox{$E^c$}}
\end{picture}
\end{minipage}
& $i\sqrt{2}g'e^{-i\phi_1/2}P_L$
\end{tabular}

\begin{tabular}{ll}
\begin{minipage}[c]{60mm}
\begin{picture}(80,90)(0,0)
\ArrowLine(0,20)(40,20)
\Text(0,10)[]{\mbox{$\tilde{h}^-$}}
\Vertex(40,20){2}
\ArrowLine(40,20)(80,20)
\Text(80,10)[l]{\mbox{$e$}}
\DashArrowLine(40,60)(40,20){4}
\Text(50,60)[l]{\mbox{$\tilde{\nu}_0$}}
\end{picture}
\end{minipage}
& $-iY_eP_L$
\end{tabular}

\vskip 5mm

\begin{tabular}{ll}
\begin{minipage}[c]{60mm}
\begin{picture}(80,40)(0,0)
\Line(0,20)(40,20)
\Text(0,10)[]{\mbox{$\tilde{h}_1^0$}}
\Line(37,17)(43,23)
\Line(43,17)(37,23)
\Line(40,20)(80,20)
\Text(80,10)[c]{\mbox{$\tilde{W}^0,\tilde{B}^0$}}
\end{picture}
\end{minipage}
& \begin{tabular}{r}
$-\frac{1}{2\sqrt{2}} 
\left\{\begin{array}{c} g \\ -g' \end{array}\right\}(v_1-v_2)
\left(e^{-i(\phi_{\mu} + \phi_{1,2})/2}P_L\right.$\\ 
$\left.-e^{i(\phi_{\mu} + \phi_{1,2})/2}P_R\right)$
\end{tabular}
\end{tabular}

\vskip 5mm

\begin{tabular}{ll}
\begin{minipage}[c]{60mm}
\begin{picture}(80,40)(0,0)
\Line(0,20)(40,20)
\Text(0,10)[]{\mbox{$\tilde{h}_2^0$}}
\Line(37,17)(43,23)
\Line(43,17)(37,23)
\Line(40,20)(80,20)
\Text(80,10)[c]{\mbox{$\tilde{W}^0,\tilde{B}^0$}}
\end{picture}
\end{minipage}
& \begin{tabular}{r}
$\frac{i}{2\sqrt{2}}
\left\{\begin{array}{c} g \\ -g' \end{array}\right\}(v_1+v_2)
\left(e^{-i(\phi_{\mu} + \phi_{1,2})/2}P_L\right.$\\ 
$\left. + e^{i(\phi_{\mu} + \phi_{1,2})/2}P_R\right)$
\end{tabular}
\end{tabular}

\vskip 5mm

\begin{tabular}{ll}
\begin{minipage}[c]{60mm}
\begin{picture}(80,40)(0,0)
\ArrowLine(0,20)(40,20)
\Text(0,10)[]{\mbox{$\tilde{h}^-$}}
\Line(37,17)(43,23)
\Line(43,17)(37,23)
\ArrowLine(40,20)(80,20)
\Text(80,10)[l]{\mbox{$\tilde{W}^-$}}
\end{picture}
\end{minipage}
& $\frac{ig}{\sqrt{2}}\left(v_1 P_L  + v_2 e^{i(\phi_{\mu}+\phi_2)}P_R\right)$
\end{tabular}

\vskip 5mm

\begin{tabular}{ll}
\begin{minipage}[c]{60mm}
\begin{picture}(80,40)(0,0)
\DashArrowLine(40,20)(0,20){4}
\Text(0,10)[]{\mbox{$E$}}
\Line(37,17)(43,23)
\Line(43,17)(37,23)
\DashArrowLine(40,20)(80,20){4}
\Text(80,10)[l]{\mbox{$E^c$}}
\end{picture}
\end{minipage}
& $-\frac{i}{\sqrt{2}}(v_1 A_e + v_2 \mu^{\star}Y_e)$
\end{tabular}

As a cross check of a correctness of the calculations and a comparison
of exact result with direct calculation of diagrams with mass
insertions one can also expand the exact results in the regime of
almost degenerate sfermion masses and large $\mu$ and $M_{1,2}$.

For sfermion masses we assume that they are almost degenerate and
expand them around the central value ($X=\nu,L,D,U$): 
\bea
m^2_{\tilde{X}_i} &=& m^2_{\tilde{X}} + \delta m^2_{\tilde{X}_i}
\label{eq:sfmassexp}
\eea
We expand also the loop integrals depending on the sfermion masses:
\bea
f(m^2_{\tilde{X}_i}, m^2_{\tilde{X}_j},\ldots)&\approx& 
f(m^2_{\tilde{X}},m^2_{\tilde{X}},\ldots)+
(m^2_{\tilde{X}_i} - m^2_{\tilde{X}}) \left.{\partial
    f(m^2_{\tilde{X}_i}, m^2_{\tilde{X}},\ldots)\over\partial
    m^2_{\tilde{X}_i}}\right|_{m^2_{\tilde{X}_i}=m^2_{\tilde{X}}}\nonumber\\
&+& (m^2_{\tilde{X}_j} - m^2_{\tilde{X}}) \left.{\partial
    f(m^2_{\tilde{X}}, m^2_{\tilde{X}_j},\ldots)\over\partial
    m^2_{\tilde{X}_j}}\right|_{m^2_{\tilde{X}_j}=m^2_{\tilde{X}}} + \ldots
\label{eq:loopexp}
\eea
and use the definitions of the mixing matrices $Z_X$ and their
unitarity to simplify expression containing combinations of the
sfermion mixing angles:
\bea
Z_X^{ik}Z_X^{jk\star} = \delta^{ij} \hskip 2cm 
Z_X^{ik}Z_X^{jk\star}m^2_{\tilde{X}_k} = {\cal M}_{\tilde{X}}^{ij}
\label{eq:zxdef}
\eea

In order to see the direct (although approximate) dependence of the
matrix elements with chargino exchanges involved on the input
parameters such as $\mu$ and $M_2$, one should expand also chargino
masses and mixing matrices. Such an expansion can be done assuming
that $M_2$, $\mu$ and also their difference are of the order of some
scale $|\mu|,|M_2|,|\mu - M_2|\sim\Lambda\gg M_Z$. Actually
corrections to the most physical quantities start from the order
${\cal O}(M_Z^2/\Lambda^2)$, so already $\Lambda\sim 2M_Z$ leads to
reasonably good approximation.

Matrices $Z^-$ and $Z^+$, diagonalizing the chargino mass matrix
(eq.~(\ref{eq:chargmass})) may be chosen as
\bea
Z^+ \approx \left[
\begin{array}{cc}
1 & -\alpha^{\star}\\
\alpha & 1\\
\end{array}
\right]
\hskip 4cm
Z^- \approx \left[
\begin{array}{cc}
e^{-i\phi_2} & -\beta^{\star}e^{-i\phi_{\mu}}\\
\beta e^{-i\phi_2} & e^{-i\phi_{\mu}}\\
\end{array}
\right]
\label{eq:zpmexp}
\eea
where
\bea
\alpha = \frac{e}{\sqrt{2}s_W} {v_2 M_2 + v_1 \mu^{\star} \over |M_2|^2 -
  |\mu|^2}
\hskip 4cm
\beta = \frac{e}{\sqrt{2}s_W} {v_1 M_2 + v_2 \mu^{\star} \over |M_2|^2 -
  |\mu|^2}
\label{eq:zpmab}
\eea
(in the above expressions we do not assume $M_2$ to be real). Phases
in $Z^+,Z^-$ are chosen to keep physical chargino masses $m_{C_1}$,
$m_{C_2}$ in eq.~(\ref{eq:chargmass}) real and positive: physical
results do not depend on this particular choice.

Eqs.~(\ref{eq:chargmass},\ref{eq:zpmexp}) give $|M_2|$ and $|\mu|$ as the
approximate masses for charginos:
\bea
\left(Z^-\right)^T{\cal M}_CZ^+ \approx \left[
\begin{array}{cc}
|M_2| & 0\\
0 &  |\mu|
\end{array}
\right] + M_Z{\cal O}\left({M_Z^2\over \Lambda^2}\right)
\eea

Expansion of the neutralino mass matrix is more complicated and tricky
as two Higgsinos are in the lowest order degenerate in mass. The
appropriate expressions can be found in~\cite{HMR}. In this case
direct mass insertion calculation using the Feynman rules given in
this Appendix is usually easier.

Below we list mass insertion approximation expressions for the
electric and chromoelectric dipole moments of $d$ and $u$ quarks. We
assume equal masses of the left and right sfermion of the first
generation $m_U\approx m_{U^c}\approx m_D\approx m_{D^c}\equiv m_Q$
and neglect small terms proportional to higher powers of the Yukawa
couplings of light quarks. In such a case, corresponding dipole
moments can be approximately written down as:
\bea
E_d &\approx&{eg^2m_d\over (4\pi)^2}\mathrm{Im}(M_2\mu)\tan\beta
\left(2{C_{11}(|\mu|^2,m_Q^2) -C_{11}(|M_2|^2,m_Q^2)
\over |\mu|^2 - |M_2|^2}\right.\nonumber\\
&+&\f{1}{2}\left.{C_{12}(m_Q^2,|\mu|^2) - C_{12}(m_Q^2,|M_2|^2)
\over |\mu|^2 - |M_2|^2}\right)\nonumber\\
&-& {e{g'}^2m_d\over 6(4\pi)^2} \mathrm{Im}(M_1\mu)\tan\beta
{C_{12}(m_Q^2,|\mu|^2) - C_{12}(m_Q^2,|M_1|^2)\over |\mu|^2 - |M_1|^2}
\nonumber\\
&-& {e{g'}^2m_d\over 27(4\pi)^2} \mathrm{Im}\left[M_1(\mu\tan\beta +
A_d^{\star})\right]{\partial C_{12}(m_Q^2,|M_1|^2)\over\partial m_Q^2}
\nonumber\\
&+&{8eg_s^2 m_d\over 9(4\pi)^2}\mathrm{Im}
\left[M_3(\mu\tan\beta + A_d^{\star})\right]{\partial
C_{12}(m_Q^2,|M_3|^2)\over\partial m_Q^2}
\label{eq:edm_d_exp}
\eea
\bea
E_u &\approx&-{2eg^2m_u\over (4\pi)^2}\mathrm{Im}(M_2\mu)\cot\beta
{C_{11}(|\mu|^2,m_Q^2) -C_{11}(|M_2|^2,m_Q^2) \over |\mu|^2 -
|M_2|^2}\nonumber\\
&+& {e{g'}^2m_u\over 3(4\pi)^2} \mathrm{Im}(M_1\mu)\cot\beta
{C_{12}(m_Q^2,|\mu|^2) - C_{12}(m_Q^2,|M_1|^2)\over |\mu|^2 - |M_1|^2}
\nonumber\\
&-& {4e{g'}^2m_u\over 27(4\pi)^2} \mathrm{Im}\left[M_1(\mu\cot\beta +
A_u^{\star})\right]{\partial C_{12}(m_Q^2,|M_1|^2)\over\partial m_Q^2}
\nonumber\\
&-&{16eg_s^2 m_u\over 9(4\pi)^2}\mathrm{Im}
\left[M_3(\mu\cot\beta + A_u^{\star})\right]{\partial
C_{12}(m_Q^2,|M_3|^2)\over\partial m_Q^2}
\label{eq:edm_u_exp}
\eea
Analogously, chromoelectric dipole moments of quarks read
approximately as:
\bea
C_d &\approx&{3g^2g_s m_d\over 2(4\pi)^2}\mathrm{Im}
(M_2\mu)\tan\beta {C_{12}(m_Q^2,|\mu|^2) -C_{12}(m_Q^2,|M_2|^2)
\over |\mu|^2 - |M_2|^2}\nonumber\\
&+& {{g'}^2g_s m_d\over 2(4\pi)^2} \mathrm{Im}(M_1\mu)\tan\beta
{C_{12}(m_Q^2,|\mu|^2) -
C_{12}(m_Q^2,|M_1|^2)\over |\mu|^2 - |M_1|^2}
\nonumber\\
&+& {{g'}^2g_s m_d\over 9(4\pi)^2} \mathrm{Im}\left[M_1(\mu\tan\beta
+ A_d^{\star})\right]{\partial C_{12}(m_Q^2,|M_1|^2)\over\partial
m_Q^2}
\nonumber\\
&+&{2g_s^3 m_d\over (4\pi)^2}\mathrm{Im}
\left[M_3(\mu\tan\beta + A_d^{\star})\right]
\left(3 {\partial
C_{11}(|M_3|^2,m_Q^2) \over \partial m_Q^2} + \frac{1}{6}
{\partial C_{12}(m_Q^2,|M_3|^2) \over \partial m_Q^2}\right)
\label{eq:cdm_d_exp}
\eea
\bea
C_u &\approx&{3g^2g_s m_u\over 2(4\pi)^2}\mathrm{Im}
(M_2\mu)\cot\beta {C_{12}(m_Q^2,|\mu|^2)
-C_{12}(m_Q^2,|M_2|^2) \over |\mu|^2 - |M_2|^2}\nonumber\\
&+& {{g'}^2g_s m_u\over 2(4\pi)^2} \mathrm{Im}(M_1\mu)\cot\beta
{C_{12}(m_Q^2,|\mu|^2) - C_{12}(m_Q^2,|M_1|^2)\over |\mu|^2 - |M_1|^2}
\nonumber\\
&-& {2{g'}^2g_s m_u\over 9(4\pi)^2} \mathrm{Im}\left[M_1(\mu\cot\beta
+ A_u^{\star})\right]{\partial C_{12}(m_Q^2,|M_1|^2)\over\partial
m_Q^2}
\nonumber\\
&+&{2g_s^3 m_u\over (4\pi)^2}\mathrm{Im}
\left[M_3(\mu\cot\beta + A_u^{\star})\right]\left( 3
{\partial C_{11}(|M_3|^2,m_Q^2)\over\partial m_Q^2} +
\frac{1}{6}{\partial C_{12}(m_Q^2,|M_3|^2)\over\partial
m_Q^2}\right)
\label{eq:cdm_u_exp}
\eea

\end{document}